\documentclass[a4paper, aps, prd, showpacs, superscriptaddress, groupedaddress, preprintnumbers, nofootinbib, twocolumn]{revtex4-1} 
\usepackage[dvipdfmx]{graphicx}     
\usepackage{amsmath, amssymb, amscd, latexsym, bm, braket}   
\usepackage{url}
\DeclareMathAlphabet{\mathpzc}{OT1}{pzc}{m}{it}
\newcommand{\Slash}[1]{{\ooalign{\hfil/\hfil\crcr$#1$}}}

\allowdisplaybreaks

\begin{document}
\title{Quark and Gluon Production from a Boost-invariantly Expanding Color Electric Field}
\preprint{UT-KOMABA/16-6}
\author{Hidetoshi Taya}
\email{After this work was completed, the author changed the affiliation.  The current affiliation is {\it iTHEMS Program, RIKEN, Wako 351-0198, Japan} and the current e-mail address is {\tt hidetoshi.taya@riken.jp}.  }
\affiliation{Institute of Physics, The University of Tokyo, Komaba, Tokyo 153-8902, Japan}
\affiliation{Department of Physics, The University of Tokyo, Hongo, Tokyo 113-0033, Japan}

\date{\today}

\begin{abstract}
Particle production from an expanding classical color electromagnetic field is extensively studied, motivated by the early stage dynamics of ultra-relativistic heavy ion collisions.  We develop a formalism at one-loop order to compute the particle spectra by canonically quantizing quark, gluon and ghost fluctuations under the presence of such an expanding classical color background field; the canonical quantization is done in the $\tau$-$\eta$ coordinates in order to take into account manifestly the expanding geometry. As a demonstration, we model the expanding classical color background field by a boost-invariantly expanding homogeneous color electric field with lifetime $T$, for which we obtain analytically the quark and gluon production spectra by solving the equations of motion of QCD non-perturbatively with respect to the color electric field.  In this paper we study (i) finite lifetime effect which is found to modify significantly the particle spectra from those expected from the Schwinger formula; (ii) the difference between the quark and gluon production; and (iii) the quark mass dependence of the production spectra.  Implications of these results to ultra-relativistic heavy ion collisions are also discussed.  
\end{abstract}


\maketitle

\section{Introduction}

Early stage dynamics of ultra-relativistic heavy ion collisions (HIC) is a big missing piece in our current understanding of the spacetime evolution of HIC: Before a collision, two incident nuclei at very high energies are saturated with a huge number of gluons, which behave like coherent classical color fields (Color Glass Condensate picture; CGC \cite{mcl94a, mcl94b, mcl94c, kov96}) rather than incoherent particles.  A collision of these classical non-Abelian fields results in a formation of longitudinal color electromagnetic fields between the two nuclei receding from each other \cite{kov95a, kov95b, lap06}.  The strength of the longitudinal fields are very strong as $gA_{\mu} \sim Q_{\rm s} \sim {\rm a\ few\ GeV}$, where $Q_{\rm s}$ is the so-called saturation scale of CGC.  Subsequently, the color fields would decay into a huge number of particles (quarks and gluons) to form a Quark-Gluon Plasma (QGP).  However, this stage of non-equilibrium dynamics is not well understood --- the questions are (a) how the huge number of quark and gluon {\it particles} are produced from the classical gluon {\it fields} (experimentally known is that about 1000 hadrons are produced per unit rapidity), and (b) how the system thermalizes to eventually form a QGP, which behaves almost like a perfect liquid as suggested by the success of hydrodynamical models (for reviews, see \cite{kol03, gal13} for example).  In particular, applications of hydrodynamical models assume that the formation time of QGP is extremely short $\tau_{\rm form} \lesssim 1\;{\rm fm}/c$ \cite{hei05, ada10a, ada10b}.  There is no satisfactory understanding of such a short formation time starting from QCD, despite of numerous theoretical attempts.  Thus, unveiling the early stage dynamics is not only an important piece for completing our understanding of the whole spacetime evolution of HIC, but also a challenge to non-equilibrium QCD physics.

The purpose of this paper is to investigate the quark and gluon production from expanding classical color electromagnetic fields starting from QCD.

Study of the particle production from classical electromagnetic fields has a long history in Quantum ElectroDynamics (QED).  Sauter \cite{sau31} was the first who claimed that spontaneous particle (electron and positron pair) production occurs when a system is exposed to strong classical electromagnetic fields.  Some years later this particle production mechanism was theoretically formulated by Heisenberg and Euler \cite{hei36}, and by Schwinger \cite{sch51} for a static and homogeneous electric field.  They derived the vacuum persistency probability ${\mathcal P} = | \braket{\rm vac;in | vac;out } |^2 $, from which one can deduce the average number of particles produced at transverse and longitudinal momenta ${\bm p}_\perp$ and $p_z$ with respect to the electric field as \cite{nik70b} 
\begin{align}
	\frac{d^3 N^{(e^-)} }{ d^2{\bm p}_{\perp} dp_z } = \frac{d^3 N^{(e^+)} }{ d^2{\bm p}_{\perp} dp_z } = \frac{V}{(2\pi)^3} \exp \left[ -\pi \frac{ m_{e}^2 +{\bm p}_{\perp}^2}{|eE|}  \right], \label{eqq1}
\end{align}
where $m_{e}$ is the electron mass, $e$ is the coupling constant of QED, $E$ is the electric field strength, and $V$ is the system volume.  The formula Eq.~(\ref{eqq1}), often called the {\it Schwinger formula}, depends on $eE$ inversely in the exponential, and hence one can understand that the particle production from a static electric field is a non-perturbative phenomenon.  This is in contrast to usual perturbative phenomena, whose dependence on $eE$ always appears with positive powers.

The Schwinger formula was generalized to QCD case \cite{amb83, yil80, bat77, gyu85}, and then applied to early stage phenomenology of HIC, e.g. the color flux tube model \cite{cas79, cas80, kaj85, gat87, rug15a, rug15b}.  However, these preceding studies may be problematic because the situation in HIC is much more complicated than the static and homogeneous field which the Schwinger formula assumes.  Thus, the particle production will be different from the naive estimate of the Schwinger formula and therefore one needs to formulate the particle production in such a more dynamic situation starting from the first principle, i.e., QCD.

In particular, we consider the following effects on particle production in QCD, which are missing in previous studies:

(i) {\it Effects of longitudinal expansion}: In HIC, two highly Lorentz-contracted nuclei pass through each other at almost the speed of light, and color electromagnetic fields are formed between the two receding nuclei with approximate boost-invariance in the beam direction (Bjorken expansion \cite{bjo83}).  Here, the longitudinal extent of the fields is finite and increasing with time, which is obviously very different from that the Schwinger formula assumes.  Hence, the applicability of the Schwinger formula must be reconsidered, and one has to deal with particle production from space- and time-dependent color electromagnetic fields.  Recently, there is progress in a theoretical treatment of particle production from such an expanding electromagnetic field within scalar QED by Tanji \cite{tan11}.  We will extend this study to the case of quark and gluon production from expanding color electric fields in QCD.

(ii) {\it Finite lifetime effects}: The color electromagnetic fields decay in time according to the classical Yang-Mills equation.  The typical scale of their lifetime is very short, whose order would be given by the inverse of the saturation scale $1/Q_{\rm s}$ \cite{lap06}.  Such a short lifetime of the fields should significantly affect the particle production mechanism.  Indeed, for a non-expanding electric field, Refs.~\cite{bre70, tay14} have shown that there is an interplay between perturbative particle production at shorter lifetimes and Schwinger's non-perturbative particle production at longer lifetimes.  As a result, the particle spectra will heavily depend on the lifetime of the fields; in particular, production of heavy particles, such as charm quarks, from a pulse field is significantly enhanced compared to the value of the Schwinger formula \cite{tay14, lev10}.  It is thus phenomenologically important to understand finite lifetime effects on particle production.  No studies have paid much attention to them so far, though there are several studies which discussed particle production from an expanding (color) electromagnetic field in QED \cite{coo93, tan11, mih08, mih09} and in QCD (but quark production only) \cite{gel06}.

In order to examine the above-mentioned points, we study quark and gluon production from a given homogeneous classical color electric background field applied for finite duration (lifetime) from $\tau=0$ to $T$ with longitudinally expanding geometry.  We solve mode equations for fluctuations non-perturbatively with respect to the classical field, and compute the Bogoliubov coefficients among creation/annihilation operators at asymptotic times ($t \rightarrow \pm\infty$).  We ignore backreaction from produced particles on the electric field, and fix the electric field strength constant during its lifetime.  For the sake of clarity, we ignore here a possible existence of color magnetic fields, which may bring interesting effects including the chiral magnetic effect.  Effects of the backreaction and of scatterings between produced particles will be decisive for thermalization of the system, but we leave it for our future study.

This paper is organized as follows: In Section \ref{sec2}, the general formalism for particle production from classical color electromagnetic fields employed in this work is explained.  Our formalism is based on a canonical quantization under the presence of classical color background fields \cite{nik70, tan09, amb83}, where a non-expanding system was treated.  We extend it to quark and gluon production in an expanding system by following Ref.~\cite{tan11}.  In Section \ref{sec3}, we model the classical field by a boost-invariantly expanding homogeneous color electric field with lifetime $T$ as a demonstration of our formalism.  In such a field configuration, one can analytically obtain quark and gluon production spectra and can investigate physical consequences of the longitudinal expansion and the finite lifetime effects in detail.  We also discuss some implications to the early stage dynamics of HIC of these results.  Section \ref{sec4} is devoted to a summary and an outlook of this work.  In Appendix \ref{appA}, details of analytical solutions of equations of motion of QCD are presented.

\section{General Formalism} \label{sec2}

Let us explain the general formalism employed in this work for particle production from a boost-invariant classical gauge field in QCD.  We consider a classical background field satisfying the classical Yang-Mills equation and quantum fluctuations of quark, gluon, and ghost around the classical field.  By assuming that the Abelian dominance holds for the classical field, we linearize equations of motion for fluctuations and solve them non-perturbatively with respect to the classical field.  Then, we adopt a canonical quantization procedure in the $\tau$-$\eta$ coordinates, instead of in the Cartesian coordinates, in order to treat the boost-invariant expansion of the system properly.  Thereby, we directly compute expectation values of number operators of quark, gluon, and ghost.

We work in the Heisenberg picture throughout this paper.  We implicitly take summation over repeated indices $m,n,\ldots$ and $\mu,\nu,\ldots$ for {\it spacetime only}, and not for other repeated indices, for instance, color labels $a,b,\ldots$, spin labels $s, s', \ldots$, and so on.

\subsection{$\tau$-$\eta$ Coordinates} \label{tau-eta}

Let us begin with a brief review on the $\tau$-$\eta$ coordinates.  It is very convenient to work in the $\tau$-$\eta$ coordinates $x^{\mu} = (\tau, x, y, \eta)$, instead of the usual Cartesian coordinates $\xi^m = (t, x, y, z)$, in order to treat the boost-invariant expansion of the system properly.  We use Latin (Greek) indices $m,n,\ldots$ ($\mu, \nu, \ldots$) for the Cartesian ($\tau$-$\eta$) coordinates throughout this paper.

The $\tau$-$\eta$ coordinates are defined by the following change of variables
\begin{align}
		\tau = \sqrt{t^2 - z^2},\ \eta = \frac{1}{2} {\rm ln} \frac{t + z}{ t - z}. \label{eq_1}
\end{align}
The line element $ds^2$ is then expressed as 
\begin{align}
	ds^2 = \eta_{mn} d\xi^m d\xi^n = g_{\mu\nu} dx^{\mu}dx^{\nu}, 
\end{align}
where 
\begin{align}
	\eta_{mn} 	&= {\rm diag}(1, -1, -1, -1), \\
	g_{\mu\nu} 	&= {\rm diag}(1, -1, -1, -\tau^2)
\end{align}
are the metric of the Cartesian coordinates and the $\tau$-$\eta$ coordinates, respectively.

For later discussions, it is convenient to introduce a {\it viervein} matrix $e^m_{\ \ \mu}$ \cite{Birrell_Davies}, which relates the Cartesian coordinates $\xi^m$ and the $\tau$-$\eta$ coordinates $x^{\mu}$ as  
\begin{align}
	d\xi^m 	=  e^m_{\ \ \mu} dx^{\mu} 
\end{align}
with
\begin{align}
	e^m_{\ \ \mu}	
			\equiv \frac{d\xi^m}{dx^{\mu}} 
			= \begin{pmatrix} 
				\cosh \eta 	& 0 & 0 & \tau \sinh \eta 	\\ 
				0			& 1	& 0 & 0					\\
				0			& 0	& 1	& 0					\\				
				\sinh \eta	& 0	& 0	& \tau \cosh \eta
			  \end{pmatrix}.  \label{eq6}
\end{align}
The inverse matrix of $e^{m}_{\ \ \mu}$, which we write $e^{\mu}_{\ m}$, is  
\begin{align}
	e^{\mu}_{\ m}	
			\equiv \frac{dx^{\mu }}{d\xi^{m}} 
			= \begin{pmatrix} 
				\cosh \eta 					& 0 & 0 & -\sinh \eta 	\\ 
				0							& 1	& 0 & 0					\\
				0							& 0	& 1	& 0					\\				
				\displaystyle -\frac{\sinh \eta}{\tau}	& 0	& 0	& \displaystyle \frac{\cosh \eta}{\tau} 
			  \end{pmatrix} = \eta_{mn} g^{\mu\nu} e^{n}_{\ \ \nu} .
\end{align}
With the viervein matrix introduced above, one can define a vector $X^{\mu}$ in the $\tau$-$\eta$ coordinates for any vector $X^m$ in the Cartesian coordinates as
\begin{align}
	X_{\mu} &\equiv e^{m}_{\ \ \mu} X_{m}, \label{eq33}\\
	X^{\mu}	&\equiv e^{\mu}_{\ m} X^{m} = g^{\mu\nu} X_{\nu}. \label{eq34}
\end{align}
From these definitions, Eqs.~(\ref{eq33}) and (\ref{eq34}), one readily finds, for example, 
\begin{align}
	\partial_{\tau} &= \cosh \eta \;\partial_t + \sinh \eta \;\partial_z,\\ 
	\partial_{\eta} &= \tau \sinh \eta \;\partial_t + \tau \cosh \eta \;\partial_z 
\end{align}
for derivatives $\partial_{\mu}$, 
\begin{align}
	\gamma^{\tau} &= \gamma^t \cosh \eta - \gamma^z \sinh \eta,
\label{eq13}
\\
	\gamma^{\eta} &= -\gamma^t \frac{\sinh \eta}{\tau} + \gamma^z \frac{\cosh \eta}{\tau}
\label{eq14}
\end{align}
for gamma matrices $\gamma^{\mu}$, 
\begin{align}
	A_{\tau} &=  A_t \cosh \eta + A_z \sinh \eta,\\
	A_{\eta} &=  A_t \tau \sinh \eta + A_z \tau \cosh \eta 
\end{align}
for vector fields $A_{\mu}$.  One can also generalize these definitions, Eqs.~(\ref{eq33}) and (\ref{eq34}), to general tensors as $X^{\mu \cdots}_{\ \ \ \ \nu \cdots} = e^{\mu}_{\ m} \cdots e^{n}_{\ \ \nu}\cdots X^{m \cdots}_{\ \ \ \ \ n \cdots}$

We also introduce a covariant derivative $\nabla_{\mu}$ for curvilinear coordinates, $\nabla_{\mu} T^{\nu \cdots}_{\ \ \ \ \rho \cdots} = \partial_{\mu} T^{\nu \cdots}_{\ \ \ \ \rho \cdots} + \Gamma^{\nu}_{\mu \lambda} T^{\lambda \cdots}_{\ \ \ \ \rho \cdots} + \cdots - \Gamma^{\lambda}_{\mu \rho} T^{\nu \cdots}_{\ \ \ \ \lambda \cdots} - \cdots$.  Here, $\Gamma^{\mu}_{\nu \rho}$ is the Christoffel symbol, whose non-zero elements in the $\tau$-$\eta$ coordinates are
\begin{align}
	\Gamma^{\eta}_{\eta \tau} = \Gamma^{\eta}_{\tau \eta} = 1/\tau, \ \Gamma^{\tau}_{\eta\eta} &= \tau.  
\end{align}

\subsection{Classical Background Field} \label{sec:background}

We consider a classical background field $\bar{A}_{\mu}$ satisfying the SU($N_{\rm c}$) classical Yang-Mills equation with an external classical source $\bar{J}^{\mu}$ as
\begin{align}
	\bar{J}^{\nu} = \bar{D}_{\mu} \bar{F}^{\mu\nu}.   \label{eq0}
\end{align}
Here, $\bar{D}_{\mu}$ is the covariant derivative with respect to the classical field $\bar{A}_{\mu}$, i.e., $\bar{D}_{\mu} = \nabla_{\mu} + ig [ \bar{A}_{\mu} , ~]$, and $\bar{F}^{\mu\nu}$ is the classical field strength tensor $\bar{F}^{\mu\nu} = \partial^{\mu} \bar{A}^{\nu} - \partial^{\nu} \bar{A}^{\mu} + ig [ \bar{A}^{\mu}, \bar{A}^{\nu} ]$.  Equation (\ref{eq0}) does not fix the gauge completely and there still remains a residual gauge freedom.  In the following discussion, we fix the residual gauge freedom by $\bar{A}_{\tau} = 0$ (temporal gauge), which is convenient for the canonical quantization procedure we adopt in Section \ref{sec:quantization}.  As a boundary condition of Eq.~(\ref{eq0}), we require that $\bar{A}_{\mu}$ becomes a pure gauge $\bar{A}_{\mu} = {\rm const.}$ at the asymptotic times ($t \rightarrow \pm\infty$), i.e., we assume that there is no external source $\bar{J}^{\mu}$ nor classical color electromagnetic field at the asymptotic times.

As we will see in Section \ref{sec:abelianization}, in order to ease some difficulties coming from the non-Abelian nature of QCD, we furthermore assume that the color direction of the classical source $\bar{J}^{\mu}$ and the classical field $\bar{A}_{\mu}$ is constant, i.e., it is independent of spacetime coordinates $x$ and the spacetime vector index $\mu$.  For this case, there always exists a constant color vector $n^a$ such that
\begin{align}
	\bar{A}_{\mu}(x) = \tilde{A}_{\mu}(x) \sum_{a=1}^{N_{\rm c}^2-1} n^a t^a.  \label{eq_20}
\end{align}
Here, $\tilde{A}_{\mu}$ is a scalar in the color space.  The matrix $t^a$ ($a=1,\cdots,N_{\rm c}^2-1$) is a generator of SU($N_{\rm c}$), and $n^a$ (normalized as $\sum_{a=1}^{N_{\rm c}^2-1} n^a n^a=1$) characterizes the color direction of the classical field $\bar{A}_{\mu}$.  Under this assumption, the commutators of $\bar{A}_{\mu}$ exactly vanish as $[\bar{A}_{\mu}, \bar{A}_{\nu}]=0$, and only the Abelian part of the classical field strength $\bar{F}_{\mu\nu}$ becomes nonvanishing as
\begin{align}
	\bar{F}_{\mu\nu} = (\partial_{\mu} \tilde{A}_{\nu} - \partial_{\nu} \tilde{A}_{\mu})  \sum_{a=1}^{N_{\rm c}^2-1} n^a t^a \equiv \tilde{F}_{\mu\nu}  \sum_{a=1}^{N_{\rm c}^2-1} n^a t^a. 
\end{align}
Thus, our assumption is essentially the same as the Abelian dominance assumption: $[\bar{A}_{\mu}, \bar{A}_{\nu}] \sim 0$ and $\bar{F}_{\mu\nu} \sim \partial_{\mu} \bar{A}_{\nu} - \partial_{\nu} \bar{A}_{\mu}$.  

Notice that we have made no restrictions on the spacetime $x^{\mu}$-dependence of $\tilde{A}_{\mu}$ as long as it satisfies the classical Yang-Mills equation Eq.~(\ref{eq0}).

\subsection{Lagrangian}

Let us consider the QCD Lagrangian with $N_{\rm c}$ colors and $N_{\rm f}$ flavors of quarks in the presence of the classical background field $\bar{A}_{\mu}$ described in Section \ref{sec:background}.  By separating the (total) gauge field $A_{\mu}$ into the classical field $\bar{A}_{\mu}$ and quantum fluctuations around it ${\mathcal A}_{\mu}$ as $A_{\mu} = \bar{A}_{\mu} + {\mathcal A}_{\mu}$, we obtain the QCD Lagrangian for the fluctuation in the $\tau$-$\eta$ coordinates as%
\footnote{In general curved spacetime coordinates, there is an additional term coming from spin connections $\Gamma_{\mu}$ in the fermion covariant derivative, which is zero in the $\tau$-$\eta$ coordinates. }
\begin{align}
	{\mathcal L} 
		&= \bar{\psi} \left[  i \Slash{\partial} - g\Slash{A}  - M \right] \psi -\frac{1}{2} {\rm tr}_{\rm c}F_{\mu \nu} F^{\mu \nu} + 2{\rm tr}_{\rm c} \bar{J}^{\mu} A_{\mu} \nonumber\\
		&\quad - \frac{1}{\alpha}{\rm tr}_{\rm c} ( \bar{D}_{\mu} {\mathcal A}^{\mu} )^2  -  2i{\rm tr}_{\rm c}(\bar{D}^{\mu} \bar{c})(D_{\mu} c).  \label{eq1}
\end{align}
Here, $\psi$ is the fermion field, and $c$ and $\bar{c}$ are the ghost and  anti-ghost fields to be quantized.  $\Slash{X} \equiv \gamma^{\mu} X_{\mu}$ is the Feynman slash notation, and $\bar{\psi}$ is a shorthand for $\bar{\psi} \equiv \psi^{\dagger} \gamma^t$.  ${\rm tr}_{\rm c}$ is the trace operator in the color space.  $M$ represents fermion masses, which is given by an $N_{\rm f} \times N_{\rm f}$ diagonal matrix $M = {\rm diag}(m_1, m_2, \ldots, m_{N_{\rm f}})$ in the flavor space.  $D_{\mu}$ is the covariant derivative with respect to the total gauge field $A_{\mu}$: $D_{\mu} = \nabla_{\mu} + ig [ A_{\mu} , ~]$.  The total field strength tensor $F_{\mu \nu}$ is given by $F_{\mu\nu} = \partial_{\mu} A_{\nu} - \partial_{\nu} A_{\mu} + ig[ A_{\mu} , A_{\nu} ]$.  The term $(1/\alpha){\rm tr}_{\rm c}(\bar{D}_{\mu} {\mathcal A}^{\mu})^2$ is a covariant background gauge fixing term \cite{amb83}.  Hereafter, we shall take $\alpha=1$ for simplicity.  One can show that a choice of the gauge parameter $\alpha$ is irrelevant to the particle spectra \cite{coo06}.

We further expand the Lagrangian Eq.~(\ref{eq1}) up to the quadratic order in the quantum fluctuations to obtain 
\begin{align}
	{\mathcal L} =& \bar{\psi} [ i\Slash{\partial} -  g\bar{\Slash{A}} - M  ]\psi - 2i {\rm tr}_{\rm c} (\bar{D}_{\mu} \bar{c})(\bar{D}^{\mu} c) \nonumber\\
				 &- {\rm tr}_{\rm c} \left[ \frac{1}{2} ( \bar{D}_{\mu} {\mathcal A}_{\nu} - \bar{D}_{\nu} {\mathcal A}_{\mu} )^2 + ( \bar{D}_{\mu} {\mathcal A}^{\mu} )^2 + 2ig \bar{F}_{\mu \nu} {\mathcal A}^{\mu} {\mathcal A}^{\nu} \right],  \label{eq3}
\end{align}
where constant and surface terms are omitted.  Here, we treat the interactions with the classical field $\bar{A}_{\mu}$ non-perturbatively.  This treatment is justified when the quantum fluctuations $\psi, {\mathcal A}_{\mu}, c$ and $\bar{c}$ are small enough compared to the strength of the classical field $\bar{A}_{\mu}$.  The ignored terms ${\mathcal O}(g \bar{\psi} {\mathcal A} \psi, g {\mathcal A}^3, g \bar{c} c {\mathcal A})$ are responsible for screening of the classical field $\bar{A}_{\mu}$ by produced particles and elastic ${\rm gg} \leftrightarrow {\rm gg}$ and inelastic ${\rm g} \leftrightarrow {\rm gg}, {\rm g} \leftrightarrow {\rm q \bar{q}}, {\rm q} \leftrightarrow {\rm qg}$ scattering processes of produced particles.  It is very interesting to see how the quark and/or gluon production is modified when these higher order quantum corrections are included; see Section \ref{sec4} for the discussion.  We also note that the classical source $\bar{J}^{\mu}$ does not directly couple to the quantum fluctuations; it couples to them only indirectly through the classical field $\bar{A}_{\mu}$ which is generated by the classical Yang-Mills equation sourced by $\bar{J}_{\mu}$ (Eq.~(\ref{eq0})).  In this sense, particle production mechanism is not directly affected by the presence of the classical source $\bar{J}^{\mu}$.

\subsection{Abelianization} \label{sec:abelianization}

It is difficult to handle the Lagrangian Eq.~(\ref{eq3}) as it is because of its non-Abelian nature.  Indeed, the equation of motion of the Lagrangian Eq.~(\ref{eq3}) are complicated matrix equations in the color space.  With the help of the Abelian dominance assumption for the classical field $\bar{A}_{\mu}$ made in Section \ref{sec:background}, one can Abelianize, i.e., diagonalize the Lagrangian Eq.~(\ref{eq3}) in the color space and obtain a set of Abelian equations of motion as below \cite{gyu85}:   

Firstly, we diagonalize the classical field $\bar{A}_{\mu} = \tilde{A}_{\mu}  \sum_{a=1}^{N_{\rm c}^2-1} n^a t^a$ in the color space.  Since $ \sum_{a=1}^{N_{\rm c}^2-1} n^a t^a$ is a constant hermitian matrix in the color space, there always exists a {\it global} unitary transformation $U$ which diagonalizes $n^a t^a$ as 
\begin{align}
	 \sum_{a=1}^{N_{\rm c}^2-1} n^a t^a \rightarrow U^{-1} \left( \sum_{a=1}^{N_{\rm c}^2-1} n^a t^a  \right) U = \sum_{\alpha = 1}^{N_{\rm c}-1} w^{\alpha} H^{\alpha}, \label{eq_10}  
\end{align}
where $w^{\alpha}$ is constant normalized as $1 = \sum_{\alpha} | w^{\alpha} |^2$.  $H^{\alpha}$ is a diagonal matrix which belongs to the Cartan subalgebra of SU($N_{\rm c}$) such that $[H^{\alpha}, H^{\beta}]=0$ with a normalization ${\rm tr}_{\rm c}[ H^{\alpha} H^{\beta} ] = \delta^{\alpha \beta}/2$.  In accordance with this transformation $U$, let us also redefine the quantum fluctuations $\psi, {\mathcal A}_{\mu}, c$ and $\bar{c}$ as
\begin{align}
	U^{\dagger} \psi &\rightarrow \psi, \\
	U^{\dagger} {\mathcal A}_{\mu} U &\rightarrow {\mathcal A}_{\mu}, \\
	U^{\dagger} \begin{pmatrix} c \\ \bar{c} \end{pmatrix} U &\rightarrow \begin{pmatrix} c \\ \bar{c} \end{pmatrix}.  
\end{align}

Secondly, we expand the color space by the Cartan-Weyl basis of SU($N_{\rm c}$): $\{ H^{\alpha}, E^{\pm A} \}$ ($\alpha = 1,\ldots, N_{\rm c}-1$; $A = 1,\ldots,N_{\rm c}(N_{\rm c}-1)/2$), where $E^{A}$ is an off-diagonal matrix satisfying the following algebra: 
\begin{align}
	E^{A \dagger} 						&= E^{-A}, \\
	{\rm tr}[E^{A} E^{B \dagger}] 	&= \frac{\delta^{AB}}{2}, \\
	[ H^{\alpha}, E^{\pm A} ] 			&= \pm (v^{\alpha})^A E^{\pm A}, 
\end{align}
where $(v^{\alpha})^A$ is the root vector of SU($N_{\rm c}$).  By using this Cartan-Weyl basis, instead of the generator $t^a$, we expand the gluon field ${\mathcal A}_{\mu}$, and ghost and anti-ghost field $c$ and $\bar{c}$ as (Cartan decomposition): 
\begin{widetext}
\begin{align}
	{\mathcal A}_{\mu} 
				&\equiv \sum_{\alpha=1}^{N_{\rm c}-1} {\mathcal W}_{\mu ,\alpha} H^{\alpha} + \sum_{A=1}^{\frac{N_{\rm c}(N_{\rm c}-1)}{2}} \left[ W_{\mu, A} E^{+A} + W_{\mu, A}^{\dagger} E^{-A}  \right], \\
	\begin{pmatrix} c \\ \bar{c} \end{pmatrix}	
				&\equiv \sum_{\alpha=1}^{N_{\rm c}-1} \begin{pmatrix} {\mathcal C}_{\alpha} \\ \bar{\mathcal C}_{\alpha} \end{pmatrix} H^{\alpha} + \sum_{A=1}^{\frac{N_{\rm c}(N_{\rm c}-1)}{2}} \left[ \begin{pmatrix} C_A \\ \bar{C}_A \end{pmatrix} E^{+A} + \begin{pmatrix} C_A^{\dagger} \\ \bar{C}_A^{\dagger} \end{pmatrix} E^{-A}  \right].
\end{align}

After completing these two steps, one can rewrite the Lagrangian Eq.~(\ref{eq3}) in an Abelianized form as
\begin{align}
	{\mathcal L} =& \sum_{f=1}^{N_{\rm f}}\sum_{i=1}^{N_{\rm c}} \bar{\psi}_{i,f} [ i\Slash{\partial} - q^{({\rm q})}_i \tilde{\Slash{A}}  - m_f ] \psi_{i,f} - \sum_{\alpha=1}^{N_{\rm c}-1} \frac{1}{4} \left| \nabla_{\mu} {\mathcal W}_{\nu, \alpha} -  \nabla_{\nu} {\mathcal W}_{\mu, \alpha} \right|^2  - i\sum_{\alpha = 1}^{N_{\rm c}-1} (\nabla_{\mu} \bar{\mathcal C}_{\alpha} )( \nabla^{\mu} {\mathcal C}_{\alpha} ) \nonumber\\
				 &-     \sum_{A=1}^{\frac{N_{\rm c}(N_{\rm c}-1)}{2}} \left[  \frac{1}{2} \left| (\nabla_{\mu} + iq^{({\rm g})}_{A} \tilde{A}_{\mu}) W_{\nu, A} -  (\nabla_{\nu} + iq^{({\rm g})}_{A} \tilde{A}_{\nu}) W_{\mu, A} \right|^2  + \left| (\nabla_{\mu} + iq^{({\rm g})}_{A} \tilde{A}_{\mu}) W^{\mu}_A \right|^2 + iq^{(\rm g)}_{A} \tilde{F}_{\mu\nu} W^{\mu}_A W^{\nu \dagger}_A  \right]  \nonumber\\
				 &  -i \sum_{A=1}^{\frac{N_{\rm c}(N_{\rm c}-1)}{2}} \left[   \left( (\nabla_{\mu} + iq^{({\rm gh})}_{A} \tilde{A}_{\mu}) \bar{C}_{A} \right) \left((\nabla^{\mu} + iq^{({\rm gh})}_{A} \tilde{A}^{\mu}) C_{A} \right)^{\dagger}  + \left( (\nabla_{\mu} + iq^{({\rm gh})}_{A} \tilde{A}_{\mu}) \bar{C}_{A} \right)^{\dagger} \left( (\nabla^{\mu} + iq^{({\rm gh})}_{A} \tilde{A}^{\mu}) C_{A} \right)  \right].  \label{eq19}
\end{align}
\end{widetext}
Here, the color indices $i,j,\ldots$ and the flavor indices $f,f',\ldots$ for the quark field $\psi$ are explicitly written.  The gluon ${\mathcal W}_{\mu, \alpha}$, ghost ${\mathcal C}_{\alpha}$ and anti-ghost $\bar{{\mathcal C}}_{\alpha}$ fields, which belong to the Cartan subalgebra of SU($N_{\rm c}$) do not couple to the classical field $\bar{A}_{\mu}$.  Thus, no particle production occurs for these fluctuations, and hence we do not consider them hereafter.  On the other hand, the quark $\psi_{i,f}$, gluon $W_{\mu, A}$, and ghost $C_A$ and anti-ghost $\bar{C}_A$ fields do couple to the classical field $\bar{A}_{\mu}$, whose effective color charges, $q^{({\rm q})}_{i}$, $q^{({\rm g})}_{A}$ and $q^{({\rm gh})}_{A}$, respectively, are given by
\begin{align}
	q^{({\rm q})}_{i} &=	g \sum_{\alpha=1}^{N_{\rm c}-1} w^{\alpha} (H^{\alpha})_{ii}, \\
	q^{({\rm g})}_{A} = q^{({\rm gh})}_{A}&=	g \sum_{\alpha=1}^{N_{\rm c}-1}  w^{\alpha} (v^{\alpha})^A.  
\end{align}
The ghost charge is identical to the gluon charge $q^{({\rm gh})}_{A} = q^{({\rm g})}_{A}$ because both gluon $W_{\mu, A}$ and ghost $C_A, \bar{C}_A$ fields belong to the adjoint representation of SU($N_{\rm c}$).  Although the effective color charges, $q^{({\rm q})}_{i}$, $q^{({\rm g})}_{A}$ and $q^{({\rm gh})}_{A}$, depend on the color direction $n^a$ and the gauge-choice of the background field $\bar{A}_{\mu}$, the traces of the squared charges are independent of them: 
\begin{align}
	\sum_{i=1}^{N_{\rm c}} | q^{({\rm q})}_{i} |^2 &= \frac{g^2}{2}, \label{eq--36} \\
	\sum_{A=1}^{ \frac{N_{\rm c}(N_{\rm c}-1)}{2}} | q^{({\rm g})}_{A} |^2 = \sum_{A=1}^{ \frac{N_{\rm c}(N_{\rm c}-1)}{2}} | q^{({\rm gh})}_{A} |^2 &= \frac{g^2N_{\rm c}}{2}.  
\label{eq--35}
\end{align}
The trace of the squared charge in the adjoint representation is $N_{\rm c}$ times as large as that in the fundamental representation.  These relations are generalization of the SU(3) results \cite{coo06, nay05a, nay05b, tan10}.

One readily obtains Abelianized equations of motion from the Lagrangian Eq.~(\ref{eq19}).  They read
\begin{align}
	[ i \Slash{\partial} - q^{({\rm q})}_{i} \tilde{\Slash{A}} - m_f ] \psi_{i,f} &= 0, \label{eq46}\\
	\left[ ( \nabla_{\rho} + iq^{({\rm g})}_{A} \tilde{A}_{\rho} )^2 g^{\mu\nu} + 2iq^{({\rm g})}_{A} \tilde{F}^{\mu\nu}  \right] W_{\nu, A} &= 0, \label{eq48}\\  
	( \nabla_{\nu} + iq^{({\rm gh})}_{A} \tilde{A}_{\nu} )^2  \begin{pmatrix} C_A \\ \bar{C}_A \end{pmatrix} &= 0. 	\label{eq50}  
\end{align}

\subsection{Quantization and particle spectrum} \label{sec:quantization}

Now, we canonically quantize the fluctuations, $\psi_{i,f}, W_{\mu, A}, C_A$ and $\bar{C}_A$, under the classical background field $\bar{A}_{\mu}$, and compute particle spectra produced from the classical field.

To be more concrete, we first define positive/negative mode functions at the asymptotic times ($t \rightarrow \pm\infty$) for the fluctuations.  At the asymptotic times, as the classical field $\bar{A}_{\mu}$ becomes merely a pure gauge and no interaction occurs (see the assumptions made in Section \ref{sec:background}), one can {\it uniquely}%
\footnote{
One can quantize the fluctuations even if there are interactions in principle, however, the definition of positive/negative mode functions, i.e., the notion of particle becomes ambiguous.  
}
define the positive/negative frequency mode functions at the corresponding asymptotic time by plane wave solutions.  With this boundary condition at $t \rightarrow \pm\infty$, we solve the equations of motion, Eqs.~(\ref{eq46})-(\ref{eq50}), non-perturbatively with respect to the classical field and hereby we obtain the positive/negative mode functions at the corresponding asymptotic time.  By expanding the fluctuations with the mode functions and imposing canonical commutation relations, one obtains creation/annihilation operators for the positive/negative frequency modes at each asymptotic time ($t \rightarrow \pm\infty$).  An important point here is that the mode functions do fully include multiple interactions with the classical field and hence the positive (or negative) frequency mode at $t \rightarrow -\infty$ will evolve into a linear combination of the positive and negative frequency modes at $t \rightarrow \infty$.  This linear relation is described by a Bogoliubov transformation, and we will see that the particle spectrum which will be observed at $t \rightarrow \infty$ evolved from a given initial state at $t \rightarrow -\infty$ is determined by the Bogoliubov coefficients.  In the following, we shall assume that the initial state is given by a vacuum for simplicity, although one can equally formulate more generic initial states as well.

We remark that our formalism, which takes into account the interactions with the classical field $\bar{A}_{\mu}$ non-perturbatively by fully solving the equations of motion, does include perturbative contributions which can be computed by, for instance, the usual diagrammatic techniques of the $S$-matrix \cite{fuk09}.  For a specific type of electric fields, one can explicitly check this \cite{tay14, gel16}.

\subsubsection{Quark}
We canonically quantize the quark field $\psi_{i,f}$ at the asymptotic times ($t \rightarrow \pm\infty$) in order to compute the quark spectrum produced from the classical field.

To do this, we first expand the quark fields $\psi_{i,f}$ with the mode functions as
\begin{align}
	\psi_{i,f}(x) 	
		&= \sum_s \int d{\bm p}_{\perp}^2 dp_{\eta} \nonumber\\
		&\ \ \ \ \ \left[ {}_+ \psi^{({\rm as})}_{i,f,{\bm p}_{\perp},p_{\eta},s}(x) a^{({\rm as})}_{i,f,{\bm p}_{\perp},p_{\eta},s}  \right.\nonumber\\
		&\left. \ \ \ \ \ \ \ \ \ \ \ \ \ + {}_- \psi^{({\rm as})}_{i,f,{\bm p}_{\perp},p_{\eta},s}(x)   b^{({\rm as})\dagger}_{i,f,-{\bm p}_{\perp},-p_{\eta},s}   \right]. \label{eq__43}
\end{align}
Here, ${\rm as}={\rm in,out}$ specifies the asymptotic time $t \rightarrow \pm\infty$, respectively, at which we define a particle picture by employing the canonical quantization.  The subscripts $\pm$ specify the positive and the negative frequency modes.  The momentum labels ${\bm p}_{\perp}$ and $p_{\eta}$ are the Fourier conjugate to the positions ${\bm x}_{\perp}$ and $\eta$, respectively; we label the longitudinal momentum by $p_{\eta}$, instead of $p_z$ conjugate to $z$, so as to treat the longitudinal expansion of the system manifestly with $\eta$ coordinate.   The label $s = 1,2$ is for the spin degree of freedom.  We identify the mode functions ${}_{\pm} \psi_{i,f,{\bm p}_{\perp},p_{\eta}, s}^{({\rm in})}$ (${}_{\pm} \psi_{i,f,{\bm p}_{\perp},p_{\eta}, s}^{({\rm out})}$) with plane wave solutions with positive/negative frequency at $t\rightarrow -\infty$ ($t\rightarrow\infty$): 
\begin{align}
	{}_{\pm} \psi_{i,f,{\bm p}_{\perp},p_{\eta}, s}^{({\rm in})} &\xrightarrow[t \rightarrow -\infty]{} {}_{\pm} \psi_{i,f,{\bm p}_{\perp},p_{\eta}, s}^{{(\rm free})} [\bar{A}_{\mu}(t \rightarrow -\infty)], \\
	{}_{\pm} \psi_{i,f,{\bm p}_{\perp},p_{\eta}, s}^{({\rm out})} &\xrightarrow[t \rightarrow \infty]{} {}_{\pm} \psi_{i,f,{\bm p}_{\perp},p_{\eta}, s}^{({\rm free})} [\bar{A}_{\mu}(t \rightarrow \infty)] , 
\end{align}
where the plane wave solutions ${}_{\pm} \psi_{i,f,{\bm p}_{\perp},p_{\eta}, s}^{({\rm free})}[\breve{A}_{\mu}]$ satisfy the free field equation of motion under a pure gauge background $\breve{A}_{\mu} = \bar{A}_{\mu}(\tau \rightarrow \pm\infty)$.  For details of the plane wave solutions ${}_{\pm} \psi_{i,f,{\bm p}_{\perp},p_{\eta}, s}^{({\rm free})}$, see Appendix~\ref{appA:quark}.  We also normalize the positive/negative frequency mode functions ${}_{\pm} \psi_{i,f,{\bm p}_{\perp},p_{\eta}, s}^{({\rm as})}$ for each ${\rm as}={\rm in,out}$ as 
\begin{align}
	( {}_{\pm} \psi_{i,f,{\bm p}_{\perp},p_{\eta},s}^{({\rm as})} |{}_{\pm} \psi_{i,f,{\bm p}'_{\perp},p'_{\eta},s'}^{({\rm as})} )_{\rm F} &= \delta_{ss'} \delta^2({\bm p}_{\perp} - {\bm p}'_{\perp}) \delta (p_{\eta} - p'_{\eta}) \label{eq_48} \\
	( {}_{\pm} \psi_{i,f,{\bm p}_{\perp},p_{\eta},s}^{({\rm as})} | {}_{\mp} \psi_{i,f,{\bm p}'_{\perp},p'_{\eta},s'}^{({\rm as})} )_{\rm F} &= 0 \label{eq_49},   
\end{align}
where the inner product for fermion fields $( \psi_1 | \psi_2 )_{\rm F}$ in the $\tau$-$\eta$ coordinates is given by 
\begin{align}
	( \psi_1 | \psi_2 )_{\rm F} = \tau \int_{\tau = {\rm const.}} d^2{\bm x}_{\perp} d\eta \; \bar{\psi}_1 \gamma^{\tau} \psi_2.   \label{eq_50}
\end{align}

Next, we impose canonical commutation relations to complete the canonical quantization.  Since we are working in the $\tau$-$\eta$ coordinates, we impose canonical commutation relations on an equal $\tau$-surface, instead on an equal $t$-surface as in the Cartesian coordinates: 
\begin{align}
	&\{ \psi_{i,f} (\tau, {\bm x}_{\perp}, \eta) , \pi_{i',f'} (\tau, {\bm x}'_{\perp}, \eta') \} \nonumber\\
	&\ \ \ \ \ \ = i \delta_{i i'} \delta_{ff'} \delta^2 ({\bm x}_{\perp} - {\bm x}'_{\perp} ) \frac{\delta (\eta - \eta')}{\tau}, \label{eq_53}\\
	&\{ \pi_{i,f} (\tau, {\bm x}_{\perp}, \eta) , \pi_{i',f'} (\tau, {\bm x}'_{\perp}, \eta') \} \nonumber\\
	&\ \ \ \ \ \ = \{ \psi_{i,f} (\tau, {\bm x}_{\perp}, \eta) , \psi_{i',f'} (\tau, {\bm x}'_{\perp}, \eta') \} = 0 \label{eq_54}, 
\end{align}
where the canonical conjugate $\pi_{i,f}$ to the quark field $\psi_{i,f}$ is given by $\pi_{i,f} = \delta {\mathcal L}/\delta ( \partial_{\tau} \psi_{i,f}) = i \bar{\psi}_{i,f} \gamma^{\tau}$.  The factor $1/\tau$ in Eq.~(\ref{eq_53}) comes from the Jacobian $\sqrt{-g} = \tau$ of the $\tau$-$\eta$ coordinates.  The canonical commutation relations, Eqs.~(\ref{eq_53}) and (\ref{eq_54}), are equivalent to require that the operators $a^{({\rm as})}_{i,f,{\bm p}_{\perp}, p_{\eta}, s}, b^{({\rm as})}_{i,f,{\bm p}_{\perp}, p_{\eta}, s}$ anti-commute as  
\begin{align}
	&\{ a^{({\rm as})}_{i,f,{\bm p}_{\perp}, p_{\eta}, s} \;,\; a^{ ({\rm as}) \dagger}_{i',f',{\bm p}'_{\perp}, p'_{\eta}, s'}   \} \nonumber\\
	&\ \ \ \ \ \ =   \{ b^{({\rm as})}_{i,f,{\bm p}_{\perp}, p_{\eta}, s} \;,\; b^{ ({\rm as})\dagger}_{i',f',{\bm p}'_{\perp}, p'_{\eta}, s'}   \} \nonumber\\
	&\ \ \ \ \ \ \ \ \ \ \ = \delta_{ii'} \delta_{ff'} \delta_{ss'} \delta^2({\bm p}_{\perp} - {\bm p}'_{\perp}) \delta(p_{\eta} - p'_{\eta}) , \label{eq59} \\
	&({\rm otherwise}) = 0.  \label{eq_60}
\end{align}
From these anti-commutation relations, Eqs.~(\ref{eq59}) and (\ref{eq_60}), one can understand as usual that the operator $a^{({\rm as})}_{i,f,{\bm p}_{\perp}, p_{\eta}, s}$ ($b^{({\rm as})}_{i,f,{\bm p}_{\perp}, p_{\eta}, s}$) acts as an annihilation operator of a quark (an anti-quark) at the corresponding asymptotic time with the momentums ${\bm p}_{\perp}, p_{\eta}$, the spin $s$, the color charge $q^{\rm{(q)}}_i$ ($-q^{\rm{(q)}}_i$), and the flavor $f$.

As is stated in the beginning of this section, the creation/annihilation operators for different asymptotic times do not coincide with each other because of the interactions with the classical field.  The linear relation is described by the following Bogoliubov transformation: 
\begin{widetext}
\begin{align}
	\begin{pmatrix} 
		a^{(\rm out)}_{i,f,{\bm p}_{\perp}, p_{\eta}, s} \\ 
		b^{(\rm out)\dagger}_{i,f,-{\bm p}_{\perp}, -p_{\eta}, s} 
	\end{pmatrix} 
		&=
	\begin{pmatrix}
		( {}_{+} \psi_{i,f,{\bm p}_{\perp},p_{\eta}, s}^{({\rm out})} | \psi_{i,f} )_{\rm F} \\
		( {}_{-} \psi_{i,f,{\bm p}_{\perp},p_{\eta}, s}^{({\rm out})} | \psi_{i,f} )_{\rm F}
	\end{pmatrix} \nonumber\\
	&=
	\sum_{s'} \int d^2{\bm p}'_{\perp} dp'_{\eta}
	\begin{pmatrix} 
		( {}_{+} \psi_{i,f,{\bm p}_{\perp},p_{\eta}, s}^{({\rm out})} | {}_{+}\psi_{i,f,{\bm p}'_{\perp},p'_{\eta}, s'}^{({\rm in})} )_{\rm F}	&	( {}_{+} \psi_{i,f,{\bm p}_{\perp},p_{\eta}, s}^{({\rm out})} | {}_{-}\psi_{i,f,{\bm p}'_{\perp},p'_{\eta}, s'}^{({\rm in})} )_{\rm F} \\ 
		( {}_{-} \psi_{i,f,{\bm p}_{\perp},p_{\eta}, s}^{({\rm out})} | {}_{+}\psi_{i,f,{\bm p}'_{\perp},p'_{\eta}, s'}^{({\rm in})} )_{\rm F}	&	( {}_{-} \psi_{i,f,{\bm p}_{\perp},p_{\eta}, s}^{({\rm out})} | {}_{-}\psi_{i,f,{\bm p}'_{\perp},p'_{\eta}, s'}^{({\rm in})} )_{\rm F}
	\end{pmatrix}
	\begin{pmatrix} 
		a^{(\rm in)}_{i,f,{\bm p}'_{\perp}, p'_{\eta}, s'} \\ 
		b^{(\rm in)\dagger}_{i,f,-{\bm p}'_{\perp}, -p'_{\eta}, s'} 
	\end{pmatrix} .  \label{eq_64}
\end{align}

In order to obtain the (anti-)quark spectrum at $t \rightarrow \infty$ produced from the background field $\bar{A}_{\mu}$, let us introduce a (anti-)quark number density operator $n^{({\rm q})}_{i,f,{\bm p}_{\perp}, p_{\eta}, s}$ ($n^{(\bar{\rm q})}_{i,f,{\bm p}_{\perp}, p_{\eta}, s}$) by
\begin{align}
	n^{({\rm q})}_{i,f,{\bm p}_{\perp}, p_{\eta}, s} \equiv a^{(\rm out)\dagger}_{i,f,{\bm p}_{\perp}, p_{\eta}, s} a^{(\rm out)}_{i,f,{\bm p}_{\perp}, p_{\eta}, s}, \ 
	n^{(\bar{\rm q})}_{i,f,{\bm p}_{\perp}, p_{\eta}, s} \equiv b^{(\rm out)\dagger}_{i,f,{\bm p}_{\perp}, p_{\eta}, s} b^{(\rm out)}_{i,f,{\bm p}_{\perp}, p_{\eta}, s}.  
\end{align}
The quark and anti-quark spectra are derived as an expectation value of the number density operators by a given initial state at $t \rightarrow -\infty$.  Hereafter, let us assume that the initial state is given by a vacuum $\ket{{\rm vac; in}}$.  By noting that the initial vacuum is a state which is annihilated by the annihilation operators at $t \rightarrow -\infty$ as $0 = a^{(\rm in)}_{i,f,{\bm p}_{\perp}, p_{\eta}, s} \ket{{\rm vac; in}} = b^{(\rm in)}_{i,f,{\bm p}_{\perp}, p_{\eta}, s} \ket{{\rm vac; in}}$ and by using the Bogoliubov transformation Eq.~(\ref{eq_64}), one immediately obtains
\begin{align}
	\frac{d^3N^{({\rm q})}_{i,f,s}}{d{\bm p}_{\perp}^2 dp_{\eta}} 	
		&\equiv \frac{\bra{{\rm vac; in}} n^{({\rm q})}_{i,f,{\bm p}_{\perp}, p_{\eta}, s} \ket{{\rm vac; in}}}{\braket{ {\rm vac; in} | {\rm vac; in} }} =  \sum_{s'} \int d^2{\bm p}'_{\perp} dp'_{\eta} \left| ( {}_{+} \psi_{i,f,{\bm p}_{\perp},p_{\eta}, s}^{({\rm out})} | {}_{-}\psi_{i,f,{\bm p}'_{\perp},p'_{\eta}, s'}^{({\rm in})} )_{\rm F} \right|^2, \label{eq_73} \\
	\frac{d^3N^{(\bar{\rm q})}_{i,f,s}}{d{\bm p}_{\perp}^2 dp_{\eta}} 
		&\equiv \frac{\bra{{\rm vac; in}} n^{(\bar{\rm q})}_{i,f,{\bm p}_{\perp}, p_{\eta}, s} \ket{{\rm vac; in}}}{\braket{ {\rm vac; in} | {\rm vac; in} }} = \sum_{s'} \int d^2{\bm p}'_{\perp} dp'_{\eta}  \left|  ( {}_{-} \psi_{i,f, -{\bm p}_{\perp}, -p_{\eta}, s}^{({\rm out})} | {}_{+}\psi_{i,f,{\bm p}'_{\perp},p'_{\eta}, s'}^{({\rm in})} )_{\rm F} \right|^2.  \label{eq_74}
\end{align}
An important point of these formulas Eqs.~(\ref{eq_73}) and (\ref{eq_74}) is that deriving the particle spectrum is thus reduced to finding out the mode functions ${}_{\pm}\psi_{i,f,{\bm p}_{\perp},p_{\eta}, s}^{({\rm as})}$ by solving the Dirac equation Eq.~(\ref{eq46}) non-perturbatively with respect to the classical field.

As is expected from the Pauli principle, one can explicitly show that the phase space density do not exceed unity.  Indeed, the anti-commutation relation Eq.~(\ref{eq59}) yields that the Bogoliubov coefficients are normalized as
\begin{align}
	\frac{1}{(2\pi)^3} \int d^2 {\bm x}_{\perp} \int d \eta 
		&=  \sum_{s'} \int d^2{\bm p}'_{\perp} dp'_{\eta} \left[ \left| ( {}_{\pm} \psi_{i,f,{\bm p}_{\perp},p_{\eta}, s}^{({\rm out})} | {}_{\pm}\psi_{i,f,{\bm p}'_{\perp},p'_{\eta}, s'}^{({\rm in})} )_{\rm F} \right|^2  + \left| ( {}_{\pm} \psi_{i,f,{\bm p}_{\perp},p_{\eta}, s}^{({\rm out})} | {}_{\mp}\psi_{i,f,{\bm p}'_{\perp},p'_{\eta}, s'}^{({\rm in})} )_{\rm F} \right|^2 \right] ,  \label{eq__65}
\end{align}
\end{widetext}
where we have used $\delta^2 ({\bm p}_{\perp} = {\bm 0}) \delta (p_{\eta} = 0) = 1/(2\pi)^3 \int d^2 {\bm x}_{\perp} \int d \eta$.  From Eq.~(\ref{eq__65}), one immediately finds $(2\pi)^3 d^6 N^{({\rm q}, \bar{\rm q})}_{i,f,s} / d{\bm x}^2_{\perp} d\eta d{\bm p}_{\perp}^2 dp_{\eta} \leq 1$.

So far, we have characterized the longitudinal momentum of produced quarks by the label $p_{\eta}$ because it is a natural quantum number conjugate to the spacetime rapidity $\eta$ and that manifestly respects the boost invariance of the system.  Consequently, what we have obtained for the quark spectra in Eqs.~(\ref{eq_73}) and (\ref{eq_74}) are the $p_{\eta}$-spectrum.  However, what we actually observe in experiments is not the $p_{\eta}$-spectrum, but the $p_z$-spectrum and/or the momentum rapidity $y_{{\bm p}}$-spectrum, where  
\begin{align}
	y_{\bm p} \equiv \frac{1}{2} {\rm ln} \frac{ \omega_{\bm p} + p_z}{\omega_{\bm p}-p_z} \label{eq_67}
\end{align}
with $\omega_{\bm p}$ being an on-shell energy $\omega_{\bm p} \equiv \sqrt{m^2 + {\bm p}_{\perp}^2 + p_z^2}$.  The $p_z$-spectrum and/or the momentum rapidity $y_{\bm p}$-spectrum can be obtained from the $p_{\eta}$-spectrum in the following way \cite{tan11}: As in the $p_{\eta}$-spectrum Eqs.~(\ref{eq_73}) and (\ref{eq_74}), the $p_z$-spectrum and/or the momentum rapidity $y_{\bm p}$-spectrum are obtained as an expectation value of the number operators, $n^{\rm (q)}_{i,f,{\bm p}_{\perp}, p_z, s}$ and $n^{\bar{\rm (q)}}_{i,f,{\bm p}_{\perp}, p_z, s}$, which are labeled by $p_z$ instead of $p_{\eta}$ as
\begin{align}
	\frac{d^3N^{({\rm q})}_{i,f,s}}{d{\bm p}_{\perp}^2 dp_{z}} 
		&= \frac{1}{\omega_{\bm p}}\frac{d^3N^{\rm (q)}_{i,f,s}}{d{\bm p}_{\perp}^2 dy_{\bm p}} \nonumber\\
		&\equiv \frac{\bra{{\rm vac; in}}  n^{({\rm q})}_{i,f,{\bm p}_{\perp}, p_z, s}  \ket{{\rm vac; in}}}{\braket{ {\rm vac; in} | {\rm vac; in} }} , \label{eq60} \\
	\frac{d^3N^{(\bar{\rm q})}_{i,f,s}}{d{\bm p}_{\perp}^2 dp_{z}} 
		&= \frac{1}{\omega_{\bm p}}\frac{d^3N^{(\bar{\rm q})}_{i,f,s}}{d{\bm p}_{\perp}^2 dy_{\bm p}} \nonumber\\
		&\equiv \frac{\bra{{\rm vac; in}}  n^{\bar{\rm (q)}}_{i,f,{\bm p}_{\perp}, p_z, s}  \ket{{\rm vac; in}}}{\braket{ {\rm vac; in} | {\rm vac; in} }},   \label{eq61}
\end{align}
where the number operators are defined by the annihilation operators $a^{\rm (out)}_{i,f,{\bm p}_{\perp}, p_z, s}, b^{\rm (out)}_{i,f,{\bm p}_{\perp}, p_z, s}$ as
\begin{align}
	n^{\rm (q)}_{i,f,{\bm p}_{\perp}, p_z, s} \equiv a^{{\rm (out)}\dagger}_{i,f,{\bm p}_{\perp}, p_z, s} a^{\rm (out)}_{i,f,{\bm p}_{\perp}, p_z, s}  , \nonumber\\ 
	n^{\bar{\rm (q)}}_{i,f,{\bm p}_{\perp}, p_z, s} \equiv b^{{\rm (out)}\dagger}_{i,f,{\bm p}_{\perp}, p_z, s}  b^{\rm (out)}_{i,f,{\bm p}_{\perp}, p_z, s} .
\end{align}
The annihilation operators $a^{\rm (out)}_{i,f,{\bm p}_{\perp}, p_z, s}, b^{\rm (out)}_{i,f,{\bm p}_{\perp}, p_z, s}$ are defined by expanding the fermion operator $\psi_{i,f}$ in terms of positive/negative frequency mode functions ${}_{\pm} \psi^{\rm (out)}_{i,f,{\bm p}_{\perp}, p_{z}, s}$ in the Cartesian coordinates which is labeled by $p_z$ being the Fourier conjugate to $z$ as
\begin{align}
	\psi_{i,f}(x) 	&= \sum_s \int d{\bm p}_{\perp}^2 dp_{z} \nonumber\\
					&\ \ \ \ \ \ \left[ {}_+ \psi^{({\rm out})}_{i,f,{\bm p}_{\perp},p_{z},s}(x) a^{({\rm out})}_{i,f,{\bm p}_{\perp},p_{z},s} \right.\nonumber\\
					&\ \ \ \ \ \ \ \ \ \ \ \left.+ {}_- \psi^{({\rm out})}_{i,f,{\bm p}_{\perp},p_{z},s}(x)   b^{({\rm out})\dagger}_{i,f,-{\bm p}_{\perp},-p_{z},s}   \right]. \label{eq_62}
\end{align}
Here, we adopt the same boundary condition as what we have required for ${}_{\pm} \psi^{\rm (out)}_{i,f,{\bm p}_{\perp}, p_{\eta}, s}$, i.e., we require ${}_{\pm} \psi^{\rm (out)}_{i,f,{\bm p}_{\perp}, p_{z}, s}$ to coincide with the plane wave solutions at $t \rightarrow \infty$.  As is shown in Appendix~\ref{appA:quark}, the mode functions in the Cartesian coordinates ${}_{\pm} \psi^{\rm (out)}_{i,f,{\bm p}_{\perp}, p_{z}, s}$ and those in the $\tau$-$\eta$ coordinates ${}_{\pm} \psi^{\rm (out)}_{i,f,{\bm p}_{\perp}, p_{\eta}, s}$ are related with each other by an integral transformation described by%
\footnote{Strictly speaking, what we show in Appendix~\ref{appA:quark} is that the plane wave solutions in the Cartesian coordinates ${}_{\pm} \psi^{\rm (free)}_{i,f,{\bm p}_{\perp}, p_{z}, s} [\tilde{A}_m]$ and those in the $\tau$-$\eta$ coordinates ${}_{\pm} \psi^{\rm (free)}_{i,f,{\bm p}_{\perp}, p_{\eta}, s} [\tilde{A}_{\mu}]$ are related with each other by the integral transformation Eq.~(\ref{eq__62}).  One can safely say that the same integral relation equally holds for the mode functions, ${}_{\pm} \psi^{\rm (out)}_{i,f,{\bm p}_{\perp}, p_{z}, s}$ and ${}_{\pm} \psi^{\rm (out)}_{i,f,{\bm p}_{\perp}, p_{\eta}, s}$: Since the two sets of mode functions, ${}_{\pm} \psi^{\rm (out)}_{i,f,{\bm p}_{\perp}, p_{z}, s}$ and ${}_{\pm} \psi^{\rm (out)}_{i,f,{\bm p}_{\perp}, p_{\eta}, s}$, obey the same differential equation $[ i\gamma^m (\partial_m -iq_i^{\rm (q)} \tilde{A}_m ) - m_f ] {}_{\pm} \psi^{\rm (out)}_{i,f,{\bm p}_{\perp}, p_{z}, s} = [ i\gamma^{\mu} (\partial_{\mu} -iq_i^{\rm (q)} \tilde{A}_{\mu} ) - m_f ] {}_{\pm} \psi^{\rm (out)}_{i,f,{\bm p}_{\perp}, p_{\eta}, s} =0$ and that the linear relation between them is conserved in the time evolution, it is sufficient to show that the integral relation at the boundary $t, \tau \rightarrow \infty$, where both solutions become plane waves.  Hence, the integral relation actually holds.  The same argument can be applied for the integral transformation for gluons (Eq.~(\ref{eq--62})), which we will discuss in the next subsection.  }
\begin{align}
	{}_{\pm} \psi^{\rm (out)}_{i,f,{\bm p}_{\perp}, p_{\eta}, s}
		= \int dp_z \frac{ {\rm e}^{\pm i p_{\eta} y_{\bm p}} }{\sqrt{2\pi \omega_{\bm p}}} {}_{\pm} \psi^{\rm (out)}_{i,f,{\bm p}_{\perp}, p_{z}, s}.   \label{eq__62}
\end{align}
Using this integral transformation Eq.~(\ref{eq__62}) and comparing the expansion in the Cartesian coordinates Eq.~(\ref{eq_62}) with that in the $\tau$-$\eta$ coordinates Eq.~(\ref{eq__43}), one finds
\begin{align}
	 a^{({\rm out})}_{i,f,{\bm p}_{\perp},p_{z},s} 
		&= \int dp_{\eta} \frac{ {\rm e}^{ i p_{\eta} y_{\bm p}} }{\sqrt{2\pi \omega_{\bm p}}} a^{({\rm out})}_{i,f,{\bm p}_{\perp},p_{\eta},s}, \label{eq65} \\
	 b^{({\rm out})\dagger}_{i,f,{\bm p}_{\perp},p_{z},s} 
		&= \int d p_{\eta} \frac{ {\rm e}^{- i p_{\eta} y_{\bm p}} }{\sqrt{2\pi \omega_{\bm p}}} b^{({\rm out})\dagger}_{i,f,{\bm p}_{\perp},p_{\eta},s} .  \label{eq66}
\end{align}
Inserting these relations, Eqs.~(\ref{eq65}) and (\ref{eq66}), back into Eqs.~(\ref{eq60}) and (\ref{eq61}), one obtains
\begin{align}
	\frac{d^3N^{({\rm q})}_{i,f,s}}{d{\bm p}_{\perp}^2 dp_{z}} 
			&= \frac{1}{\omega_{\bm p}}\frac{d^3N^{\rm (q)}_{i,f,s}}{d{\bm p}_{\perp}^2 dy_{\bm p}} \nonumber\\ 
			&= \frac{1}{\omega_{\bm p}} \int dp_{\eta} dp'_{\eta}  \frac{ {\rm e}^{i y_{\bm p} (p_{\eta} - p'_{\eta}) } }{2\pi} \nonumber\\
			&\quad \times \frac{\bra{{\rm vac; in}}  a^{{\rm (out)}\dagger}_{i,f,{\bm p}_{\perp}, p'_{\eta}, s} a^{{\rm (out)}}_{i,f,{\bm p}_{\perp}, p_{\eta}, s}  \ket{{\rm vac; in}}}{\braket{ {\rm vac; in} | {\rm vac; in} }} , \label{eq-67} \\
		\frac{d^3N^{(\bar{\rm q})}_{i,f,s}}{d{\bm p}_{\perp}^2 dp_{z}} 
			&= \frac{1}{\omega_{\bm p}}\frac{d^3N^{(\bar{\rm q})}_{i,f,s}}{d{\bm p}_{\perp}^2 dy_{\bm p}} \nonumber\\
			&= \frac{1}{\omega_{\bm p}} \int dp_{\eta} dp'_{\eta}  \frac{ {\rm e}^{i y_{{\bm p}} (p_{\eta} - p'_{\eta}) } }{2\pi}  \nonumber\\
			&\quad \times \frac{\bra{{\rm vac; in}}  b^{{\rm (out)}\dagger}_{i,f,{\bm p}_{\perp}, p'_{\eta}, s} b^{{\rm (out)}}_{i,f,{\bm p}_{\perp}, p_{\eta}, s}  \ket{{\rm vac; in}}}{\braket{ {\rm vac; in} | {\rm vac; in} }} . \label{eq-68}
\end{align}
When the system is perfectly boost invariant, the expectation values in Eqs.~(\ref{eq-67}) and (\ref{eq-68}) for $p_{\eta} \neq p'_{\eta}$ vanish because $p_{\eta}$ is a good quantum number and it never mixes with other values of $p_{\eta}$ during the time evolution.  In this case, one can further simplify Eqs.~(\ref{eq-67}) and (\ref{eq-68}) as
\begin{align}
	\frac{d^3N^{({\rm q})}_{i,f,s}}{d{\bm p}_{\perp}^2 dp_{z}} 
		&= \frac{1}{\omega_{\bm p}}\frac{d^3N^{\rm (q)}_{i,f,s}}{d{\bm p}_{\perp}^2 dy_{\bm p}} \nonumber\\ 
		&= \frac{1}{2\pi \delta(p_{\eta} = 0) } \times \frac{1}{\omega_{\bm p}} \int dp_{\eta} \frac{d^3N^{({\rm q})}_{i,f,s}}{d{\bm p}_{\perp}^2 dp_{\eta}},  \label{eq67}\\
	\frac{d^3N^{(\bar{\rm q})}_{i,f,s}}{d{\bm p}_{\perp}^2 dp_{z}} 
		&= \frac{1}{\omega_{\bm p}}\frac{d^3N^{(\bar{\rm q})}_{i,f,s}}{d{\bm p}_{\perp}^2 dy_{\bm p}} \nonumber\\
		&= \frac{1}{2\pi \delta(p_{\eta} = 0) } \times \frac{1}{\omega_{\bm p}} \int dp_{\eta} \frac{d^3N^{(\bar{\rm q})}_{i,f,s}}{d{\bm p}_{\perp}^2 dp_{\eta}},  \label{eq68}
\end{align}
which are manifestly boost invariant in the sense that the $y_{\bm p}$-spectrum does not depend on the momentum rapidity $y_{\bm p}$.  We note that we have derived the formulas, Eqs.~(\ref{eq67}) and (\ref{eq68}), in a quantum field theoretical manner by following Ref.~\cite{tan11}, but one can also obtain the same formulas within classical mechanics \cite{mih09, coo93}, though these two derivations agree with each other only if the system is perfectly boost invariant.

\subsubsection{Gluon}
Next, we turn to the canonical quantization of the gluon field $ W_{\mu, A}$, and compute the gluon spectrum at $t \rightarrow \infty$.  We do essentially the same procedure as what we have done in the quark case although there are slight differences due to the vector nature of gluons.

First, we expand the gluon field $W_{\mu, A}$ as 
\begin{align}
	W_{\mu, A} 	&= \sum_{\sigma} \int d{\bm p}^2_{\perp} dp_{\eta} \nonumber\\
				&\ \ \ \ \ \ \ \left[ {}_+ W^{({\rm as})}_{\mu, A, {\bm p}_{\perp},p_{\eta},\sigma} c^{({\rm as})}_{A, {\bm p}_{\perp},p_{\eta},\sigma} \right.\nonumber\\
				&\ \ \ \ \ \ \ \ \ \ \ \left.+ {}_- W^{({\rm as})}_{\mu, A, {\bm p}_{\perp},p_{\eta},\sigma}  d^{({\rm as})\dagger}_{A, -{\bm p}_{\perp}, -p_{\eta},\sigma}   \right].  \label{eq__69}
\end{align}
$\sigma = 0,1,2,3$ labels the polarization and the other labels are the same as in the quark case.  The mode functions ${}_{\pm} W_{\mu, A, {\bm p}_{\perp}, p_{\eta}, \sigma}^{({\rm as})} $ are the solutions of the equations of motion Eq.~(\ref{eq48}) with the plane wave boundary condition:  
\begin{align}
	{}_{\pm} W_{\mu, A,{\bm p}_{\perp},p_{\eta}, \sigma}^{({\rm in})} &\xrightarrow[t\rightarrow -\infty]{} {}_{\pm} W_{\mu, A, {\bm p}_{\perp}, p_{\eta}, \sigma}^{{(\rm free})} [\bar{A}_{\mu}(t \rightarrow -\infty)], \label{eq__63} \\
	{}_{\pm} W_{\mu, A, {\bm p}_{\perp},p_{\eta}, \sigma}^{({\rm out})} &\xrightarrow[t\rightarrow \infty]{} {}_{\pm} W_{\mu, A,{\bm p}_{\perp},p_{\eta}, \sigma}^{({\rm free})} [\bar{A}_{\mu}(t \rightarrow \infty)], 
\end{align}
where the plane wave solutions ${}_{\pm} W_{\mu, A, {\bm p}_{\perp}, p_{\eta}, \sigma}^{{(\rm free})}[\breve{A}_{\mu}]$ satisfy the free field equation of motion under a pure gauge background field $\breve{A}_{\mu} = \bar{A}_{\mu}(t \rightarrow \pm\infty)$.  For details of the plane wave solutions $ {}_{\pm} W_{\mu, A, {\bm p}_{\perp}, p_{\eta}, \sigma}^{({\rm free})} $, see Appendix~\ref{appA:gluon}.  The positive/negative frequency mode functions ${}_{\pm} W_{\mu, A, {\bm p}_{\perp}, p_{\eta}, \sigma}^{({\rm as})} $ are normalized as  
\begin{align}
	&-g^{\mu\nu} ( {}_{\pm} W_{\mu, A, {\bm p}_{\perp}, p_{\eta}, \sigma}^{({\rm as})} |  {}_{\pm} W_{\nu, A, {\bm p}'_{\perp}, p'_{\eta}, \sigma' }^{({\rm as})} )_{\rm B}  \nonumber\\
	&\ \ \ \ \ \ \ \ \ \ \ \ \ \ \ \ \ \ = \pm \xi_{\sigma \sigma'} \delta^2({\bm p}_{\perp} - {\bm p}'_{\perp}) \delta (p_{\eta} - p'_{\eta}) , \label{eq_75} \\
	&-g^{\mu\nu} ( {}_{\pm} W_{\mu, A, {\bm p}_{\perp}, p_{\eta}, \sigma}^{({\rm as})}  |  {}_{\mp} W_{\nu, A, {\bm p}'_{\perp}, p'_{\eta}, \sigma'}^{({\rm as})}  )_{\rm B} = 0 \label{eq_76}
\end{align}
for each ${\rm as}={\rm in, out}$.  Here, the inner product for boson fields $( \phi_1 | \phi_2 )_{\rm B}$ in the $\tau$-$\eta$ coordinates is given by 
\begin{align}
	( \phi_1 | \phi_2 )_{\rm B} = i \tau \int_{\tau = {\rm const.}} d^2{\bm x}_{\perp} d\eta  \; \phi_1^*  \overset{\leftrightarrow}{\nabla}_{\tau} \phi_2 , \label{eq_77}
\end{align}
where $\overset{\leftrightarrow}{\nabla}_{\tau}  \equiv \overset{\rightarrow}{\nabla}_{\tau}  - \overset{\leftarrow}{\nabla}_{\tau} $.  The indefinite metric $\xi_{\sigma \sigma'}$ is introduced by 
\begin{align}
	\xi_{\sigma \sigma'} 
		\equiv 
			\begin{pmatrix} 
				0 & 0 & 0 & 1 \\ 
				0 & 1 & 0 & 0 \\
				0 & 0 & 1 & 0 \\
				1 & 0 & 0 & 0 \\
			\end{pmatrix},  \label{eq76}
\end{align}
which has symmetric off-diagonal elements $\xi_{03} = \xi_{30}$.  Because of this property, the 0-th and the 3-rd polarization mode of gluons become unphysical and they do not appear in the physical spectrum as we will show later.

For later convenience, we decompose the positive/negative frequency mode functions ${}_{\pm} W_{\mu, A, {\bm p}_{\perp}, p_{\eta}, \sigma}^{({\rm as})} $ by introducing a polarization vector $\varepsilon_{\mu, \sigma}$ and scalar amplitudes ${}_{\pm} \Phi^{\rm (as)}_{ A, {\bm p}_{\perp}, p_{\eta}, \sigma}$ as 
\begin{align}
	{}_{\pm} W_{\mu, A, {\bm p}_{\perp}, p_{\eta}, \sigma}^{({\rm as})} \equiv \varepsilon_{\mu, \sigma} {}_{\pm} \Phi^{\rm (as)}_{ A, {\bm p}_{\perp}, p_{\eta}, \sigma}. \label{eq.dec}  
\end{align}
It is convenient to normalize the polarization vector as
\begin{align}
	g^{\mu\nu} \varepsilon_{\mu, \sigma}^{*} \varepsilon_{\nu, \sigma'} &= -\xi_{\sigma \sigma'} , \nonumber\\
	\sum_{\sigma,\sigma'} \xi_{\sigma \sigma'} \varepsilon_{\mu, \sigma}^{*} \varepsilon_{\nu, \sigma'} &= -g_{\mu\nu}, \label{eq.norm}
\end{align}
and to require that the covariant derivatives vanish as%
\footnote{One can always construct such a polarization vector by contracting the viervein matrix $e^{m}_{\;\;\mu}$ with a constant vector $\tilde{\varepsilon}_{m, \sigma}$ normalized as
\begin{align}
	\eta^{mn} \tilde{\varepsilon}_{m, \sigma}^{*} \tilde{\varepsilon}_{n, \sigma'} &= -\xi_{\sigma \sigma'} , \nonumber\\
	\sum_{\sigma,\sigma'} \xi_{\sigma \sigma'} \tilde{\varepsilon}_{m, \sigma}^{*} \tilde{\varepsilon}_{n, \sigma'} &= -\eta_{mn}. 
\end{align}
}.  
\begin{align}
	\nabla_{\mu} \varepsilon_{\nu, \sigma} = 0.  
\end{align}
Then, the normalization conditions for ${}_{\pm} W_{\mu, A, {\bm p}_{\perp}, p_{\eta}, \sigma}^{({\rm as})} $, Eqs.~(\ref{eq_75}) and (\ref{eq_76}), can be rewritten in terms of the scalar amplitudes as
\begin{align}
	&\sum_{\sigma'} \xi_{\sigma \sigma'}( {}_{\pm} \Phi_{A, {\bm p}_{\perp}, p_{\eta}, \sigma}^{({\rm as})} |  {}_{\pm} \Phi_{A, {\bm p}'_{\perp}, p'_{\eta}, \sigma' }^{({\rm as})} )_{\rm B} \nonumber\\
	&\ \ \ \ \ \ \ \ \ \ \ \ \ \ \ \ \ \ \ \ \ \ = \pm \delta^2({\bm p}_{\perp} - {\bm p}'_{\perp}) \delta (p_{\eta} - p'_{\eta}) , \\
	&\sum_{\sigma'} \xi_{\sigma \sigma'}( {}_{\pm} \Phi_{A, {\bm p}_{\perp}, p_{\eta}, \sigma}^{({\rm as})}  |  {}_{\mp} \Phi_{A, {\bm p}'_{\perp}, p'_{\eta}, \sigma'}^{({\rm as})}  )_{\rm B} = 0 .   
\end{align}

Next, we impose canonical commutation relations to complete the canonical quantization: 
\begin{align}
	&[ W_{\mu,A} (\tau, {\bm x}_{\perp}, \eta) , \pi_{\nu, A'} (\tau, {\bm x}'_{\perp}, \eta') ] \nonumber\\
	&\ \ \ \ \ \ \ \ \ \ \ = i g_{\mu\nu} \delta_{AA'} \delta^2 ({\bm x}_{\perp} - {\bm x}'_{\perp} ) \frac{\delta (\eta - \eta')}{\tau}, \label{eq_81}\\
	&[ W_{\mu,A} (\tau, {\bm x}_{\perp}, \eta) , W_{\nu,A'} (\tau, {\bm x}'_{\perp}, \eta') ] \nonumber\\
	&\ \ \ \ \ \ \ \ \ \ \  = [ \pi_{\mu, A} (\tau, {\bm x}_{\perp}, \eta) , \pi_{\nu, A'} (\tau, {\bm x}'_{\perp}, \eta') ] = 0, \label{eq_82}
\end{align}
where the canonical conjugate field $ \pi_{\mu,A} $ to the gluon field $ W_{\mu,A} $ is given by $\pi_{\mu, A} = \delta {\mathcal L}/\delta ( \nabla_{\tau} W^{\mu}_{A}) = -\nabla_{\tau} W_{\mu, A}^{\dagger}$.  The canonical commutation relations, Eqs.~(\ref{eq_81}) and (\ref{eq_82}), are equivalent to requiring the operators $c^{({\rm as})}_{A,{\bm p}_{\perp}, p_{\eta}, \sigma}, d^{({\rm as})}_{A,{\bm p}_{\perp}, p_{\eta}, \sigma}$ to commute as  
\begin{align}
	&[ c^{({\rm as})}_{A,{\bm p}_{\perp}, p_{\eta}, \sigma} \;,\; c^{ ({\rm as}) \dagger}_{A',{\bm p}'_{\perp}, p'_{\eta}, \sigma'}   ] \nonumber\\
	&\ \ \ \ \ \ \ =   [ d^{({\rm as})}_{A,{\bm p}_{\perp}, p_{\eta}, \sigma} \;,\; d^{ ({\rm as})\dagger}_{A',{\bm p}'_{\perp}, p'_{\eta}, \sigma'}   ] \nonumber\\
	&\ \ \ \ \ \ \ \ \ \ \ \ \ \ \ = \delta_{AA'} \xi_{\sigma \sigma'} \delta^2({\bm p}_{\perp} - {\bm p}'_{\perp}) \delta(p_{\eta} - p'_{\eta}) , \label{eq_83} \\
	&({\rm otherwise}) = 0.  \label{eq_84}
\end{align}
From these commutation relations, Eqs.~(\ref{eq_83}) and (\ref{eq_84}), the operator $c^{({\rm as})}_{A,{\bm p}_{\perp}, p_{\eta}, \sigma}$ ($d^{({\rm as})}_{A,{\bm p}_{\perp}, p_{\eta}, \sigma}$) can be understood as an annihilation operator of a gluon at the corresponding asymptotic time with the momentums ${\bm p}_{\perp}, p_{\eta}$, the polarization $\sigma$ and the color charge $q^{({\rm g})}_A$ ($-q^{({\rm g})}_A$).

As in the quark case, the creation/annihilation operators at different asymptotic times do not coincide with each other and the linear relation is described by a Bogoliubov transformation given by
\begin{widetext}
\begin{align}
	\begin{pmatrix} 
		c^{(\rm out)}_{A,{\bm p}_{\perp}, p_{\eta}, \sigma} \\ 
		d^{(\rm out)\dagger}_{A,-{\bm p}_{\perp}, -p_{\eta}, \sigma} 
	\end{pmatrix} 
		&=
	\sum_{\sigma''} \xi_{\sigma \sigma''} (-g^{\mu\nu})
	\begin{pmatrix}
		( {}_{+} W_{\mu, A,{\bm p}_{\perp},p_{\eta}, \sigma''}^{({\rm out})} | W_{\nu, A} )_{\rm B} \\
		-( {}_{-} W_{\mu, A,{\bm p}_{\perp},p_{\eta}, \sigma''}^{({\rm out})} | W_{\nu, A} )_{\rm B}
	\end{pmatrix} \nonumber\\
	&=
	\sum_{\sigma'} \int d^2{\bm p}'_{\perp} dp'_{\eta} \left\{  \sum_{\sigma''} \xi_{\sigma \sigma''} (-g^{\mu\nu}) \right. \nonumber\\
	&\ \ \ \ \ \ \times\left.
	\begin{pmatrix} 
		( {}_{+} W_{\mu, A,{\bm p}_{\perp},p_{\eta}, \sigma''}^{({\rm out})} | {}_{+}W^{({\rm in})}_{\nu, A,{\bm p}'_{\perp},p'_{\eta}, \sigma'} )_{\rm B}	&	( {}_{+} W_{\mu, A,{\bm p}_{\perp},p_{\eta}, \sigma''}^{({\rm out})} | {}_{-}W^{({\rm in})}_{\nu, A,{\bm p}'_{\perp},p'_{\eta}, \sigma'} )_{\rm B} \\ 
		-( {}_{-} W_{\mu, A,{\bm p}_{\perp},p_{\eta}, \sigma''}^{({\rm out})} | {}_{+}W^{({\rm in})}_{\nu, A,{\bm p}'_{\perp},p'_{\eta}, \sigma'} )_{\rm B}	&	-( {}_{-} W_{\mu, A,{\bm p}_{\perp},p_{\eta}, \sigma''}^{({\rm out})} | {}_{-}W^{({\rm in})}_{\nu, A,{\bm p}'_{\perp},p'_{\eta}, \sigma'} )_{\rm B}
	\end{pmatrix}
	\begin{pmatrix} 
		c^{(\rm in)}_{A,{\bm p}'_{\perp}, p'_{\eta}, \sigma'} \\ 
		d^{(\rm in)\dagger}_{A,-{\bm p}'_{\perp}, -p'_{\eta}, \sigma'} 
	\end{pmatrix} \right\} .  \label{eq_85}
\end{align}

In order to obtain the gluon spectrum at $t \rightarrow \infty$, let us introduce a gluon number density operator $n^{({\rm g})}_{\pm A,{\bm p}_{\perp}, p_{\eta}, \sigma}$ by
\begin{align}
	n^{({\rm g})}_{A,{\bm p}_{\perp}, p_{\eta}, \sigma} \equiv c^{(\rm out)\dagger}_{A,{\bm p}_{\perp}, p_{\eta}, \sigma} c^{(\rm out)}_{A,{\bm p}_{\perp}, p_{\eta}, \sigma}, \ 
	n^{({\rm g})}_{-A,{\bm p}_{\perp}, p_{\eta}, \sigma} \equiv d^{(\rm out)\dagger}_{A,{\bm p}_{\perp}, p_{\eta}, \sigma} d^{(\rm out)}_{A,{\bm p}_{\perp}, p_{\eta}, \sigma}.    
\end{align}
As in the quark spectrum (Eqs.~(\ref{eq_73}) and (\ref{eq_74})), one can derive the gluon spectrum as an expectation value of the number density operators in the initial state as
\begin{align}
	\frac{d^3 N^{({\rm g})}_{\pm A, \sigma}}{d{\bm p}_{\perp}^2 dp_{\eta}} 	
		&\equiv \frac{\bra{{\rm vac; in}} n^{({\rm g})}_{\pm A,{\bm p}_{\perp}, p_{\eta}, \sigma} \ket{{\rm vac; in}}}{ \braket{{\rm vac; in}|{\rm vac; in}} }  \nonumber\\
		&= 
		\sum_{\sigma_1 \sigma_2} \xi_{\sigma_1 \sigma_2} \int d^2{\bm p}'_{\perp} dp'_{\eta} 
		\left\{
			\sum_{\sigma'_1 \sigma'_2} \xi_{\sigma \sigma'_1} \xi_{\sigma \sigma'_2} (-g^{\mu_1 \nu_1}) (-g^{\mu_2 \nu_2})  \right.\nonumber\\
		&\ \ \ \ \ \ \ \ \ \ \ \ \ \ \ \ \ \times\left.  ( {}_{\pm} W_{\mu_1 , A,{\bm p}_{\perp},p_{\eta}, \sigma'_1}^{({\rm out})} | {}_{\mp}W^{({\rm in})}_{\nu_1, A, {\bm p}'_{\perp}, p'_{\eta}, \sigma_1} )_{\rm B} ( {}_{\mp} W_{\nu_2 , A,{\bm p}'_{\perp},p'_{\eta}, \sigma_2}^{({\rm in})} | {}_{\pm}W^{({\rm out})}_{\mu_2, A, {\bm p}_{\perp}, p_{\eta}, \sigma'_2} )_{\rm B}   \right\} .  
 \label{eq91}
\end{align}
We note that only gluons from the quantum fluctuation are counted in Eq.~(\ref{eq91}) and there are no contributions from those from the classical background field.  This treatment is justified only for the gluon spectrum at $t\rightarrow \infty$, where the classical background field is vanishing.  If one is interested in the gluon spectrum at transient times $|t|<\infty$, where the classical background field is still present, then one has to count not only quantum gluons but also classical gluons in some way.  

One can perform the polarization sum in this formula Eq.~(\ref{eq91}) with the help of the decomposition Eq.~(\ref{eq.dec}).  Inserting the decomposition Eq.~(\ref{eq.dec}) into the formula Eq.~(\ref{eq91}), one obtains 
\begin{align}
	\frac{d^3 N^{({\rm g})}_{\pm A, \sigma}}{d{\bm p}_{\perp}^2 dp_{\eta}}
		&= \sum_{\sigma_1 \sigma_2 \sigma'_1 \sigma'_2} \int d^2{\bm p}'_{\perp} dp'_{\eta} \xi_{\sigma_1 \sigma_2} \xi_{\sigma \sigma'_1} \xi_{\sigma \sigma'_2} \xi_{\sigma_1 \sigma'_1} \xi_{\sigma_2 \sigma'_2}  ( {}_{\pm} \Phi^{\rm (out)}_{ A, {\bm p}_{\perp}, p_{\eta}, \sigma'_1 } | {}_{\mp} \Phi^{\rm (in)}_{A, {\bm p}'_{\perp}, p'_{\eta}, \sigma_1 }  )_{\rm B} ( {}_{\mp} \Phi^{\rm (in)}_{A, {\bm p}'_{\perp}, p'_{\eta}, \sigma_2 } | {}_{\pm} \Phi^{\rm (out)}_{ A, {\bm p}_{\perp}, p_{\eta}, \sigma'_2 }  )_{\rm B} , \label{eq__84}
\end{align}
where the use is made of the normalization condition for the polarization vector Eq.~(\ref{eq.norm}).  By noting that the indefinite metric $\xi_{\sigma \sigma'}$ has an off-diagonal structure as defined in Eq.~(\ref{eq76}), one finally finds
\begin{align}
	\frac{d^3 N^{({\rm g})}_{\pm A, \sigma}}{d{\bm p}_{\perp}^2 dp_{\eta}}
		&= 
		\left\{ 
			\begin{array}{ll}
				\displaystyle  \int d^2{\bm p}'_{\perp} dp'_{\eta}  \left| ( {}_{\pm} \Phi^{\rm (out)}_{ A, {\bm p}_{\perp}, p_{\eta}, \sigma } | {}_{\mp} \Phi^{\rm (in)}_{ A, {\bm p}'_{\perp}, p'_{\eta}, \sigma }  )_{\rm B} \right|^2 & {\rm for}\ \sigma=1,2 \\
				&\\
				0 & {\rm for}\ \sigma=0,3 
			\end{array}
		\right. . \label{eq__90}
\end{align}
It is now evident that gluons with the 0-th and the 3-rd polarizations vanish.  This is consistent with our expectation that only two out of four polarization modes of gluons are physical.  We stress that deriving the particle spectrum is thus reduced to finding out the mode functions ${}_{\pm} W_{\mu, A, {\bm p}_{\perp}, p_{\eta}, \sigma}^{\rm (as)}$, or ${}_{\pm} \Phi_{A, {\bm p}_{\perp}, p_{\eta}, \sigma}^{\rm (as)}$, by solving the equation of motion Eq.~(\ref{eq48}) non-perturbatively with respect to the classical field.

Unlike the quark case, the phase space density can exceed unity because bosons are not subject to the Pauli principle.  Indeed, one can show from the normalization condition for ${}_{\pm} \Phi^{\rm (as)}_{A, {\bm p}_{\perp}, p_{\eta}, \sigma}$ that the inner products between ${}_{\pm} \Phi^{\rm (in)}_{A, {\bm p}_{\perp}, p_{\eta}, \sigma}$ and ${}_{\pm} \Phi^{\rm (out)}_{A, {\bm p}_{\perp}, p_{\eta}, \sigma}$ are normalized as
\begin{align}
	\frac{1}{(2\pi)^3} \int d^2{\bm x}_{\perp} \int d\eta
		&= \int d^2{\bm p}'_{\perp} dp'_{\eta} \left[   \left| ( {}_{\pm} \Phi^{\rm (out)}_{ A, {\bm p}_{\perp}, p_{\eta}, \sigma } | {}_{\pm} \Phi^{\rm (in)}_{ A, {\bm p}'_{\perp}, p'_{\eta}, \sigma }  )_{\rm B} \right|^2 - \left| ( {}_{\pm} \Phi^{\rm (out)}_{ A, {\bm p}_{\perp}, p_{\eta}, \sigma } | {}_{\mp} \Phi^{\rm (in)}_{ A, {\bm p}'_{\perp}, p'_{\eta}, \sigma }  )_{\rm B} \right|^2 \right] \label{eqb__65}
\end{align}
for the physical polarization modes $\sigma=1,2$.  One finds that the inner products $\left| ( {}_{\pm} \Phi^{\rm (out)}_{ A, {\bm p}_{\perp}, p_{\eta}, \sigma } | {}_{\mp} \Phi^{\rm (in)}_{ A, {\bm p}'_{\perp}, p'_{\eta}, \sigma' }  )_{\rm B} \right|^2$, or the phase space density is not bounded because of the $-$ sign in Eq.~(\ref{eqb__65}).  Notice that for the quark case Eq.~(\ref{eq__65}), we have $+$ sign, which reflects the statistics of particles, i.e., $+$ for fermions and $-$ for bosons.  
\end{widetext}

Finally, let us connect the $p_{\eta}$-spectrum Eq.~(\ref{eq91}) to the $p_z$-spectrum and/or the momentum rapidity $y_{\bm p}$-spectrum as was done in the quark case.  The $p_z$-spectrum and/or the momentum rapidity $y_{\bm p}$-spectrum are obtained as an expectation value of the number operator $n^{\rm (g)}_{\pm A,{\bm p}_{\perp}, p_z, \sigma}$, which are labeled by $p_z$ instead of $p_{\eta}$, as
\begin{align}
	\frac{d^3N^{({\rm g})}_{ \pm A,\sigma}}{d{\bm p}_{\perp}^2 dp_{z}} = \frac{1}{\omega_{\bm p}}\frac{d^3N^{\rm (g)}_{ \pm A,\sigma}}{d{\bm p}_{\perp}^2 dy_{\bm p}} &\equiv \frac{\bra{{\rm vac;in}}  n^{\rm (g)}_{ \pm A,{\bm p}_{\perp}, p_z, \sigma}  \ket{{\rm vac;in}}}{\braket{ {\rm vac;in} | {\rm vac;in} }} , \label{eq-60}
\end{align}
where the number operator $n^{\rm (g)}_{\pm A,{\bm p}_{\perp}, p_z, \sigma}$ is defined by the annihilation operators $c^{\rm (out)}_{A,{\bm p}_{\perp}, p_z, \sigma}, d^{\rm (out)}_{A,{\bm p}_{\perp}, p_z, \sigma}$ as
\begin{align}
	n^{\rm (g)}_{A,{\bm p}_{\perp}, p_z, \sigma} 
		&\equiv c^{{\rm (out)}\dagger}_{A,{\bm p}_{\perp}, p_z, \sigma} c^{\rm (out)}_{A,{\bm p}_{\perp}, p_z, \sigma}  , \nonumber\\ 
	n^{\rm (g)}_{-A,{\bm p}_{\perp}, p_z, \sigma} 
		&\equiv d^{{\rm (out)}\dagger}_{A,{\bm p}_{\perp}, p_z, \sigma}  d^{\rm (out)}_{A,{\bm p}_{\perp}, p_z, \sigma} .
\end{align}
The annihilation operators $c^{\rm (out)}_{A,{\bm p}_{\perp}, p_z, \sigma}, d^{\rm (out)}_{A,{\bm p}_{\perp}, p_z, \sigma}$ are defined by expanding the gluon operator $W_{m,A} = e^{\mu}_{\;\; m} W_{\mu,A}$ in terms of positive/negative frequency mode functions ${}_{\pm} W^{\rm (out)}_{m,A,{\bm p}_{\perp}, p_{z}, \sigma}$ in the Cartesian coordinates which are labeled by $p_z$ being the Fourier conjugate to $z$ as
\begin{align}
	W_{m,A}(x) 	
		&= \sum_{\sigma} \int d{\bm p}_{\perp}^2 dp_{z} \nonumber\\
		&\ \ \ \ \ \left[ {}_+ W^{({\rm out})}_{m,A,{\bm p}_{\perp},p_{z},\sigma}(x) c^{({\rm out})}_{A,{\bm p}_{\perp},p_{z},\sigma} \right. \nonumber\\
		&\ \ \ \ \ \ \ \ \left.+ {}_- W^{({\rm out})}_{m,A,{\bm p}_{\perp},p_{z},\sigma}(x)   d^{({\rm out})\dagger}_{A,-{\bm p}_{\perp},-p_{z},\sigma}   \right]. \label{eq-62}
\end{align}
Here, we again require the plane wave boundary condition ${}_{\pm} W^{\rm (out)}_{m, A,{\bm p}_{\perp}, p_{z}, \sigma}$ at $t \rightarrow \infty$.  As is shown in Appendix~\ref{appA:gluon}, if properly normalized, the mode functions in the Cartesian coordinates ${}_{\pm} W^{\rm (out)}_{m, A,{\bm p}_{\perp}, p_{z}, \sigma}$ and those in the $\tau$-$\eta$ coordinates ${}_{\pm} W^{\rm (out)}_{\mu, A,{\bm p}_{\perp}, p_{\eta}, \sigma}$ are related with each other by an integral transformation described by
\begin{align}
	{}_{\pm} W^{\rm (out)}_{\mu ,A,{\bm p}_{\perp}, p_{\eta}, \sigma}
		= e^{m}_{\ \ \mu} \int dp_z \frac{ {\rm e}^{\pm i p_{\eta} y_{\bm p}} }{\sqrt{2\pi \omega_{\bm p}}} {}_{\pm} W^{\rm (out)}_{m,A,{\bm p}_{\perp}, p_{z}, \sigma}.   \label{eq--62}
\end{align}
Using this integral transformation Eq.~(\ref{eq--62}) and comparing the expansion in the Cartesian coordinates Eq.~(\ref{eq-62}) with that in the $\tau$-$\eta$ coordinates Eq.~(\ref{eq__69}), one finds 
\begin{align}
	 c^{({\rm out})}_{A,{\bm p}_{\perp},p_{z},\sigma} 
		&= \int dp_{\eta} \frac{ {\rm e}^{ i p_{\eta} y_{\bm p}} }{\sqrt{2\pi \omega_{\bm p}}} c^{({\rm out})}_{A,{\bm p}_{\perp},p_{\eta},\sigma}, \label{eq-65} \\
	 d^{({\rm out})\dagger}_{A,{\bm p}_{\perp},p_{z},\sigma} 
		&= \int d p_{\eta} \frac{ {\rm e}^{- i p_{\eta} y_{\bm p}} }{\sqrt{2\pi \omega_{\bm p}}} d^{({\rm out})\dagger}_{A,{\bm p}_{\perp},p_{\eta},\sigma} .  \label{eq-66}
\end{align}
Inserting these relations, Eqs.~(\ref{eq-65}) and (\ref{eq-66}), back into Eq.~(\ref{eq-60}), one obtains
\begin{align}
	\frac{d^3N^{({\rm g})}_{A,\sigma}}{d{\bm p}_{\perp}^2 dp_{z}} 
		&= \frac{1}{\omega_{\bm p}}\frac{d^3N^{\rm (g)}_{A,\sigma}}{d{\bm p}_{\perp}^2 dy_{\bm p}} \nonumber\\ 
		&= \frac{1}{\omega_{\bm p}} \int dp_{\eta} dp'_{\eta}  \frac{ {\rm e}^{i y_{\bm p} (p_{\eta} - p'_{\eta}) } }{2\pi}  \nonumber\\
		&\ \times \frac{\bra{{\rm vac; in}}  c^{{\rm (out)}\dagger}_{A,{\bm p}_{\perp}, p'_{\eta}, \sigma} c^{{\rm (out)}}_{A,{\bm p}_{\perp}, p_{\eta}, \sigma}  \ket{{\rm vac; in}}}{\braket{ {\rm vac; in} | {\rm vac; in} }} , \label{eq--67} \\
		\frac{d^3N^{({\rm g})}_{-A,\sigma}}{d{\bm p}_{\perp}^2 dp_{z}} 
		&= \frac{1}{\omega_{\bm p}}\frac{d^3N^{({\rm g})}_{-A,\sigma}}{d{\bm p}_{\perp}^2 dy_{\bm p}} \nonumber\\
		&= \frac{1}{\omega_{\bm p}} \int dp_{\eta} dp'_{\eta}  \frac{ {\rm e}^{i y_{{\bm p}} (p_{\eta} - p'_{\eta}) } }{2\pi} \nonumber\\
		&\ \times  \frac{\bra{{\rm vac; in}}  d^{{\rm (out)}\dagger}_{A,{\bm p}_{\perp}, p'_{\eta}, \sigma} d^{{\rm (out)}}_{A,{\bm p}_{\perp}, p_{\eta}, \sigma}  \ket{{\rm vac; in}}}{\braket{ {\rm vac; in} | {\rm vac; in} }} . \label{eq--68}
\end{align}
When the system is perfectly boost invariant, the expectation values in Eqs.~(\ref{eq--67}) and (\ref{eq--68}) for $p_{\eta} \neq p'_{\eta}$ vanish as in the quark case and one finally obtains
\begin{align}
	\frac{d^3N^{({\rm g})}_{\pm A,\sigma}}{d{\bm p}_{\perp}^2 dp_{z}} 
		&= \frac{1}{\omega_{\bm p}}\frac{d^3N^{\rm (g)}_{\pm A,\sigma}}{d{\bm p}_{\perp}^2 dy_{\bm p}} \nonumber\\ 
		&= \frac{1}{2\pi \delta(p_{\eta} = 0) } \times \frac{1}{\omega_{\bm p}} \int dp_{\eta} \frac{d^3N^{({\rm g})}_{\pm A, \sigma}}{d{\bm p}_{\perp}^2 dp_{\eta}}.  
\end{align}

\subsubsection{Ghost}

Finally, we consider the canonical quantization of the ghost and anti-ghost fields, $C_A$ and $\bar{C}_A$, and show that ghosts are never produced from the classical field.  We do essentially the same procedure as what we did in the previous quark and gluon cases.

We first expand the ghost and anti-ghost fields, $C_A$ and $\bar{C}_A$, as 
\begin{align}
	\begin{pmatrix} C_A \\ \bar{C}_A  \end{pmatrix} 
		&= \int d{\bm p}^2_{\perp} dp_{\eta} \nonumber\\
		&\ \ \ \ \ \ \ \left[ {}_+ \Theta^{({\rm as})}_{A, {\bm p}_{\perp},p_{\eta}} \begin{pmatrix} e^{({\rm as})}_{A, {\bm p}_{\perp},p_{\eta} } \\ \bar{e}^{({\rm as})}_{A, {\bm p}_{\perp},p_{\eta} } \end{pmatrix}  \right.\nonumber\\
		&\ \ \ \ \ \ \ \ \ \ \ \ \ \left.+  {}_- \Theta^{({\rm as})}_{A, {\bm p}_{\perp},p_{\eta}}   \begin{pmatrix} f^{({\rm as})\dagger}_{A, -{\bm p}_{\perp}, -p_{\eta} } \\ \bar{f}^{({\rm as})\dagger}_{A, -{\bm p}_{\perp}, -p_{\eta} } \end{pmatrix} \right], 
\end{align}
where the labels are the same as in the previous two cases.  The mode functions ${}_{\pm} \Theta^{({\rm as})}_{A, {\bm p}_{\perp},p_{\eta}}$ are the solutions of the equations of motion Eq.~(\ref{eq50}) with the plane wave boundary condition at $t \rightarrow \pm\infty$: 
\begin{align}
	{}_{\pm} \Theta_{A,{\bm p}_{\perp},p_{\eta}}^{({\rm in})} &\xrightarrow[t \rightarrow -\infty]{} {}_{\pm} \Theta_{A, {\bm p}_{\perp}, p_{\eta}}^{{(\rm free})} [\bar{A}_{\mu}(t \rightarrow -\infty)], \\
	{}_{\pm} \Theta_{A, {\bm p}_{\perp},p_{\eta} }^{({\rm out})} &\xrightarrow[t \rightarrow \infty]{} {}_{\pm} \Theta_{A,{\bm p}_{\perp},p_{\eta} }^{({\rm free})} [\bar{A}_{\mu}(t \rightarrow \infty)], 
\end{align}
where the plane wave solutions ${}_{\pm} \Theta_{A, {\bm p}_{\perp}, p_{\eta} }^{{(\rm free})}[\breve{A}_{\mu}]$ satisfy the free field equation of motion under a pure gauge background field $\breve{A}_{\mu} = \bar{A}_{\mu}(t \rightarrow \pm\infty)$.  For details of the plane wave solutions $ {}_{\pm} \Theta_{ A, {\bm p}_{\perp}, p_{\eta} }^{({\rm free})} $, see Appendix~\ref{appA:ghost}. The normalization conditions for the positive/negative frequency mode functions ${}_{\pm} \Theta_{ A, {\bm p}_{\perp}, p_{\eta} }^{({\rm as})} $ are 
\begin{align}
	( {}_{\pm} \Theta_{A, {\bm p}_{\perp}, p_{\eta} }^{({\rm as})} |  {}_{\pm} \Theta_{A, {\bm p}'_{\perp}, p'_{\eta} }^{({\rm as})} )_{\rm B} &= \pm \delta^2({\bm p}_{\perp} - {\bm p}'_{\perp}) \delta (p_{\eta} - p'_{\eta}) \label{eq_96} \\
	( {}_{\pm} \Theta_{ A, {\bm p}_{\perp}, p_{\eta} }^{({\rm as})}  |  {}_{\mp} \Theta_{ A, {\bm p}'_{\perp}, p'_{\eta} }^{({\rm as})}  )_{\rm B} &= 0 \label{eq_97}.  
\end{align}

Next, we canonically quantize the fluctuations by imposing canonical commutation relations: 
\begin{align}
	&\{ \overset{(-)}{C}_{A} (\tau, {\bm x}_{\perp}, \eta) , \overset{(-)}{\pi}_{A'} (\tau, {\bm x}'_{\perp}, \eta') \} \nonumber\\
	&\ \ \ \ \ \ \ \ \ \ \ \ = i \delta_{AA'} \delta^2 ({\bm x}_{\perp} - {\bm x}'_{\perp} ) \frac{\delta (\eta - \eta')}{\tau}, \label{eq_100}\\
	&\{ \overset{(-)}{C}_{A} (\tau, {\bm x}_{\perp}, \eta) , \overset{(-)}{C}_{A'} (\tau, {\bm x}'_{\perp}, \eta') \} \nonumber\\
	&\ \ \ \ \ \ \ \ \ \ \ \ = \{ \overset{(-)}{\pi}_{A} (\tau, {\bm x}_{\perp}, \eta) , \overset{(-)}{\pi}_{A'} (\tau, {\bm x}'_{\perp}, \eta') \} = 0, \label{eq101}
\end{align}
where the canonical conjugate field to the ghost and anti-ghost field, $C_{A}$ and $\bar{C}_A$, are given by $\pi_{A} = \delta {\mathcal L}/\delta ( \partial_{\tau} C_{A}) = -i \partial_{\tau} \bar{C}_A^{\dagger}$ and $\bar{\pi}_{A} = \delta {\mathcal L}/\delta ( \partial_{\tau} \bar{C}_{A}) = i \partial_{\tau} C_A^{\dagger}$, respectively.  As a result of the canonical commutation relations, one finds the following anti-commutation relations for the operators $e^{({\rm as})}_{A,{\bm p}_{\perp}, p_{\eta}}, \bar{e}^{({\rm as})}_{A,{\bm p}_{\perp}, p_{\eta}}, f^{({\rm as})}_{A,{\bm p}_{\perp}, p_{\eta}}, \bar{f}^{({\rm as})}_{A,{\bm p}_{\perp}, p_{\eta}}$ given by
\begin{align}
	&\{ e^{({\rm as})}_{A,{\bm p}_{\perp}, p_{\eta} } \;,\; \bar{e}^{ ({\rm as}) \dagger}_{A',{\bm p}'_{\perp}, p'_{\eta} }   \} \nonumber\\
	&\ \ \ \ \ \ \ \ \ \ =  \{ f^{({\rm as})}_{A,{\bm p}_{\perp}, p_{\eta}} \;,\; \bar{f}^{ ({\rm as})\dagger}_{A',{\bm p}'_{\perp}, p'_{\eta}}  \} \nonumber\\
	&\ \ \ \ \ \ \ \ \ \ \ \ \ \ \ = i\delta_{AA'} \delta^2({\bm p}_{\perp} - {\bm p}'_{\perp}) \delta(p_{\eta} - p'_{\eta}) , \label{eq_104} \\
	&({\rm otherwise}) = 0.  \label{eq_105}
\end{align}
Now, one can understand that the operators, $e^{({\rm as})}_{A,{\bm p}_{\perp},  p_{\eta}}, f^{({\rm as})}_{A,{\bm p}_{\perp}, p_{\eta}}$ ($ \bar{e}^{({\rm as})}_{A,{\bm p}_{\perp}, p_{\eta}}, \bar{f}^{({\rm as})}_{A,{\bm p}_{\perp}, p_{\eta}}$), act as annihilation operators of a ghost (an anti-ghost) at the corresponding asymptotic time with the momentums ${\bm p}_{\perp}, p_{\eta}$ and the color charge $q^{({\rm gh})}_A, -q^{({\rm gh})}_A$.

As is seen in the previous two cases, the creation/annihilation operators at different asymptotic times do not coincide with each other and the linear relation is given by the following Bogoliubov transformation:
\begin{widetext}
\begin{align}
	\begin{pmatrix} 
		\overset{(-)}{e}^{(\rm out)}_{A,{\bm p}_{\perp}, p_{\eta}} \\ 
		\overset{(-)}{f}^{(\rm out)\dagger}_{A,-{\bm p}_{\perp}, -p_{\eta} } 
	\end{pmatrix}
	&=
	\begin{pmatrix} 
		( {}_{+} \Theta_{ A,{\bm p}_{\perp},p_{\eta} }^{({\rm out})} | \overset{(-)}{C}_{A} )_{\rm B} \\ 
		-( {}_{-} \Theta_{ A,{\bm p}_{\perp},p_{\eta} }^{({\rm out})} | \overset{(-)}{C}_{A} )_{\rm B}
	\end{pmatrix} \nonumber\\
	&=
	\int d^2{\bm p}'_{\perp} dp'_{\eta} 
	\begin{pmatrix}
		( {}_{+} \Theta_{ A,{\bm p}_{\perp},p_{\eta} }^{({\rm out})} | {}_{+} \Theta_{ A,{\bm p}_{\perp},p_{\eta} }^{({\rm in})} )_{\rm B}	&	( {}_{+} \Theta_{ A,{\bm p}_{\perp},p_{\eta} }^{({\rm out})} | {}_{-} \Theta_{ A,{\bm p}_{\perp},p_{\eta} }^{({\rm in})} )_{\rm B} \\
		-( {}_{-} \Theta_{ A,{\bm p}_{\perp},p_{\eta} }^{({\rm out})} | {}_{+} \Theta_{ A,{\bm p}_{\perp},p_{\eta} }^{({\rm in})} )_{\rm B}	&	-( {}_{-} \Theta_{ A,{\bm p}_{\perp},p_{\eta} }^{({\rm out})} | {}_{-} \Theta_{ A,{\bm p}_{\perp},p_{\eta} }^{({\rm in})} )_{\rm B} 
	\end{pmatrix}
	\begin{pmatrix} 
		\overset{(-)}{e}^{(\rm in)}_{A,{\bm p}'_{\perp}, p'_{\eta}} \\ 
		\overset{(-)}{f}^{(\rm in)\dagger}_{A,-{\bm p}'_{\perp}, -p'_{\eta} } 
	\end{pmatrix} .  \label{eq_106}
\end{align}
\end{widetext}

In order to obtain the ghost and anti-ghost spectrum at $t \rightarrow \infty$, let us introduce a ghost and an anti-ghost number density operators $n^{({\rm gh})}_{\pm A,{\bm p}_{\perp}, p_{\eta}}$ and $n^{(\bar{\rm gh})}_{\pm A,{\bm p}_{\perp}, p_{\eta}}$, respectively, by
\begin{align}
	n^{({\rm gh})}_{A,{\bm p}_{\perp}, p_{\eta} } 
		\equiv e^{(\rm out)\dagger}_{A,{\bm p}_{\perp}, p_{\eta} } e^{(\rm out)}_{A,{\bm p}_{\perp}, p_{\eta} }, \ 
	n^{({\rm gh})}_{-A,{\bm p}_{\perp}, p_{\eta} } 
		\equiv f^{(\rm out)\dagger}_{A,{\bm p}_{\perp}, p_{\eta} } f^{(\rm out)}_{A,{\bm p}_{\perp}, p_{\eta} }
\end{align}
and
\begin{align}
	n^{(\bar{\rm gh})}_{A,{\bm p}_{\perp}, p_{\eta} } 
		\equiv \bar{e}^{(\rm out)\dagger}_{A,{\bm p}_{\perp}, p_{\eta} } \bar{e}^{(\rm out)}_{A,{\bm p}_{\perp}, p_{\eta} }, \  
	n^{(\bar{\rm gh})}_{-A,{\bm p}_{\perp}, p_{\eta} } 
		\equiv \bar{f}^{(\rm out)\dagger}_{A,{\bm p}_{\perp}, p_{\eta} } \bar{f}^{(\rm out)}_{A,{\bm p}_{\perp}, p_{\eta} }. 
\end{align}
As in the previous two cases, the ghost and the anti-ghost spectrum can be derived as an expectation value of the number density operators.  By using the Bogoliubov transformation Eq.~(\ref{eq_106}) and the fact that the commutation relation between $\overset{(-)}{e}^{({\rm as})}_{A,{\bm p}_{\perp}, p_{\eta} } (\overset{(-)}{f}^{({\rm as})}_{A,{\bm p}_{\perp}, p_{\eta} })$ and $ \overset{(-)}{e}^{ ({\rm as}) \dagger}_{A',{\bm p}'_{\perp}, p'_{\eta} } (\overset{(-)}{f}^{ ({\rm as}) \dagger}_{A',{\bm p}'_{\perp}, p'_{\eta}})$ vanishes because of the anti-commutation relations Eqs.~(\ref{eq_104}) and (\ref{eq_105}), one finds
\begin{align}
	\frac{d^3N^{({\rm gh})}_{\pm A}}{d{\bm p}_{\perp}^2 dp_{\eta}} &= \frac{\bra{{\rm vac;in}} n^{({\rm gh})}_{\pm A,{\bm p}_{\perp}, p_{\eta}} \ket{{\rm vac;in}}}{ \braket{{\rm vac;in}| {\rm vac;in}} } = 0, \label{eq111} \\
	\frac{d^3N^{(\bar{\rm gh})}_{\pm A}}{d{\bm p}_{\perp}^2 dp_{\eta}} &= \frac{\bra{{\rm vac;in}} n^{(\bar{\rm gh})}_{\pm A,{\bm p}_{\perp}, p_{\eta}g} \ket{{\rm vac;in}}}{ \braket{{\rm vac;in}| {\rm vac;in}} } = 0.\label{eq112}
\end{align}
That is, ghosts and anti-ghosts are never produced from the classical field $\bar{A}_{\mu}$.  This is a reasonable result because ghosts and anti-ghosts are unphysical particles and that they never appear in the physical spectrum.  In general, the RHS is always zero for any physical initial state $\ket{{\rm phys; in}}$ because any physical state does not contain ghosts or anti-ghosts.

\section{Particle Production from an Expanding Color Electric Field} \label{sec3}

In Section \ref{sec2}, we have shown, at the one-loop level quantum calculation and within the Abelian dominance assumption for the classical background field $\bar{A}_{\mu}$, that the particle spectra are obtained by solving the equations of motion of QCD non-perturbatively with respect to the classical field.  In principle, the equations of motion are solvable, i.e., the particle spectra are computable for any $\bar{A}_{\mu}$ with arbitrary spacetime dependence, for instance, by using numerical methods.  However, before going into more realistic calculations, where $\bar{A}_{\mu}$ has a complicated spacetime dependence, we consider a simple situation, where analytic solutions of the equations of motion are available.  This enables us to to get more insights on the particle production in QCD in an expanding system.  In particular, we consider a spatially homogeneous and constant classical color electric background field with finite lifetime $T$ in a boost-invariantly expanding geometry, i.e., ${\bm E} = {\bm e}_{z} E \theta(\tau) \theta(T-\tau), {\bm B}={\bm 0}$ given by a gauge potential:
\begin{align}
	\tilde{A}_{\tau}, \tilde{A}_x, \tilde{A}_y = 0, \ 
	\tilde{A}_{\eta} = 	\left\{   
							\begin{array}{ll} 
								E\tau^2 /2 	& 	(0<\tau<T) \\ 
								ET^2/2 		&	(T < \tau) 
							\end{array} 
						\right. . \label{eq114}
\end{align}  
As is explained in Appendix~\ref{appA}, the analytical formula for the particle spectra become
\begin{widetext}
\begin{align}
	\frac{d^3 N^{\rm (q)}_{i,f,s}}{d^2{\bm p}_{\perp} dy_{\bm p}}
			= \frac{d^3 N^{(\bar{\rm q})}_{i,f,s}}{d^2{\bm p}_{\perp} dy_{\bm p}}
			= \frac{S_{\perp}}{(2\pi)^3} \int dp_{\eta} \left|   A^{\rm (q)}_{i,f,{\bm p}_{\perp}, p_{\eta}-q^{\rm (q)}_iET^2/2, s}(0) B^{{\rm (q)}*}_{i,f,{\bm p}_{\perp}, p_{\eta}, s}(T) - B^{{\rm (q)}*}_{i,f,{\bm p}_{\perp}, p_{\eta}-q^{\rm (q)}_i ET^2/2, s}(0) A^{{\rm (q)}}_{i,f,{\bm p}_{\perp}, p_{\eta}, s}(T) \right|^2
\end{align}
for quarks and anti-quarks, 
\begin{align}
	\frac{d^3 N^{\rm (g)}_{\pm A, \sigma}}{d^2{\bm p}_{\perp} dy_{\bm p}}
			= \frac{S_{\perp}}{(2\pi)^3} \int dp_{\eta} \left|   A^{\rm (g)}_{A,{\bm p}_{\perp}, p_{\eta}-q^{\rm (g)}_A ET^2/2, \sigma}(0) B^{{\rm (g)}*}_{A,{\bm p}_{\perp}, p_{\eta}, \sigma}(T) - B^{{\rm (g)}*}_{A,{\bm p}_{\perp}, p_{\eta}-q^{\rm (g)}_A ET^2/2, \sigma}(0) A^{{\rm (g)}}_{A,{\bm p}_{\perp}, p_{\eta}, \sigma}(T) \right|^2
\end{align}
\end{widetext}
for physical gluons ($\sigma = 1,2$).  Here, we have used $\delta^2({\bm p}_{\perp}={\bm 0}) = S_{\perp}/(2\pi)^2$ with  $S_{\perp}$ being the transverse area.  For the explicit expressions for the Bogoliubov coefficients $A^{\rm (q)}, B^{\rm (q)}$ and $A^{\rm (g)}, B^{\rm (g)}$, see Appendix~\ref{appC:quark} and Appendix~\ref{appC:gluon}, respectively.  Notice that unphysical gluons ($\sigma=0,3$) and ghosts are never produced as shown in Eq.~(\ref{eq__90}), and in Eqs.~(\ref{eq111}) and (\ref{eq112}), and hence we do not consider them hereafter.  In the following, we numerically carry out the $p_{\eta}$-integration and show the momentum-rapidity $y_{\bm p}$-spectra for quarks and gluons.

\subsection{Features of particle production}

In this subsection, we investigate specific features of quark and gluon production focusing on impacts of the longitudinal expansion and of finite lifetime effects.  For this purpose, we treat the quark mass $m_{\rm f}$, the coupling $g$, the field strength $E$, the lifetime $T$ as free parameters, and compute the quark and gluon spectra without taking the summation of colors, $i$ and $A$.

We stress that the particle spectra for a fixed color discussed here are very useful in understanding the specific features of the particle production.  However, the spectra are apparently gauge-dependent and hence one has to take the color summation in order to get a physically meaningful results, which are discussed in Section \ref{sec:phen}.

\subsubsection{Transverse Distribution $d^3N/dy_{\bm p} d{\bm p}_{\perp}^2$}

\begin{figure*}
\begin{center}
\includegraphics[clip, width=0.495\textwidth]{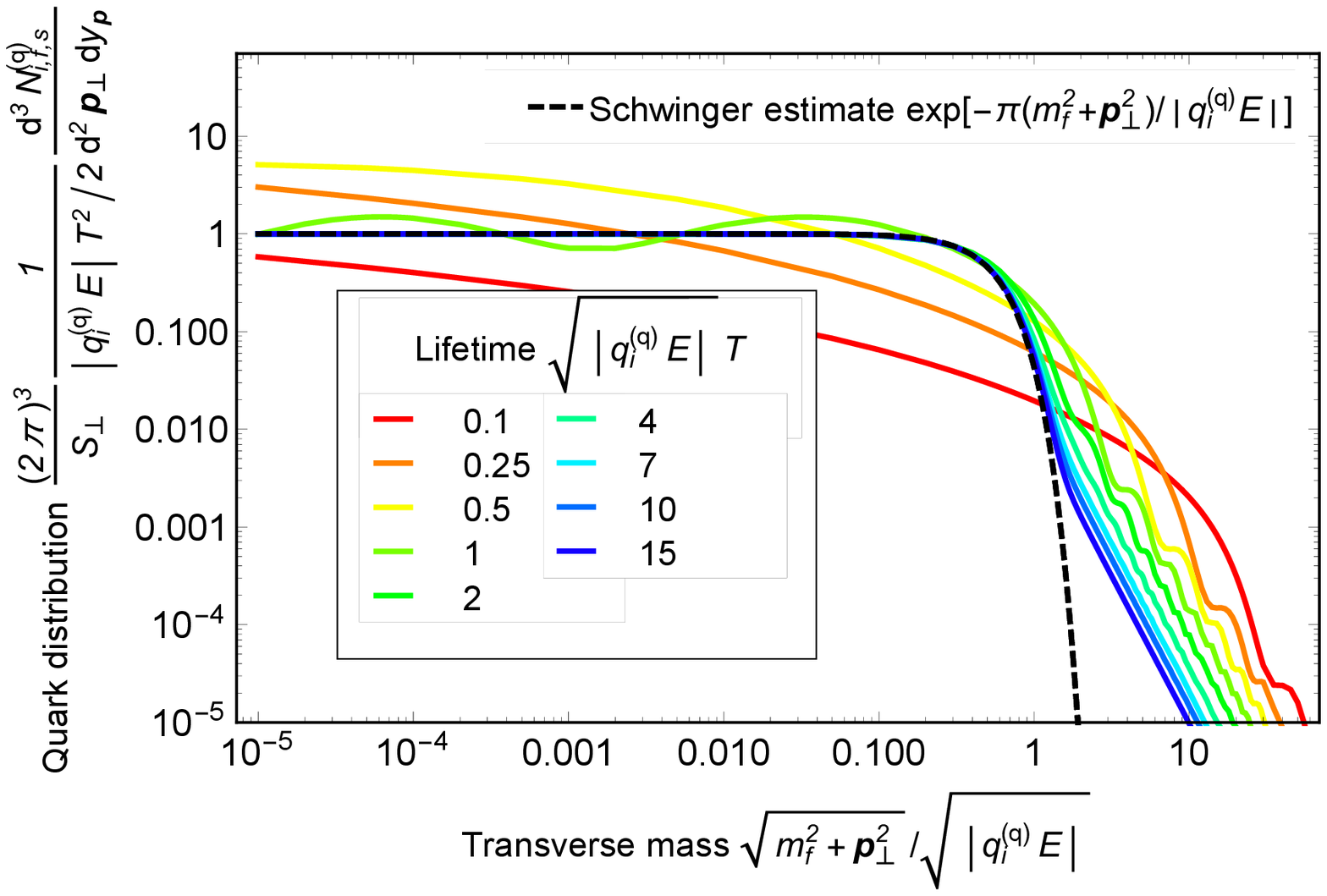}
\includegraphics[clip, width=0.495\textwidth]{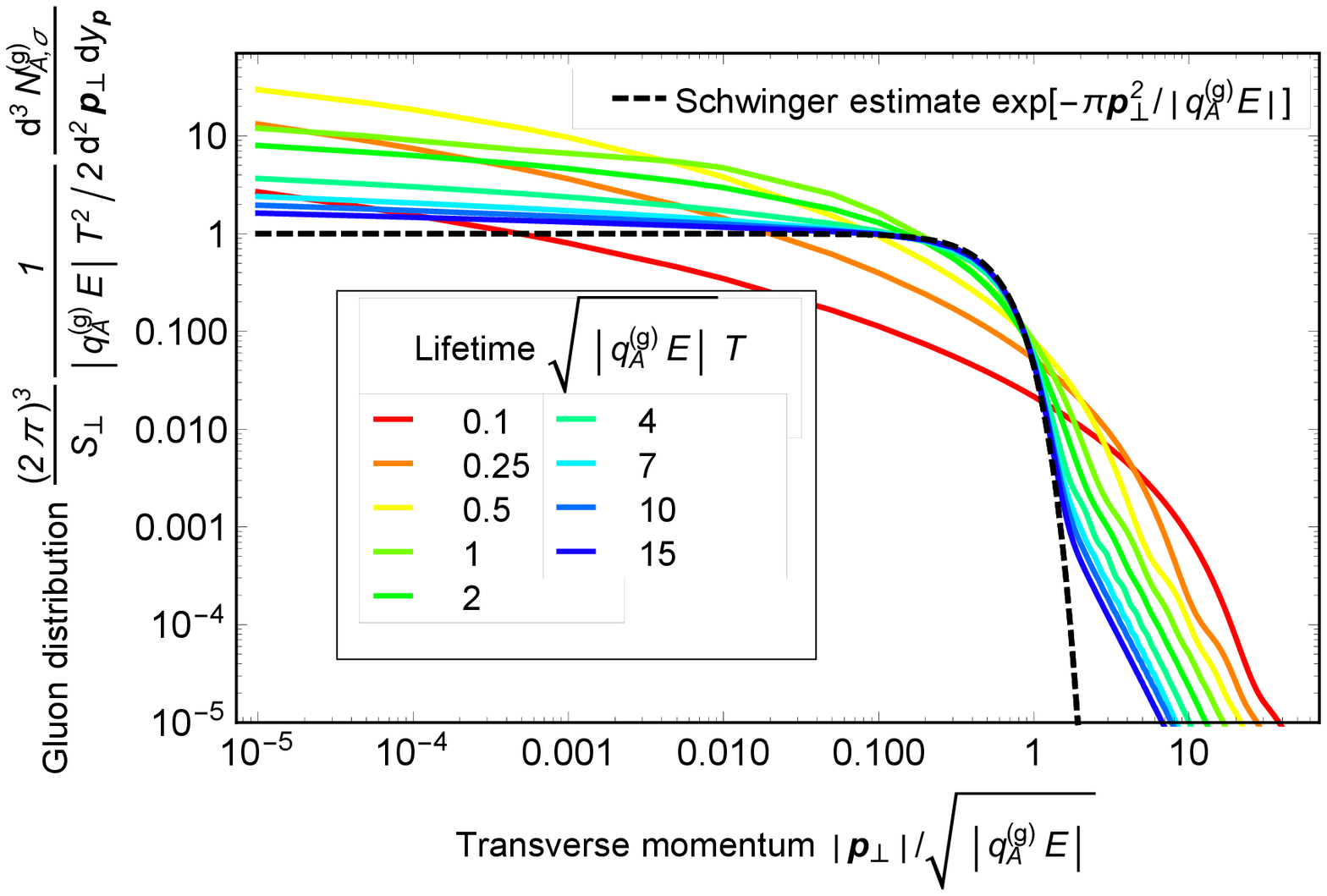}
\caption{\label{fig1} (color online) Transverse distribution of quarks (left) and gluons (right) for various lifetimes $T$.  The dashed lines are expectations from the Schwinger formula Eqs.~(\ref{eq__121}) and (\ref{eq__122}).  }
\end{center}
\end{figure*}

Figure~\ref{fig1} shows the transverse spectrum of quarks $d^3 N^{\rm (q)}_{i,f,s}/d^2{\bm p}_{\perp} dy_{\bm p}$ (left) and of gluons $d^3 N^{\rm (g)}_{A,\sigma}/d^2{\bm p}_{\perp} dy_{\bm p}$ (right).  We observe that, for long lifetimes $\sqrt{|q^{\rm (q)}_i E|}T, \sqrt{|q^{\rm (g)}_A E|}T \gtrsim 1$, both spectra approach Gaussian distributions multiplied by a square of the lifetime $T$.  This is consistent with what we naively expect from the Schwinger formula: \begin{align}
	\left.\frac{d^3 N^{\rm (q)}_{i,f,s}}{d^2{\bm p}_{\perp} dp_z}\right|_{\rm Schwinger} 
		&= \frac{V}{(2\pi)^3} \exp\left[-\pi \frac{m_{\rm f}^2 +{\bm p}_{\perp}^2}{|q_i^{\rm (q)} E|}\right], \\
	\left.\frac{d^3 N^{\rm (g)}_{A,\sigma}}{d^2{\bm p}_{\perp} dp_z}\right|_{\rm Schwinger} 
		&= \frac{V}{(2\pi)^3} \exp\left[-\pi \frac{{\bm p}_{\perp}^2}{|q_A^{\rm (g)} E|}\right].   
\end{align}
Indeed, for large values of $T$, the produced particles are sufficiently accelerated by the electric field as $\omega_{\bm p} \sim |p_z| \sim |q A_z| $ ($q=q_i^{\rm (q)}$ for quarks and $q=q_A^{\rm (g)}$ for gluons), and $A_z \sim A_{\eta}/T = ET/2$ and $V \sim S_{\perp} T$ hold.  Thus, we find
\begin{align}
	\left.\frac{d^3 N^{\rm (q)}_{i,f,s}}{d^2{\bm p}_{\perp} dy_{\bm p}}\right|_{\rm Schwinger} 
		&\sim \frac{S_{\perp}}{ (2\pi)^3} \frac{| q_i^{\rm (q) } E| T^2  }{2} \exp\left[ -\pi \frac{m_{\rm f}^2 +{\bm p}_{\perp}^2}{|q_i^{\rm (q)}E|} \right], \label{eq__121}\\
	\left. \frac{d^3 N^{\rm (g)}_{A,\sigma}}{d^2{\bm p}_{\perp} dy_{\bm p}} \right|_{\rm Schwinger} 
		&\sim \frac{S_{\perp}}{ (2\pi)^3} \frac{| q_A^{\rm (g)} E| T^2  }{2} \exp\left[ -\pi \frac{{\bm p}_{\perp}^2}{|q_A^{\rm (g)}E|} \right], \label{eq__122}
\end{align}
which are plotted in the dashed lines as ``Schwinger estimate" in Fig.~\ref{fig1}.  On the other hand, for short lifetimes $\sqrt{|q^{\rm (q)}_i E|}T, \sqrt{|q^{\rm (g)}_A E|}T \lesssim 1$, the spectra are harder compared to those for larger lifetimes and do not decay exponentially in $|{\bm p}_{\perp}|$ because the typical frequency $\omega \sim 1/T$ of the classical electric field is hard enough to excite hard particles.  In other words, a naive application of the Schwinger formula is valid only for large values of the lifetime $T$ , while one should take care of finite lifetime effects for small values of $T$.

In the low momentum region $|{\bm p}_{\perp}| \lesssim \sqrt{|q_i^{\rm (q)} E|}, \sqrt{|q_A^{\rm (g)} E|}$, gluons are more abundant than quarks.  This is because the quark production is subject to the Pauli principle but the gluon production is not.  The gluon spectrum shows a weak divergence for $|{\bm p}_{\perp}| \rightarrow 0$ but its inverse power is smaller than one and it approaches zero with increasing the lifetime $T$.

\subsubsection{Number Density $dN/dy$}

We numerically integrate the transverse distributions over ${\bm p}_{\perp}$ to compute the total number of produced particles per unit rapidity for quarks $d N^{\rm (q)}_{i,f,s}/dy_{\bm p}$ and for gluons $dN^{\rm (g)}_{A,\sigma}/dy_{\bm p}$.

\begin{figure*}
\begin{center}
\includegraphics[clip, width=0.495\textwidth]{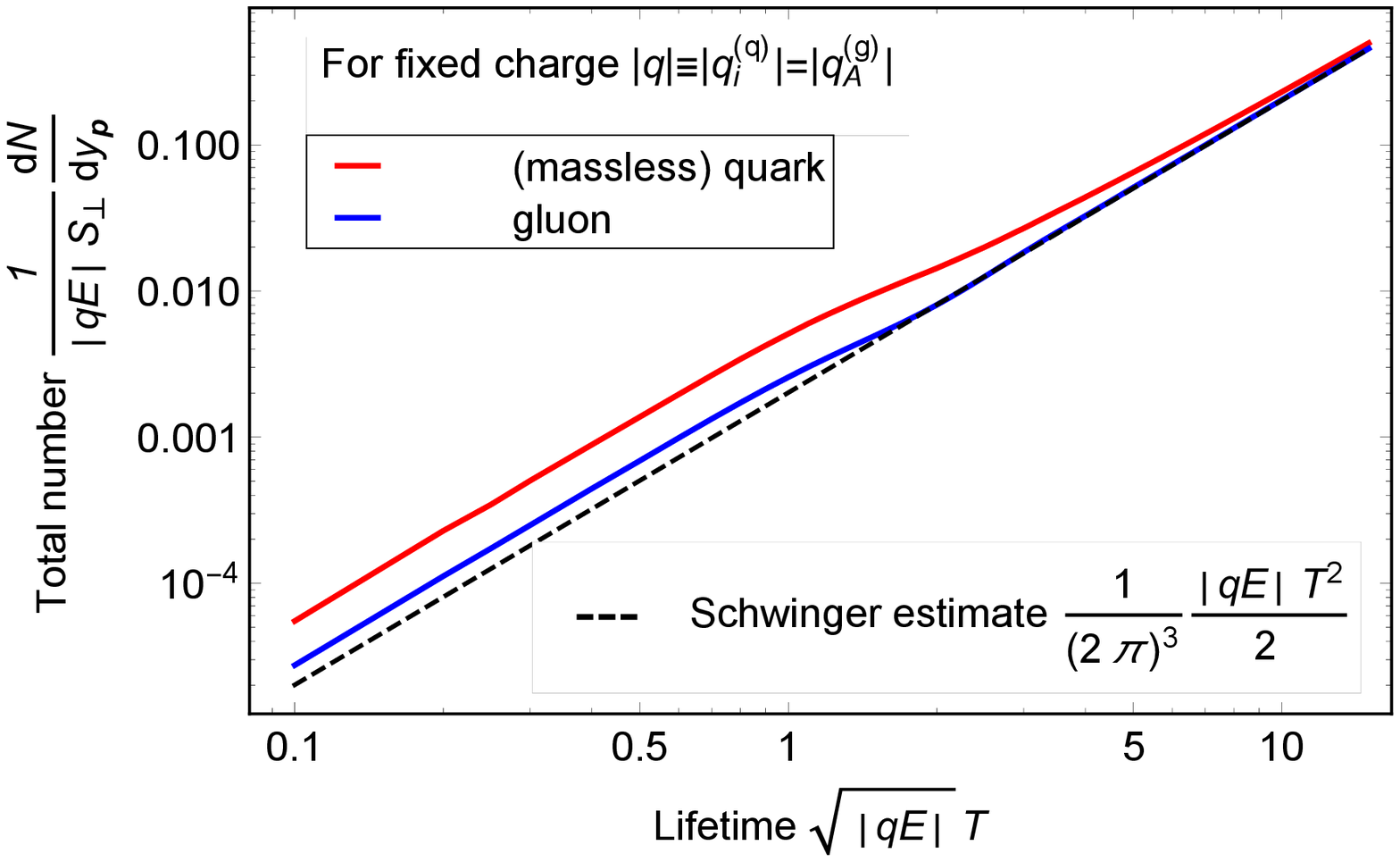}
\includegraphics[clip, width=0.495\textwidth]{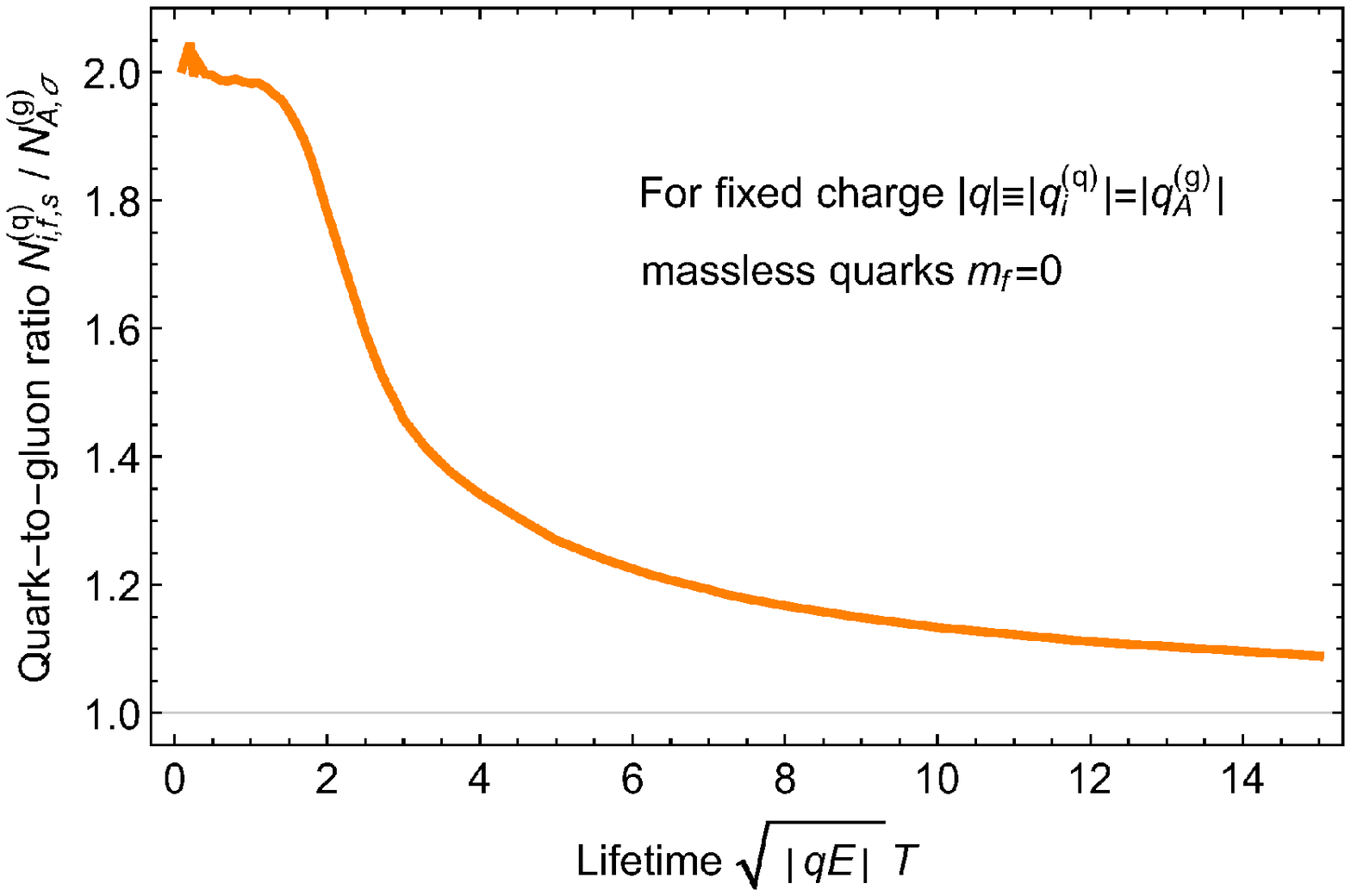}
\caption{\label{fig2}(color online) [Left] Total number of massless quarks (red) and of gluons (blue) per unit rapidity for $|q_{i}^{\rm (q)}| = |q_{A}^{\rm (g)}|$.  The dashed line is an expectation from the Schwinger formula Eqs.~(\ref{eq123}) and (\ref{eq124}).  [Right] A ratio of total number of produced massless quarks to that of gluons $N_{i,f,s}^{\rm (q)}/N_{A,\sigma}^{\rm (g)}$ for $|q_{i}^{\rm (q)}| = |q_{A}^{\rm (g)}|$.  }
\end{center}
\end{figure*}

The left panel of Fig.~\ref{fig2} shows the total number of massless quarks $d N^{\rm (q)}_{i,f,s}/dy_{\bm p}$ and of gluons $dN^{\rm (g)}_{A,\sigma}/dy_{\bm p}$.  Here, we artificially set $|q| = |q_{i}^{\rm (q)}| = |q_{A}^{\rm (g)}|$ for the quark and gluon charges.  For long lifetimes $\sqrt{|q E|}T \gtrsim 1$, one finds that the quark and gluon productions are consistent with the Schwinger formula: By integrating Eqs.~(\ref{eq__121}) and (\ref{eq__122}) over ${\bm p}_{\perp}$, the Schwinger formula gives
\begin{align}
\left.\frac{d N^{\rm (q)}_{i,f,s}}{dy_{\bm p}} \right|_{\rm Schwinger}
			&\sim \frac{S_{\perp}}{ (2\pi)^3} \frac{ |q_i^{\rm (q)} E|^2 T^2  }{2} \exp \left[ -\pi \frac{m_{\rm f}^2}{ |q_i^{\rm (q)} E| } \right] , \label{eq123}\\
\left.\frac{d N^{\rm (g)}_{\pm A, \sigma}}{dy_{\bm p}} \right|_{\rm Schwinger}
			&\sim \frac{S_{\perp}}{ (2\pi)^3} \frac{ |q_A^{\rm (g)} E|^2 T^2  }{2} \label{eq124}, 
\end{align}
which are plotted in the dashed lines in the left panel of Fig~\ref{fig2}.  For short lifetimes $\sqrt{|q E|}T \lesssim 1$, one observes that the quark and gluon production are more abundant than the Schwinger estimates.  This is because the typical frequency $\omega \sim 1/T$ of the electric field for such small values of $T$ becomes so hard that a large amount of hard particles are produced as was discussed in Fig.~\ref{fig1}, for which the phase space is larger than those for soft particles expected from the Schwinger formula.  For classical background fields with such hard frequencies, perturbative particle production from a single classical background field gives a better description than Schwinger's non-perturbative particle production mechanism \cite{tay14}, and hence this enhancement is purely a perturbative effect.

For all values of $T$, we observe that the massless quark production is more abundant than the gluon production for $|q_{i}^{\rm (q)}| = |q_{A}^{\rm (g)}|$.  This aspect is more clearly illustrated in the right panel of Fig.~\ref{fig2}: For long lifetimes $\sqrt{|q E|}T \gtrsim 1$, the ratio of the produced quarks to that of gluons $N^{\rm (q)}_{i,f,s}/N^{\rm (g)}_{A,\sigma}$ approaches unity because both quarks and gluons are produced via Schwinger's non-perturbative particle production mechanism, in which the statistics of particles are irrelevant as is seen in Eqs.~(\ref{eq123}) and (\ref{eq124}).  For short lifetimes $\sqrt{|q E|}T \lesssim 1$, however, the ratio deviates from unity.  This is because, for such small values of $T$, Schwinger's non-perturbative particle production mechanism is not efficient but perturbative particle production occurs, which depends on the statistics of particles in general.  It is interesting to point out that the ratio is always larger than unity so that quarks are more abundantly produced than gluons.  This is because quark spectrum is harder than gluon one for small values of $T$ due to the statistics of particles as we saw in Fig.~\ref{fig1} and the phase space for produced quarks becomes larger than that of gluons.

In the $T \rightarrow 0$ limit, the ratio amounts to nearly two.  In order to convince ourselves that this number ``two" is correct and that this enhancement is indeed a perturbative phenomenon due to the finite lifetime effects, we consider a non-expanding, spatially homogeneous but time-dependent electric field, ${\bm E} = E(t) {\bm e}_z$ as an example for a moment.  In this case, one can analytically compute $S$-matrix elements, $\braket{ {\rm q}_{i,f,{\bm p}_{\perp},p_z,s}\bar{{\rm q}}_{i',f',{\bm p}'_{\perp},p'_z,s'};{\rm in}  | S|  {\rm vac;in} }$ and $\braket{ {\rm g}_{A,{\bm p}_{\perp},p_z,\sigma} {\rm g}_{A',{\bm p}'_{\perp},p'_z,\sigma'} ;{\rm in} |S| {\rm vac;in} }$, in the lowest order perturbation theory with respect to the classical background field $E(t)$.  After some manipulations, one obtains 
\begin{align}
\frac{N^{\rm (q)}_{i,f,s}}{V} 
	&= 
		\sum_{i',f',s'} \int d^3{\bm p}_{\perp} dp_z \int d^2{\bm p}'_{\perp} dp'_z \nonumber\\
		&\ \ \ \ \ \ \ \ \  | \bra{ {\rm q}_{i,f,{\bm p}_{\perp},p_z,s}\bar{{\rm q}}_{i',f',{\bm p}'_{\perp},p'_z,s'};{\rm in}  } S  \ket{  {\rm vac;in} }|^2  \nonumber\\
	&=
		\frac{1}{24\pi^2} \int_{2m_{\rm f}}^{\infty} d\omega \sqrt{1-\frac{4m_{\rm f}^2}{\omega^2}} \left( 1+\frac{2m^2_{\rm f}}{\omega^2} \right) | q_{i}^{\rm (q)}\tilde{E}(\omega) |^2 \nonumber\\
	&\xrightarrow[m_{\rm f}=0]{} \frac{1}{24\pi^2} \int_0^{\infty} d\omega | q_{i}^{\rm (q)}\tilde{E}(\omega) |^2
\end{align}
and 
\begin{align}
\frac{N^{\rm (g)}_{A,\sigma}}{V}
	&=
		\sum_{\pm A',\sigma'} \int d^3{\bm p}_{\perp} dp_z  \int d^2{\bm p}'_{\perp} dp'_z \nonumber\\
		&\ \ \ \ \ \ \ \ \ |\braket{ {\rm g}_{A,{\bm p}_{\perp},p_z,\sigma} {\rm g}_{A',{\bm p}'_{\perp},p'_z, \sigma'};{\rm in}  | S|  {\rm vac;in} }|^2 \nonumber\\
	&=
		\frac{1}{48\pi^2} \int_0^{\infty} d\omega | q_{A}^{\rm (g)}\tilde{E}(\omega) |^2,
\end{align}
where $\tilde{E}(\omega)$ is the Fourier transformation of the electric field $\tilde{E}(\omega) \equiv \int dt E(t) {\rm e}^{i\omega t}$.  Thus, $ N^{\rm (q)}_{i,f,s}/ N^{\rm (g)}_{A,\sigma} = 2|q_{i}^{\rm (q)}|^2/|q_{A}^{\rm (g)}|^2$ holds for massless quarks.

\begin{figure*}
\begin{center}
\includegraphics[clip, width=0.495\textwidth]{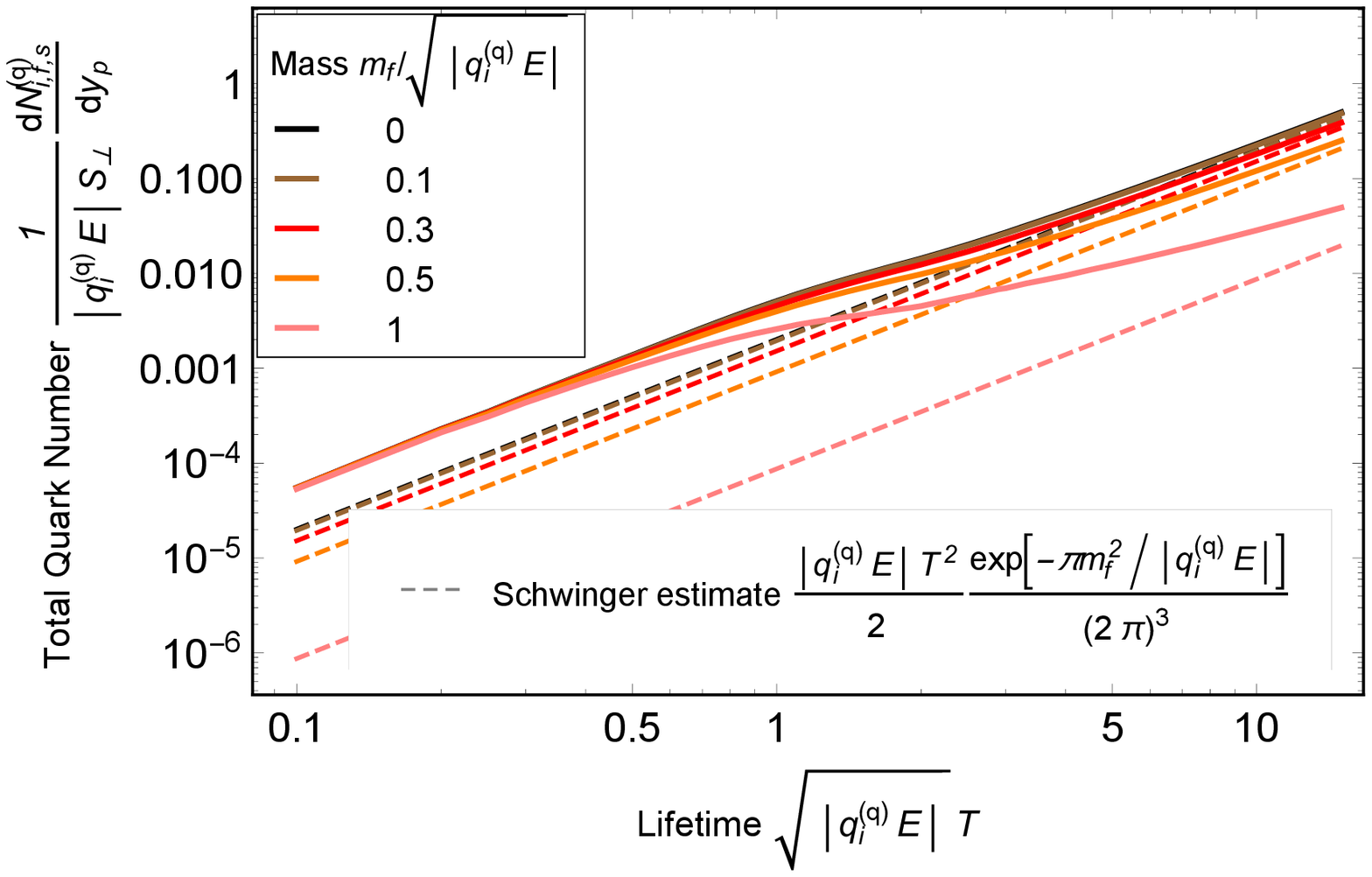}
\includegraphics[clip, width=0.495\textwidth]{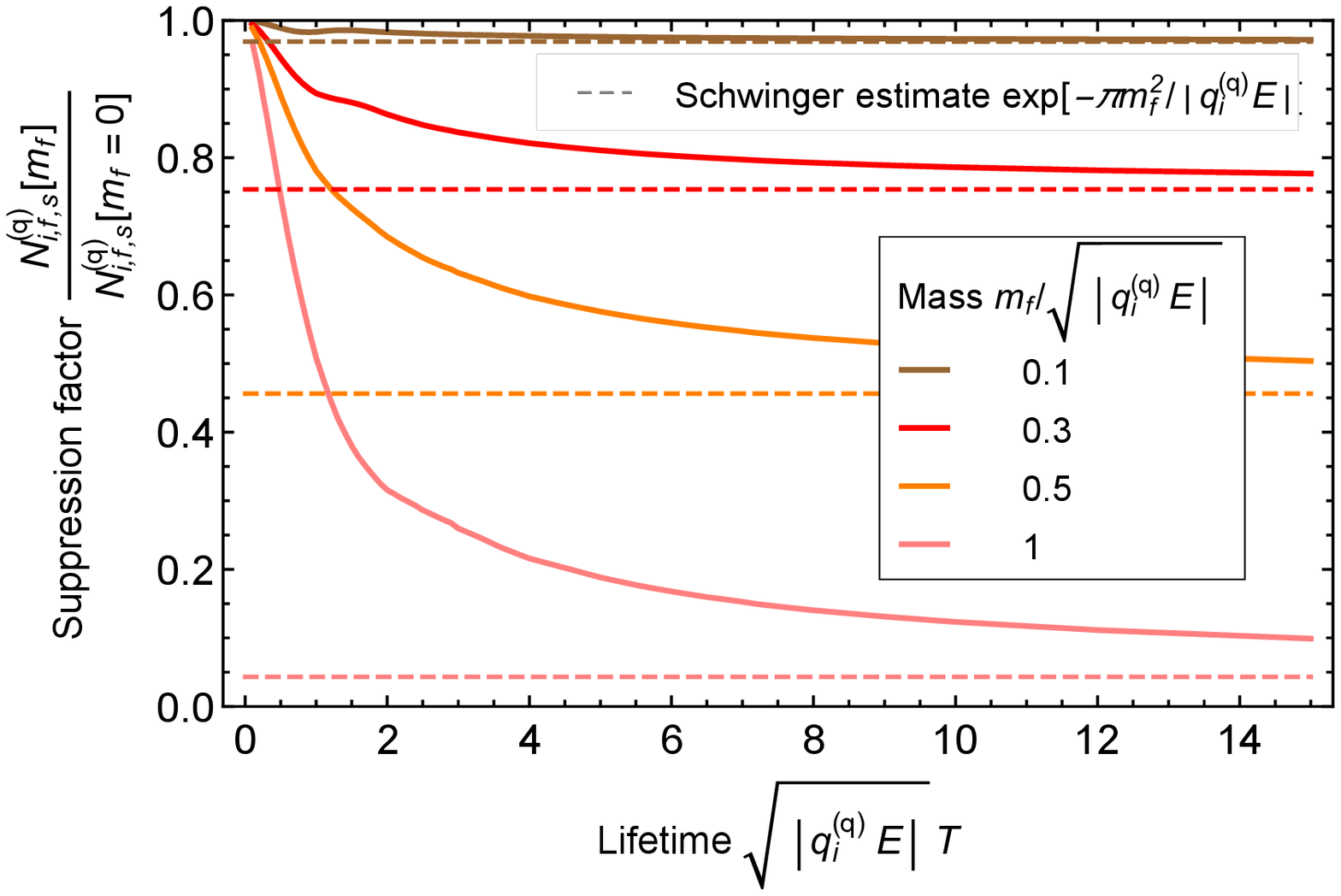}
\caption{\label{fig3}(color online) [Left] Total number of produced quarks per unit rapidity for various quark masses.  The dashed lines are expectations from the Schwinger formula Eq.~(\ref{eq123}).  [Right] The ratio of total number of produced massive quarks to that of massless quarks $N_{i,f,s}^{\rm (q)}[m_{\rm f}]/N_{i,f,s}^{\rm (q)}[m_{\rm f}=0]$.  The dashed lines are expectations from the Schwinger formula: $\exp[-\pi m_{\rm f}^2/|q_i^{\rm (q)}E|]$.  }
\end{center}
\end{figure*}

Figure~\ref{fig3} shows quark mass dependences of the quark production: The total quark number $d N^{\rm (q)}_{i,f,s}/dy_{\bm p}$ (left) and the ratio of the total number of massive quarks to that of massless quarks (right) for several different values of the quark mass are plotted.  One finds the following:

(i) For short lifetimes $\sqrt{|q_i^{\rm (q)} E|}T \lesssim 1$, the total quark production number becomes independent of the quark mass and the ratio comes close to one because the typical energy scale of the classical electric field, which is characterized by its typical frequency $\omega \sim 1/T$, is much larger the quark mass scale.

(ii) For long lifetimes $\sqrt{|q_i^{\rm (q)} E|}T \gtrsim 1$, the total quark production number approaches the expectation of the Schwinger formula Eq.~(\ref{eq123}) and the ratio starts to be suppressed exponentially with respect to the quark mass as $\exp[-\pi m_{\rm f}^2/|q_i^{\rm (q)} E|]$.  Notice that the quark production is still always larger than Schwinger's value.  This is because it needs a long lifetime $T$ to justify Schwinger's non-perturbative particle production mechanism because of the finite lifetime effects.

(iii) The larger lifetime $T$ is required for heavier quark production to converge to Schwinger's estimate, compared to that required for lighter quarks.  One can understand this observation in terms of the Keldysh parameter $\gamma_{\rm Keldysh} = q_i^{\rm (q)} E T/m_{\rm f}$ \cite{tay14, bre70}, which is one of the dimensionless parameters characterizing the interplay between Schwinger's non-perturbative particle production ($\gamma_{\rm Keldysh} \gg 1$) and perturbative particle production ($\gamma_{\rm Keldysh} \ll 1$): The Keldysh parameter $\gamma_{\rm Keldysh}$ becomes smaller for larger values of $m_{\rm f}$ and thus it requires larger lifetimes $T$ to realize $\gamma_{\rm Keldysh} \gg 1$.

\subsection{Phenomenology of particle production} \label{sec:phen}

For discussions on more phenomenological implications, let us consider particle production with physical parameter settings: $N_{\rm c} = 3$, and $m_{\rm u}, m_{\rm d} = 0~{\rm GeV}, m_{\rm s} = 0.1~{\rm GeV}$ and $m_{\rm c} = 1.2~{\rm GeV}$ representing the mass of up, down, strange and charm quark, respectively.  We set $gE = 1\;{\rm GeV}^2$ as a typical value at RHIC energy scale.  Under this setting, we consider the inclusive particle production by summing up the color degrees of freedom, $i$ and $A$.  Here, we assume for simplicity that the Abelianized classical electric field (see Eq.~(\ref{eq_10})) is always directing to the $t^3$-direction in the color space.  The particle spectra depend on this color direction in general, however, one can numerically demonstrate that its dependence is rather small \cite{tan10, coo08}.

Before showing results, let us make some remarks on the validity of our results to the early stage dynamics of HIC: 
\begin{itemize}
\item Our formalism assumes the Abelian dominance (see Section \ref{sec:background}).  This assumption is non-trivial because one can naively expect in HIC that the non-Abelian part for the classical field strength $g \bar{A}\bar{A} \sim Q^2_{\rm s}/g$ is about the same order as the Abelian one $\partial \bar{A} \sim Q_{\rm s}^2/g$.  Nevertheless, it is known that the full numerical simulation of the classical Yang-Mills evolution (Eq.~(\ref{eq0})) \cite{lap06} can be understood well within the Abelian dominance assumption, i.e., effects of the non-Abelian part is rather small \cite{fuj08}.  Hence, it may be good to assume the Abelian dominance for the first approximation.  

\item Our formalism neglects higher order quantum effects beyond one-loop order.  Hence, one cannot treat scatterings and screening effects of produced particles, which are essential for the thermalization of the system.  Strictly speaking, this treatment works fine when the lifetime $T$ is not so long, where the fluctuations are small enough compared to the strength of the classical field.  

\item The classical field configuration Eq.~(\ref{eq114}) is very simple compared to the one in realistic situations: 

(i) The classical field is assumed to be constant in time for $\tau<T$ and suddenly switched off at $\tau=T$.  A realistic classical field is also finite in time, however, it should smoothly decay in time and not experience such a sudden switching-off; 

(ii) We only consider a purely longitudinal electric field.  In realistic situations, however, not only a longitudinal electric field but also a longitudinal magnetic field can exist.  

(iii) The spatial homogeneity is assumed for the classical field.  A realistic classical field, however, should have spatial structure with a typical length scale $\sim 1/Q_{\rm s}$ due to CGC;

Thus, the quark and gluon spectrum presented below are just a first-order approximation.  To get a more reliable results for the phenomenology, one has to consider the above points for the field configuration.  This is numerically possible within our formalism, although we leave it for a future study.  Nevertheless, we stress that the simple field configuration Eq.~(\ref{eq114}) does capture some essential features of the strong color electromagnetic field exists just after a collision such as the boost-invariance, the finite lifetime, and the existence of longitudinal color electric field.  

\end{itemize}

\subsubsection{Transverse Distribution $d^3N/dy_{\bm p} d{\bm p}_{\perp}^2$}

\begin{figure}
\begin{center}
\includegraphics[clip, width=0.5\textwidth]{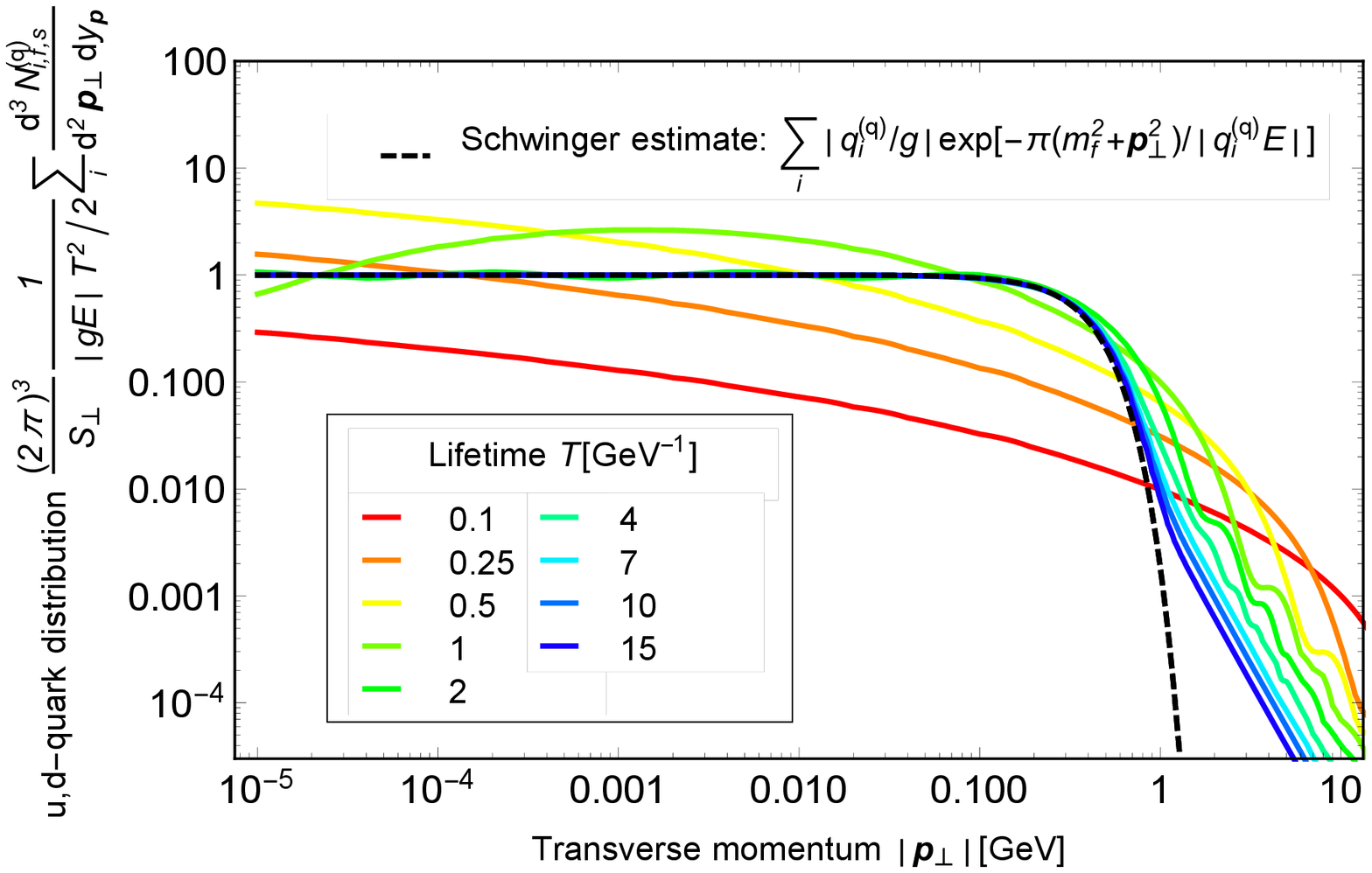} \\
\includegraphics[clip, width=0.5\textwidth]{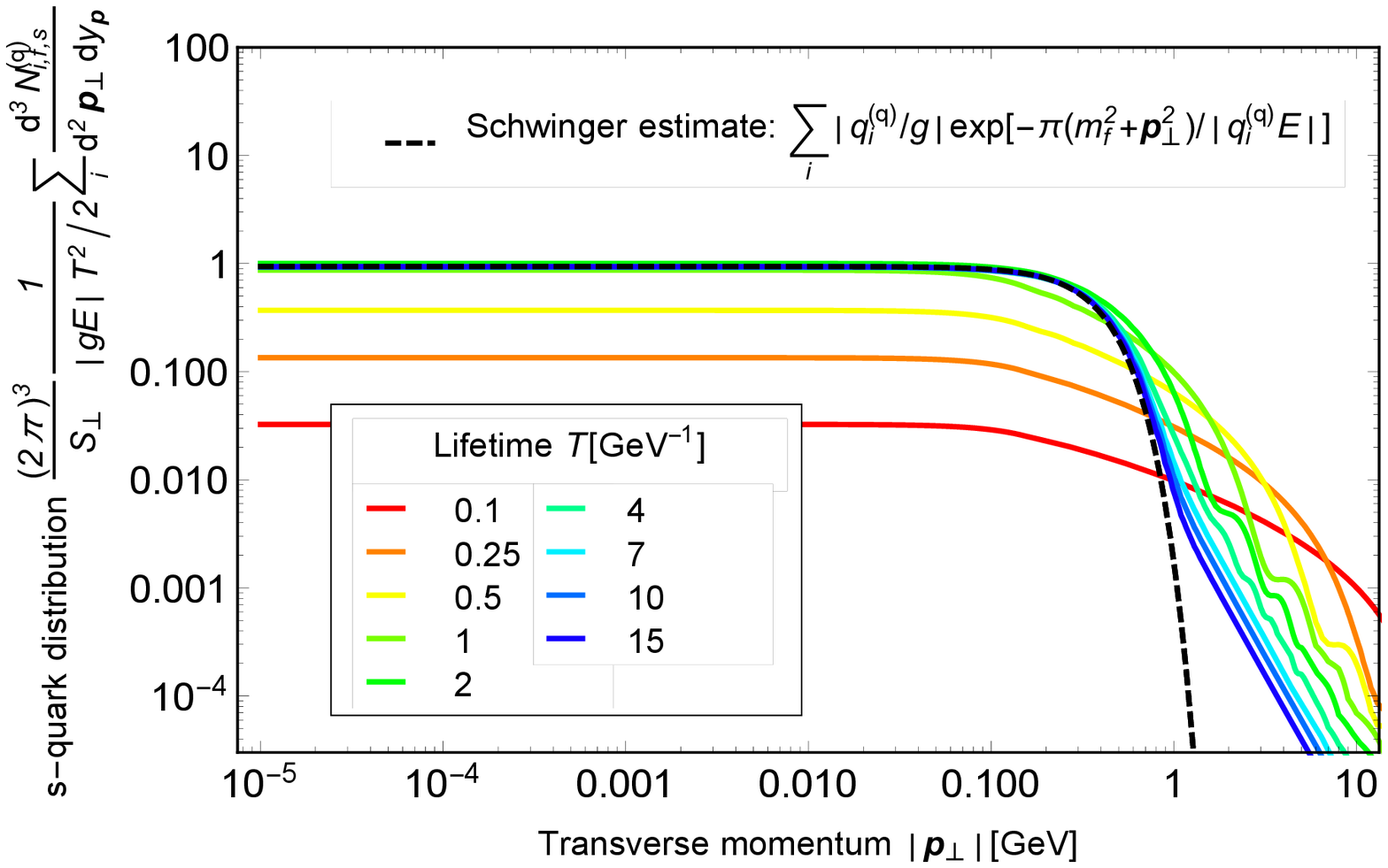} \\
\includegraphics[clip, width=0.5\textwidth]{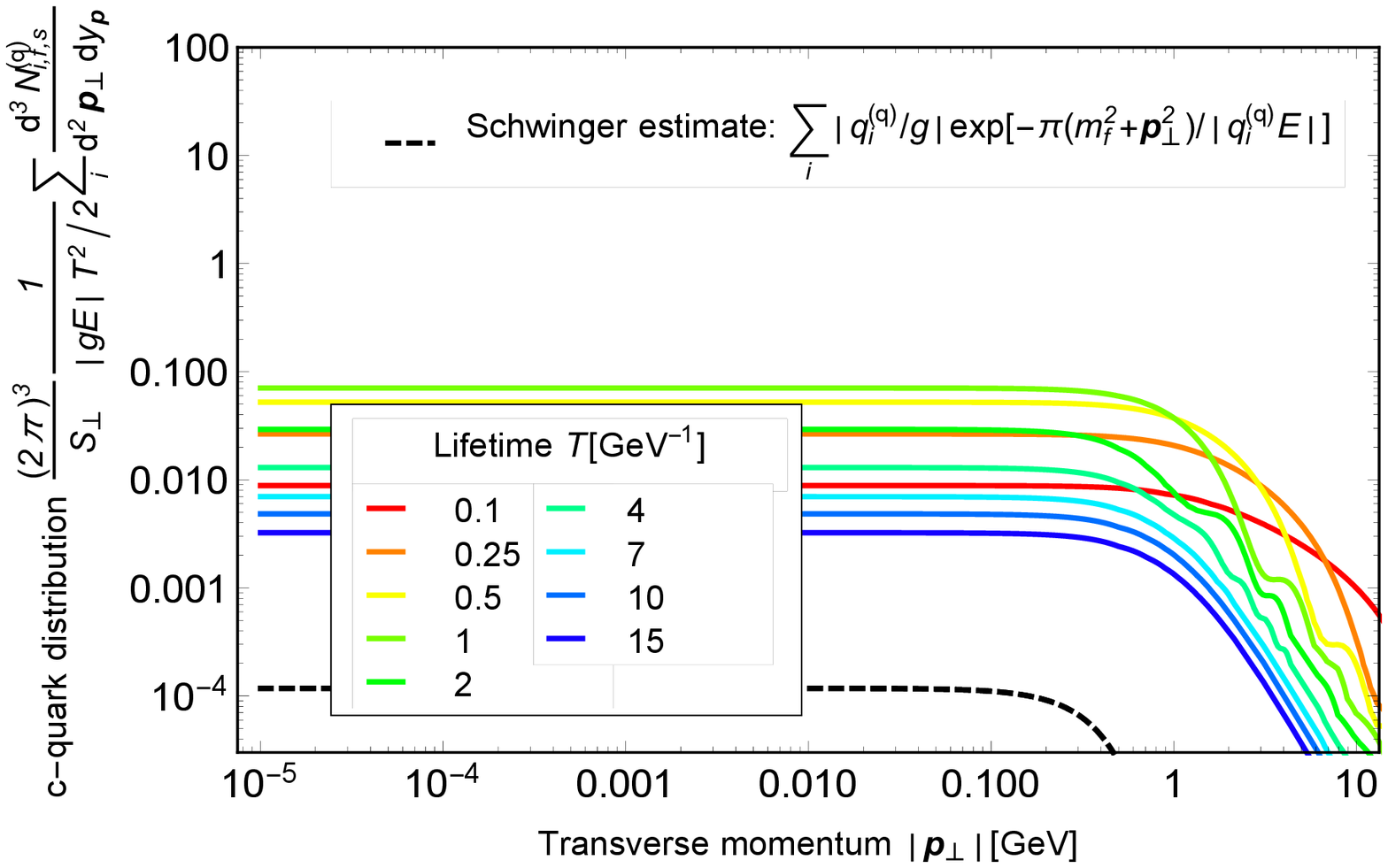}
\caption{\label{fig4}(color online) Transverse distribution of quarks (top for up and down, middle for strange, and bottom for charm) for various lifetimes $T$.  The dashed lines are expectations from the Schwinger formula Eq.~(\ref{eq__121}).  }
\end{center}
\end{figure}

\begin{figure}
\begin{center}
\includegraphics[clip, width=0.5\textwidth]{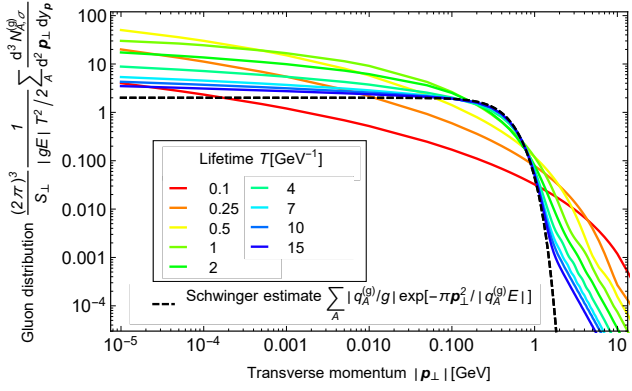}
\caption{\label{fig5}(color online) Transverse distribution of gluons for various lifetimes $T$.  The dashed line is an expectation from the Schwinger formula Eq.~(\ref{eq__122}).  }
\end{center}
\end{figure}

Figures~\ref{fig4} and \ref{fig5} show the transverse momentum spectrum of up and down (top), strange (middle), and charm quark (bottom) $\sum_{i} d^3 N^{\rm (q)}_{i,f,s}/d^2{\bm p}_{\perp} dy_{\bm p}$, and that of gluons $\sum_{A} d^3 N^{\rm (g)}_{A,\sigma}/d^2{\bm p}_{\perp} dy_{\bm p}$, respectively.  The dashed line in the figures represents the expectation from the Schwinger formula, Eqs.~(\ref{eq__121}) and (\ref{eq__122}).  We again recognize the interplay between Schwinger's non-perturbative particle production (long lifetimes $T\gtrsim 1{\rm \; GeV}^{-1}$) and perturbative particle production (short lifetimes $T \lesssim 1{\rm \;GeV}^{-1}$).  This implies that finite lifetime effects are very relevant to the early stage dynamics of HIC, where the typical lifetime of the strong field is short as $T \sim 1/Q_{\rm s} \lesssim 1{\rm \;GeV}^{-1}$.  

One also finds that the quark mass value largely affects the transverse spectrum:

(i) For small transverse momentum $|{\bm p}_{\perp}| \lesssim m_{\rm f}$, the spectra become constant in $|{\bm p}_{\perp}|$.  This is because the ${\bm p}_{\perp}$-dependence of the particle production always appears in the combination of the transverse mass $\sqrt{m_{\rm f}^2 + {\bm p}_{\perp}^2}$ in homogeneous systems (See the explicit expressions of the mode functions given in Appendix \ref{appC:quark}).  Thus, one can neglect the ${\bm p}_{\perp}$-dependence and that the spectra are determined solely by the quark mass $m_{\rm f}$ for $|{\bm p}_{\perp}| \lesssim m_{\rm f}$.

(ii) For large transverse momentum $|{\bm p}_{\perp}| \gtrsim m_{\rm f}$, the spectra become independent of the quark mass $m_{\rm f}$ because now the transverse mass is determined by $|{\bm p}_{\perp}|$ only and hence the $m_{\rm f}$-dependence can be neglected.

(iii) The larger lifetime $T$ is required for the heavier (charm) quark production spectrum to converge to the Schwinger estimate, compared to that required for lighter quarks (up, down and strange quarks) as was discussed in Fig.~\ref{fig3} for the total quark number $dN^{\rm (q)}_{i,f,s}/dy_{\bm p}$.

\subsubsection{Number Density $dN/dy$}

\begin{figure}
\begin{center}
\includegraphics[clip, width=0.5\textwidth]{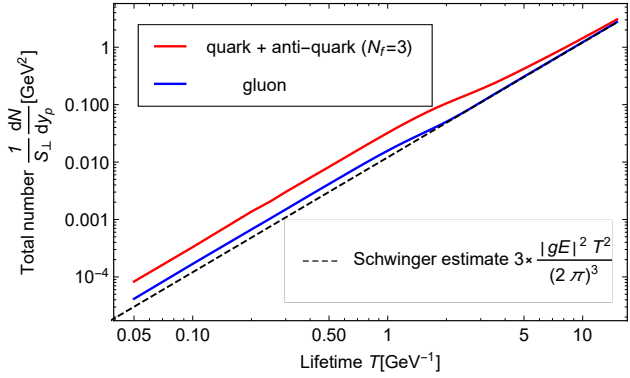}
\caption{\label{fig6}(color online)  Total number of quarks and anti-quarks (red) and gluons (blue) per unit rapidity for $N_{\rm f}=3$.  The dashed line is an expectation from the Schwinger formula Eqs.~(\ref{eq125}) and (\ref{eq126}).  }
\end{center}
\end{figure}

Figure~\ref{fig6} shows the total number of quarks and anti-quarks $\sum_{i,f,s} d(N^{\rm (q)}_{i,f,s} + N^{\rm (\bar{q})}_{i,f,s})/dy_{\bm p}$, and that of gluons $\sum_{\pm A,\sigma} dN^{\rm (g)}_{A, \sigma}/dy_{\bm p}$.  Here, we consider three flavor case ($N_{\rm f} = 3$), i.e., up, down and strange quarks are considered.  (The number-of-flavor $N_{\rm f}$-dependence of the particle production will be discussed in Fig.~\ref{fig7} below.)  As in Fig.~\ref{fig2}, we observe the following points:

(i) For long lifetimes $T \gtrsim 1{\rm \;GeV}^{-1}$, the total number of quarks and anti-quarks, and that of gluons approaches the Schwinger estimates, which are given by
\begin{align}
	\sum_{i,f,s,{\rm q\bar{q}}}\left.\frac{d N^{\rm (q)}_{i,f,s}}{dy_{\bm p}} \right|_{\rm Schwinger}
			&\sim N_s N_{\rm q\bar{q}} \times \sum_{i,f} \frac{S_{\perp}}{ (2\pi)^3} \frac{ |q_i^{\rm (q)} E|^2 T^2  }{2} \nonumber\\
			&\ \ \ \ \ \ \ \ \ \ \ \ \ \ \ \ \ \times \exp \left[ -\pi \frac{m_{\rm f}^2}{ |q_i^{\rm (q)} E| } \right] \nonumber\\
			&\sim N_s N_{\rm q\bar{q}} N_{\rm lq} \times \frac{S_{\perp}}{ (2\pi)^3} \frac{ |g E|^2 T^2  }{4}  , \label{eq125}\\
	\sum_{\pm A, \sigma} \left.\frac{d N^{\rm (g)}_{\pm A, \sigma}}{dy_{\bm p}} \right|_{\rm Schwinger}
			&\sim 2 N_{\rm c} N_{\sigma}  \times \frac{S_{\perp}}{ (2\pi)^3} \frac{ |g E|^2 T^2  }{4},   \label{eq126}
\end{align}
where $N_{s} = 2, N_{\rm q\bar{q}}=2, N_{\sigma} = 2$ count the number of the spin, the quark and anti-quark degeneracy, and the physical polarization of gluons.  $N_{\rm lq}$ represents the number of ``light" quarks satisfying $m^2_{\rm f} \ll |g E|$.  We have used the color charge formulas, Eqs.~(\ref{eq--36}) and (\ref{eq--35}).  As the strange quark mass is much smaller than the strength of the electric field, $m_{\rm s}^2 \ll |gE|$, we regard the strange quark as a ``light" quark and set $N_{\rm lq} = 3$ (see the middle panel of Fig.~\ref{fig7} for justification of this consideration). Then, the Schwinger estimates for quark, and anti-quark production and for gluon production accidentally coincide with each other for $N_{\rm c} = 3$ because the prefactors in both cases give the same value, $ N_s N_{\rm q\bar{q}} N_{\rm lq} = 2 N_{\rm c} N_{\sigma} = 12$.

(ii) For short lifetimes $T \lesssim 1 {\rm \;GeV}^{-1}$, quarks and anti-quarks are more abundantly produced than gluons because of the finite lifetime effects.  For details of the ratio of the total number of quarks and anti-quarks to that of gluons, see the bottom panel of Fig.~\ref{fig7}, which we will discuss later.

(iii) The particle production is very {\it fast}, which is consistent with an earlier work on the quark production \cite{gel06}.  For the typical value of the transverse area in HIC $S_{\perp} \sim \pi (7 {\rm \; fm})^{2}$, about 1000 particles per unit rapidity (650 quarks and anti-quarks plus 350 gluons) are produced at about $T \sim 0.5\;{\rm fm}/c$.

\begin{figure}
\begin{center}
\includegraphics[clip, width=0.5\textwidth]{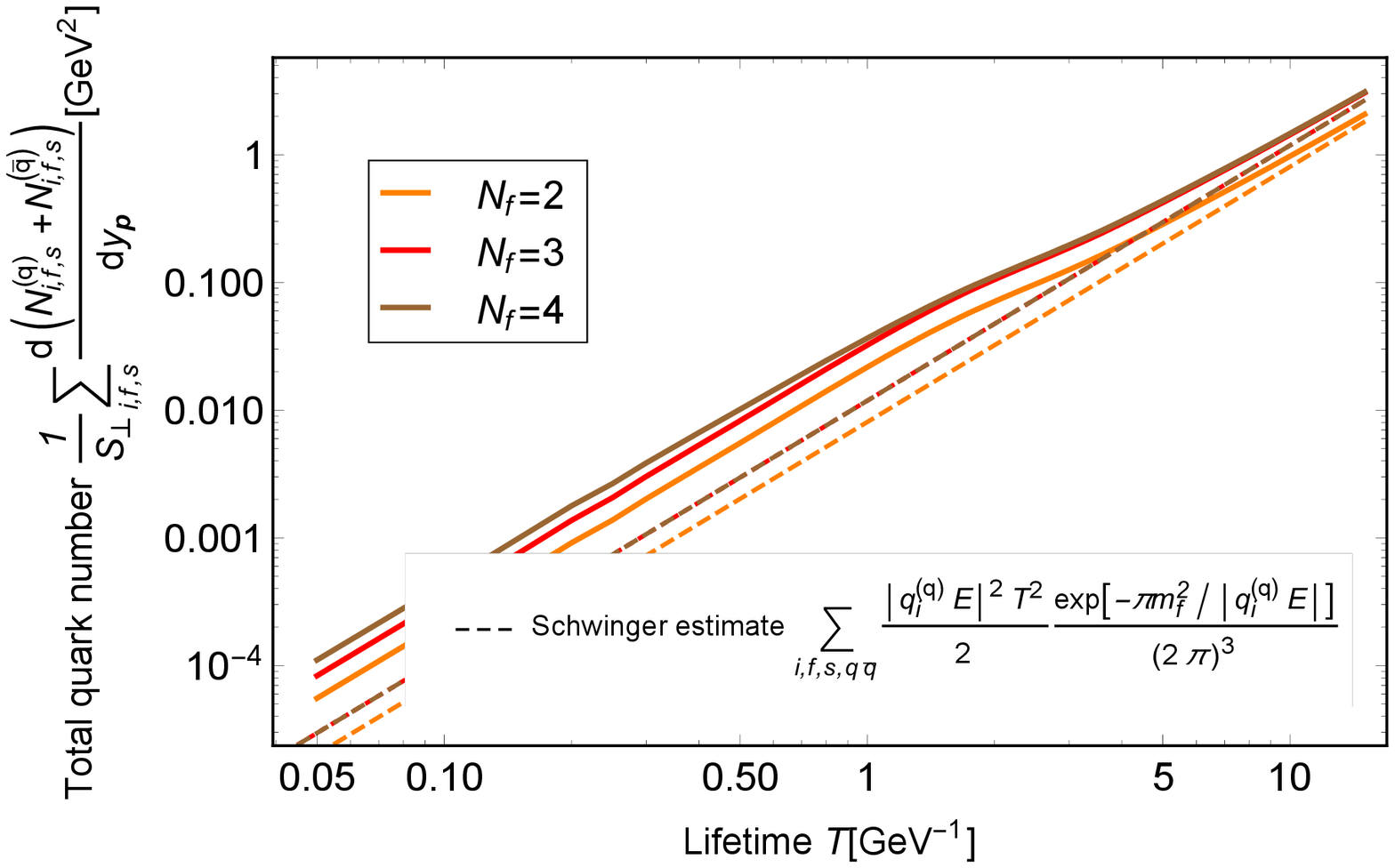} \\
\includegraphics[clip, width=0.5\textwidth]{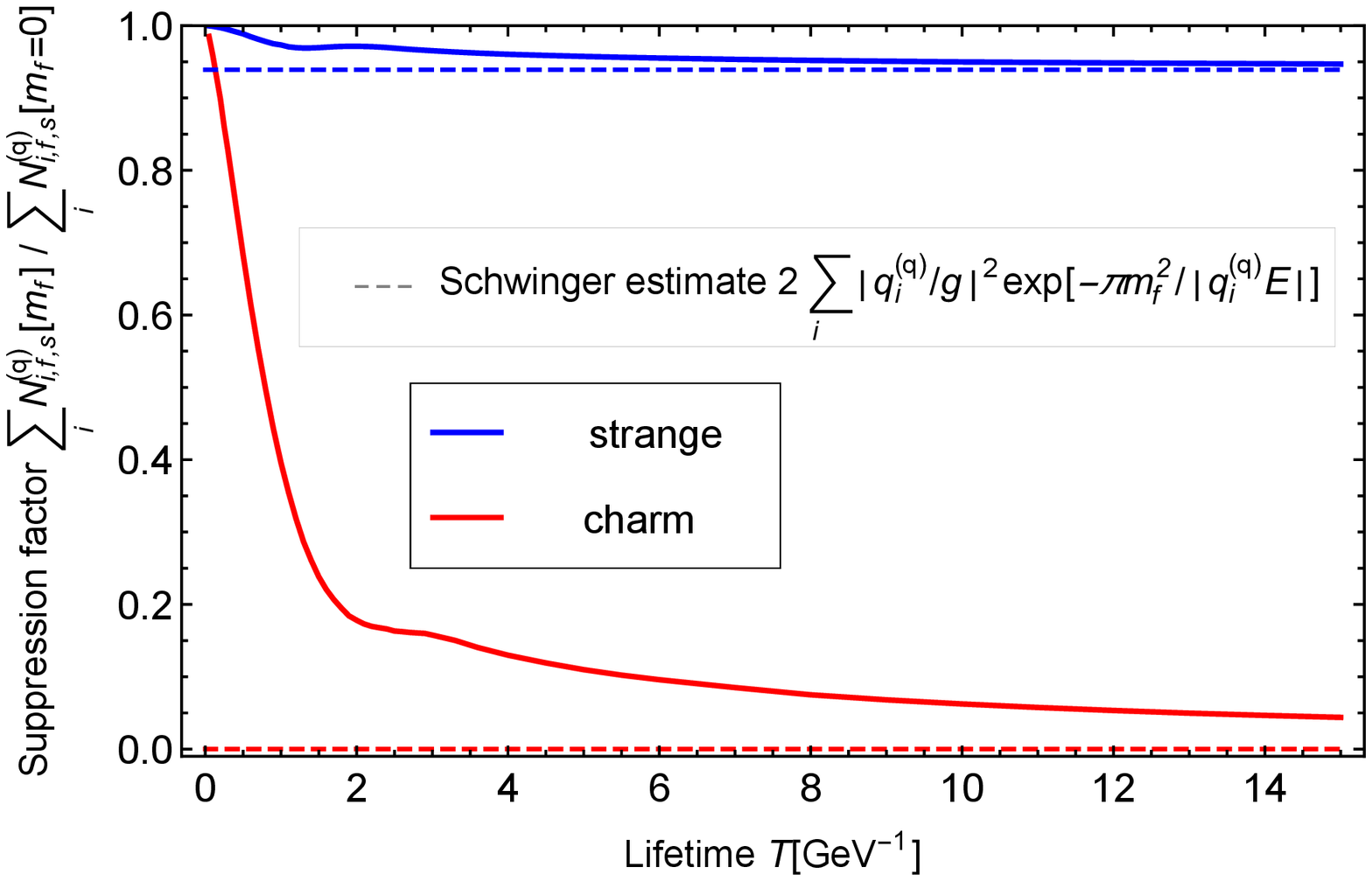} \\
\includegraphics[clip, width=0.5\textwidth]{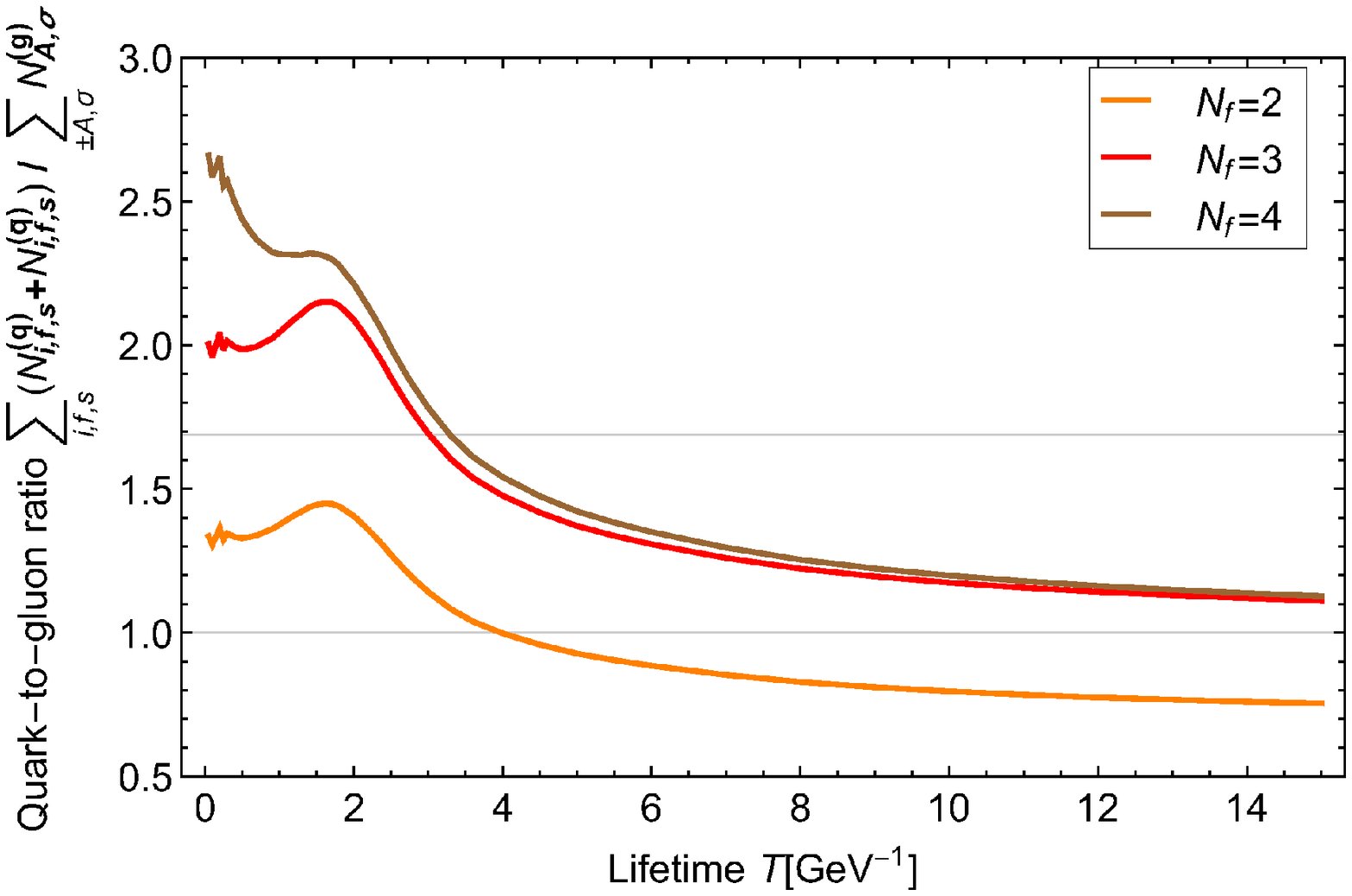}
\caption{\label{fig7} (color online) The number-of-flavor $N_{\rm f}= 2$ (u,d), 3 (u,d,s), 4 (u,d,s,c) dependences: [Top] The total quark and anti-quark number.  The dashed lines are expectations from the Schwinger formula Eq.~(\ref{eq125}).  [Middle] A ratio of the total number of massive charm- and strange quarks to that of massless up- and down quarks $\sum_i N^{\rm (q)}_{i,f,s}[m_{\rm f}\neq 0]/\sum_i N^{\rm (q)}_{i,f,s}[m_{\rm f}=0]$.  The dashed lines are expectations from the Schwinger formula Eq.~(\ref{eq__128}).  [Bottom] The ratio of the total number of quarks and anti-quarks to that of gluons $\sum_{i,f,s} (N^{\rm (q)}_{i,f,s} + N^{\rm (\bar{q})}_{i,f,s} )/ \sum_{\pm A, \sigma} N^{\rm (g)}_{A,\sigma}$.  The upper horizontal line indicates the thermal ratio $27/16$.  }
\end{center}
\end{figure}

Figure~\ref{fig7} shows dependence of particle production on the number of flavors, $N_{\rm f}= 2$ (u,d), 3 (u,d,s), 4 (u,d,s,c): The total quark and anti-quark number, $\sum_{i,f,s} d( N^{\rm (q)}_{i,f,s} + N^{\rm (\bar{q})}_{i,f,s} )/dy_{\bm p} $, for several different values of $N_{\rm f}$ is plotted in the top panel.  There is a significant change from $N_{\rm f} = 2$ to $N_{\rm f} = 3$ for all values of $T$.  This means that the inclusion of strangeness degree of freedom is inevitable in understanding the early stage dynamics of HIC quantitatively.  Whereas, the change of the quark multiplicity from $N_{\rm f} = 3$ to $N_{\rm f} = 4$, i.e., by inclusion of charm quarks, is negligible (noticeable) for long (short) lifetimes $T$.

In order to see more clearly how this difference appears, we plot a ratio of the total number of massive charm and strange quarks to that of massless up and down quarks, $ \sum_{i} N^{\rm (q)}_{i,f,s} [ m_{\rm f} \neq 0 ] / \sum_{i}  N^{\rm (q)}_{i,f,s} [ m_{\rm f} = 0 ]$, in the middle panel.  From this panel, one can understand that the strange quark production is comparable to the production of up and down quarks (the ratio is almost unity) for all values of $T$ because the strange quark mass is sufficiently ``light" compared to the strength of the electric field $m_{\rm s}^2 \ll |gE|$.  On the other hand, one finds for the charm quark production that it is comparable to the production of massless quarks for smaller values of $T \lesssim 1{\rm \; GeV}^{-1}$ because of the perturbative enhancement of the particle production discussed in Fig.~\ref{fig2}, while it is negligible for larger values of $T \gtrsim 1{\rm \; GeV}^{-1}$ because Schwinger's non-perturbative particle production is strongly suppressed by the mass effect:
\begin{align}
	&\left.\frac{ \sum_{i} N^{\rm (q)}_{i,f,s} [ m_{\rm f} \neq 0 ] }{ \sum_{i}  N^{\rm (q)}_{i,f,s} [ m_{\rm f} = 0] }\right|_{\rm Schwinger} \nonumber\\
		&\ \ \ \ \ \ \ \ \ \ \sim 2 \sum_i \left| \frac{q^{\rm (q)}_i E}{gE} \right|^2 \exp\left[ -\pi \frac{m_{\rm f}^2}{|q^{\rm (q)}_i E|}\right].  \label{eq__128}
\end{align}
We note that the enhancement of the charm quark production from a pulsed electric field in a non-expanding system was previously discussed in Refs.~\cite{tay14, lev10}.

Finally, in the bottom panel of Fig.~\ref{fig7}, we plot a ratio of the total number of quarks and anti-quarks to that of gluons, $ \sum_{i,f,s} (N^{\rm (q)}_{i,f,s} + N^{\rm (\bar{q})}_{i,f,s}) / \sum_{\pm A ,\sigma} N^{\rm (g)}_{A,\sigma}$, for several different values of $N_{\rm f}$.  As was discussed in the top and middle panel of Fig.~\ref{fig7}, there is a significant change from $N_{\rm f} = 2$ to $N_{\rm f} = 3$ for all values of $T$, while the change from $N_{\rm f} = 3$ to $N_{\rm f} = 4$ is negligible (noticeable) for long (short) lifetimes $T$ depending on the strange and charm quark masses.  In the short lifetime limit $T \rightarrow 0$, the ratio approaches $2N_{\rm f}/N_{\rm c}$.  This is because the quark masses become irrelevant to the quark production for such small values of $T$ and that the (lowest order) perturbative particle production, which becomes efficient for small values of lifetimes $T$, says that quarks will be produced twice as many as gluons as was discussed in Fig.~\ref{fig3}.  On the other hand, in the long lifetime limit $T\rightarrow \infty$, the ratio approaches the value $N_{\rm lq}/N_{\rm c}$ because particle production is dominated by Schwinger's non-perturbative mechanism whose contribution is estimated by the formulas Eqs.~(\ref{eq125}) and (\ref{eq126}).  One of the important points here is that the ratio is always larger than unity for realistic values, $N_{\rm f} \geq 3, N_{\rm lq} = 3,  N_{\rm c}=3$, i.e., quarks and anti-quarks in total are more abundant than gluons.  This means that not only gluons but also quarks are important in understanding the early stage dynamics of HIC.  We note that this result is based on our one-loop order treatment as was remarked in the beginning of this section.  Thus, the above result does not take into account effects of scatterings such as ${\rm g} \rightarrow {\rm q}{\rm \bar{q}}$, which is important in the chemical equilibration of the system and could substantially change the ratio to $\sum_{i,f,s} (N^{\rm (q)}_{i,f,s} + N^{\rm (\bar{q})}_{i,f,s}) / \sum_{\pm A ,\sigma} N^{\rm (g)}_{A,\sigma} |_{\rm chem.\ eq.} \sim 27/16$.

In the figure, we see bumps at around $T \sim 0.5\;{\rm fm}/c$ while we did not find such bumps in the right panel of Fig.~2.  These bumps appear because gluons typically have a larger effective charge than quarks have.  Indeed, one can estimate the typical magnitude of the effective charge of a quark $\braket{|q^{\rm (q)}|}$ and a gluon $\braket{|q^{\rm (g)}|}$ from Eqs.~(\ref{eq--36}) and (\ref{eq--35}) as
\begin{align}
	\braket{|q^{\rm (q)}|} 
		&\equiv \sqrt{ \frac{1}{N_{c}} \sum_{i=1}^{N_{c}} | q^{\rm (q)}_i |^2 } 
		= g \times \frac{1}{\sqrt{2N_{c}}},  \\
	\braket{|q^{\rm (g)}|} 
		&\equiv \sqrt{ \frac{1}{N_{c}(N_{c}-1)/2} \sum_{i=1}^{N_{c}(N_{c}-1)/2} | q^{\rm (g)}_A |^2 } \nonumber\\ 
		&= g \times \frac{1}{\sqrt{N_{c}-1}} > \braket{|q^{\rm (q)}|} .    
\end{align} 
As was discussed in Sec.~III-A, quark and gluon production number approach the expectation of the Schwinger formula quicker for larger values of $|q| T$.  Thus, we can say that gluon production number approaches the expectation of the Schwinger formula quicker than quark production number does because $\braket{|q^{\rm (g)}|} > \braket{|q^{\rm (q)}|}$.  In other words, the interplay between the perturbative and non-perturbative particle production mechanism for quarks and gluons occurs not at the same time and the time needed for gluon's interplay is smaller than that for quark's one.  (This is not the case in Fig.~2, where $|q^{\rm (q)}|=|q^{\rm (g)}|$ is assumed.  )  By noting that Schwinger's non-perturbative particle production mechanism produces smaller numbers of particles compared to that the perturbative particle production mechanism does, we understand that production number of gluons is relatively smaller compared to that of quarks at transient times $T \sim 0.5\;{\rm fm}/c$, where the perturbative (non-perturbative) particle production mechanism dominates for quarks (gluons).  This is the reason why the bumps appear.

\section{Summary and Outlook} \label{sec4}

We have extensively studied the quark and gluon production from an expanding classical color electric field, motivated by the early stage dynamics of HIC.  Firstly, we have formulated the particle production from classical color electromagnetic fields in an expanding system in the one-loop level quantum calculation within the Abelian dominance assumption for the classical fields.

Then, we compute the quark and gluon spectra within this formalism for the simplest case of the classical background field in an expanding geometry.  That is, the classical field is assumed to be purely electric and boost-invariantly expanding, homogeneous and constant within finite duration (lifetime) $T$, ${\bm E} = {\bm e}_{z} E \theta(\tau) \theta(T-\tau)$.  In this setup, analytical solutions for the equations of motion of QCD are available; this enables us to develop a clear understanding of the particle production from the classical fields in an expanding system in QCD.

In this way, we have explicitly demonstrated for the first time in an expanding system that there is a significant interplay between Schwinger's non-perturbative particle production (long lifetimes $T$) and perturbative particle production (short lifetimes $T$), which results in that the transverse momentum ${\bm p}_{\perp}$-spectrum becomes harder (softer) for smaller (larger) values of $T$ and in an enhancement of the particle production for small values of $T$, compared to the estimate of the Schwinger formula.

In addition to this, we have studied the difference in the production of quarks and of gluons.  We have found that quarks are more abundantly produced than gluons, and that the difference of the statistics results in the increase of soft gluons and in an efficiency of the perturbative enhancement $N^{\rm (q)}_{i,f,s}/N^{\rm (g)}_{A,\sigma} \sim 2$ for small values of $T$.

The quark mass dependence of the quark production is also studied by examining the ratio $N^{\rm (q)}_{i,f,s}[m_{\rm f} \neq 0]/ N^{\rm (q)}_{i,f,s}[m_{\rm f} = 0]$: We found that it varies from one for short lifetimes $T$ to the expected value of the Schwinger formula $\exp[ -\pi m_{\rm f}^2/|q_i^{\rm (q)} E|]$ for long lifetimes $T$, and that it needs longer lifetime $T$ for heavier quark production to be described by the Schwinger formula.

As implications to the heavy ion phenomenology, we have argued that (i) the naive use of the Schwinger formula may be inappropriate in describing the early stage dynamics of HIC because of the finite lifetime effects; (ii) very {\it fast} particle production occurs, which results in about 1000 particles per unit rapidity (650 quarks and anti-quarks plus 350 gluons) produced at about $T \sim 0.5\;{\rm fm}/c$ at the RHIC energy scale; (iii) since quark production is more abundant than gluon production from the classical electric fields in the early times, it is very important to study the dynamics, not of the pure gluonic system, but of the {\it quark}-gluon system in understanding the early stage dynamics of HIC; and (iv) the strange quark production is comparable to the light (up and down) quarks for any value of the lifetime $T$, while the charm quark production rate heavily depends on the lifetime $T$, which is noticeable (negligible) for small (large) values of $T$.

There are many possible future directions of this work:

The first direction is to improve our formalism to include the higher order quantum corrections.  These terms are responsible for scatterings and screening effects, which are essential in understanding thermalization, i.e, isotropization, hydrodynamization and chemical equilibration of the system.  Besides, it is discussed vigorously that momentum exchanges due to the scatterings induce spectral cascades (for a recent review covering this topic, see \cite{fuk06}), which result in some interesting behaviors such as a formation of gluonic Bose-Einstein condensates \cite{bla14, bla16, bla12, bla13}.  We note for completeness that recently the screening effects from quark currents in a non-expanding system were discussed in Refs.~\cite{tan10, gelf16}.

Another direction is to improve configurations of the classical field: In realistic situations, the classical field has finite extent in the transverse direction and has random fluctuations with a typical transverse correlation length $\sim Q_{\rm s}^{-1}$.  Time-dependence of the classical field should also affect the particle spectra.  Besides, it is known that the classical color field has magnetic components in addition to electric components in the longitudinal direction \cite{lap06}.  The existence of longitudinal magnetic fields may enhance particle production rate \cite{tan09, tan10, nik70, hid11a, hid11b}.  In addition to this, such field configuration is known to invoke the Nielsen-Olesen type instability \cite{nie78a, nie78b, cha79}, although its typical timescale is rather slow.  It is interesting to study the particle production under the presence of such instabilities; for non-expanding, static ($T\rightarrow \infty$) color electromagnetic fields, it was discussed that the instability may enhance the gluon production \cite{tan12}.

The last direction which we would like to mention is about the quark dynamics.  As was discussed so far, quarks are abundantly produced at very early times and hence they may have important information about and/or an important role in the early stage dynamics of HIC.  Since quarks have an U(1) electromagnetic charge, which does not suffer from the strong interactions, one can investigate the quark dynamics by using U(1) electromagnetic probes such as photons \cite{tan15} and dileptons \cite{bia85, asa91}.  Another interesting topic involving the quark dynamics is an existence of strong U(1) electromagnetic fields just after a collision of nuclei \cite{den12, bzd12}.  Although such strong U(1) electromagnetic fields die away immediately after a collision within less than 1\;fm/$c$, they could significantly influence the quark dynamics because the strong U(1) electromagnetic fields are as strong as the pion mass scale and that the quark production is fast enough.  Thus, one can expect some experimental traces of them, for instance, in U(1) charge dependences in observables.  In particular, a U(1) charge dependent directed flow $v_1^{\pm}$ in asymmetric heavy ion collisions \cite{hir14, vor14} are recently measured by the STAR collaboration \cite{nii16}.  This should provide important insights in the quark production, i.e, the early stage dynamics of HIC, although theoretical understanding of this observable is still lacking.  Another interesting physics that involves the strong U(1) electromagnetic fields is the Chiral Magnetic Effect \cite{kha07}, whose real time dynamics from the microscopic point of view is still incomplete (although there are some earlier works on this topic \cite{fuk10, fuk15}) and hence is worth to be investigated further by extending our work.

\section*{Acknowledgements} 
The author thanks H.~Fujii for fruitful discussion and for carefully reading this manuscript, and T.~Matsui for the financial support in his attendance of the conference Quark Matter 2015 in Kobe, Japan (Grant-in-Aid \# 25400247 of MEXT, Japan). The author also thanks SAKURA research-exchange program between France and Japan and hospitality extended to him by Institut de Physique Theorique CEA/DSM/Saclay and CPHT Ecole Polytechnique, where part of this work was done.  This work was financially supported by Iwanami Fujukai foundation when completed.  The author is now supported by JSPS KAKENHI Grant Number 16J02712.

\appendix
\section{Analytic Solutions of the Abelianized Equation of Motion} \label{appA}

In this Appendix, we analytically obtain mode functions for the equations of motion Eqs.~(\ref{eq46})-(\ref{eq50}) under 
(a) a pure gauge background field (i.e., ${\bm E} = {\bm 0}, {\bm B}={\bm 0}$): 
\begin{align} 
	\tilde{A}_{\mu} = {\rm const.}, \label{eqpure}
\end{align}
(b) a spatially homogeneous and constant color electric background field (i.e., ${\bm E} = {\bf e}_{z} E $, ${\bm B}={\bm 0}$): 
\begin{align}
	\tilde{A}_{\tau}, \tilde{A}_x, \tilde{A}_y = 0, \ \tilde{A}_{\eta} = \tau^2 E/2 . \label{eqconstE}
\end{align}
and (c) a spatially homogeneous and constant color electric background field for a finite duration $T$ (i.e., ${\bm E} = {\bf e}_{z} E \theta(\tau-\tau_0) \theta(\tau_0+T-\tau), {\bm B}={\bm 0}$): 
\begin{align}
	\tilde{A}_{\tau}, \tilde{A}_x, \tilde{A}_y = 0, \tilde{A}_{\eta} = \left\{   \begin{array}{ll} \tau_0^2 E/2 & ( \tau<\tau_0) \\ \tau^2 E/2 & ( \tau_0<\tau<\tau_0+T) \\ (\tau_0+T)^2 E/2 & (\tau_0+T < \tau) \end{array} \right. . \label{eqEfield}
\end{align}

\subsection{Quark} 

We consider the equation of motion for the quark field $\psi$ under the Abelianized background gauge field in the $\tau$-$\eta$ coordinates (see Eq.~(\ref{eq46})): 
\begin{eqnarray}
[ i \Slash{\partial} - q\Slash{\tilde{A}} - m ] \psi(x) = 0. \label{eqa1}   
\end{eqnarray}
Here, we have suppressed the indices for color $i$ and flavor $f$ for simplicity.

To avoid complexities coming from the spinor structure of Eq.~(\ref{eqa1}), we consider a solution of a form \cite{tan09, nik70}:
\begin{align}
	\psi \equiv [ i \Slash{\partial} - q\Slash{\tilde{A}} + m ] \phi. \label{eqa4} 
\end{align}
One can readily find a differential equation for $\phi$ as 
\begin{align}
	0 	= \left[ ( \partial_{\mu} + iq \tilde{A}_{\mu} )^2 + \frac{\partial_{\tau} + iq \tilde{A}_{\tau}}{\tau}  + \frac{iq}{2} \gamma^{\mu} \gamma^{\nu} \tilde{F}_{\mu\nu}  + m^2 \right] \phi.   \label{eqa2} 
\end{align}
Since we are interested in the situations where color electric field pointing to the $z$-direction exists at most in this Appendix, one can simplify Eq.~(\ref{eqa2}) as
\begin{align}
	0	= \left[ ( \partial_{\mu} + iq \tilde{A}_{\mu} )^2 + \frac{\partial_{\tau} + iq \tilde{A}_{\tau}}{\tau}  + iqE \gamma^{t} \gamma^{z} + m^2 \right] \phi . \label{eqa3}  
\end{align}

Next, we expand $\phi$ in terms of eigenvectors%
\footnote{Strictly speaking, the matrix $\gamma^t \gamma^z$ has four eigenvectors $\Gamma_{s}$ ($s=1,2,3,4$) in total; $\Gamma_{1,2}$ with eigenvalues $\lambda_{1,2}=1$ and $\Gamma_{3,4}$ with eigenvalues $\lambda_{3,4}=-1$.  Solutions $\psi_{3,4}$ for the original equation Eq.~(\ref{eqa1}) constructed from $\phi_3 \Gamma_3$ and $\phi_4 \Gamma_4$ are linearly dependent on solutions $\psi_{1,2}$ constructed from $\phi_1 \Gamma_1$ and $\phi_2 \Gamma_2$ \cite{tan09}.  Thus, it is sufficient to consider $s=1,2$ only in order to obtain all the independent solutions of the differential equation Eq.~(\ref{eqa1}) for $\psi$ .  }
of $\gamma^t \gamma^z$ as
\begin{align}
	\phi \equiv \sum_{s=1}^2 \phi_s \Gamma_s.  
\end{align}
where $\phi_s$ are scalar functions and that the eigenvectors $\Gamma_s$ ($s=1,2$) satisfy
\begin{align}
	\gamma^t \gamma^z \Gamma_s = \lambda_s \Gamma_s, \ \ \Gamma_s^{\dagger} \Gamma_{s'} = \delta_{ss'} \label{eqa8}
\end{align}
with the eigenvalues $\lambda_s$ given by $\lambda_1=\lambda_2 = 1$.  Physically, we have defined the spin of quarks by the direction of the background field $\tilde{F}_{\mu\nu}$ because $\gamma^t \gamma^z$ is proportional to the $tz$-component of the background field as $\tilde{F}_{tz} = E \gamma^t \gamma^z$.  Now, a differential equation for $\phi_s$ reads
\begin{align}
	0 = \left[ ( \partial_{\mu} + iq \tilde{A}_{\mu} )^2 + \frac{\partial_{\tau} + iq \tilde{A}_{\tau}}{\tau}  + iqE + m^2 \right] \phi_s,  \label{eq11}
\end{align}
which are free from the cumbersome spinor structure in the original equation Eq.~(\ref{eqa1}) for $\psi$.

\begin{widetext}
\subsubsection{Under a pure gauge background field (plane wave solutions)} \label{appA:quark}

Let us construct all the mode functions for the equation of motion Eq.~(\ref{eqa1}) under a pure gauge background field $\tilde{A}_{\mu}$ given by Eq.~(\ref{eqpure}), which we write $\psi^{\rm (free)}$.  We first consider to solve the differential equation for $\phi^{\rm (free)}_s$ (Eq.~(\ref{eq11})).  For this, we make an ansatz of a form: 
\begin{align}
	\phi_s^{\rm (free)} (x) 
		\equiv \int d^2{\bm p}_{\perp} dp_{\eta}\; \phi_{{\bm p}_{\perp}, p_{\eta}, s}^{\rm (free)}(x)
		\equiv
	\int d^2{\bm p}_{\perp} dp_{\eta}\; \Omega(x)  \chi^{\rm (free)}_{{\bm p}_{\perp}, p_{\eta},s}(\tau) \frac{{\rm e}^{-\eta/2}}{\sqrt{m^2 + {\bm p}_{\perp}^2}}   \frac{{\rm e}^{i {\bm p}_{\perp}\cdot {\bm x}_{\perp} {\rm e}^{i p_{\eta} \eta} } }{ (2\pi)^{3/2} }  .
\end{align}
Here, the momentum labels ${\bm p}_{\perp}, p_{\eta}$ are introduced and $\Omega$ is a Wilson-line gauge factor denoted by 
\begin{align}
	\Omega (x) \equiv \exp \left[-iq\int^{x} dx^{\mu} \tilde{A}_{\mu} \right] .  \label{eqa_14} 
\end{align}
The factor ${\rm e}^{-\eta/2} / \sqrt{m^2+{\bm p}_{\perp}^2}$ is inserted so as to properly normalize $\psi^{\rm (free)}$ in the $\tau$-$\eta$ coordinates as we shall see.  Now, one can readily find that $\chi^{\rm (free)}_{{\bm p}_{\perp}, p_{\eta},s}$ satisfies the Bessel differential equation,
\begin{align}
	0 	&= \Biggl[  \tau^2 \partial_{\tau}^2  +  \tau \partial_{\tau}  + \left\{ \left( \sqrt{m^2+{\bm p}_{\perp}^2} \tau \right)^2 -   (ip_{\eta} + 1/2)^2     \right\}  \Biggl]  \chi^{\rm (free)}_{{\bm p}_{\perp}, p_{\eta},s}.  \label{eqa14}
\end{align}
Since the differential equation Eq.~(\ref{eqa14}) is a second order differential equation, there are two independent solutions, which we write ${}_{k} \chi^{\rm (free)}_{{\bm p}_{\perp}, p_{\eta},s}$ ($k=1,2$).  It is convenient for our purpose to choose  
\begin{align}
	\begin{pmatrix}
		{}_1 \chi^{\rm (free)}_{{\bm p}_{\perp}, p_{\eta},s} \\
		{}_2 \chi^{\rm (free)}_{{\bm p}_{\perp}, p_{\eta},s}
	\end{pmatrix}
		&\equiv
	\frac{\sqrt{\pi}}{2} ( m^2 + {\bm p}_{\perp}^2 )^{1/4} {\rm e}^{\pi p_{\eta}/2} {\rm e}^{-i\pi/4} 
	\begin{pmatrix}
		H^{(2)}_{ip_{\eta} + 1/2 } (\sqrt{m^2+{\bm p}_{\perp}^2} \tau) \\
		H^{(1)}_{-ip_{\eta} - 1/2 } (\sqrt{m^2+{\bm p}_{\perp}^2} \tau) 
	\end{pmatrix}, 
\end{align}
where $H_{\nu}^{(n)} (z)$ $(n=1,2)$ are the Hankel function of the $n$-th kind, and we have normalized the solutions ${}_{k} \chi_{{\bm p}_{\perp}, p_{\eta},s}$ by 
\begin{align}
	| {}_1 \chi^{\rm (free)}_{{\bm p}_{\perp}, p_{\eta},s} |^2 + | {}_2 \chi^{\rm (free)}_{{\bm p}_{\perp}, p_{\eta},s} |^2 = 1/\tau .  \label{eqa_17}
\end{align}
It is also useful to point out that the solutions ${}_{k} \chi^{\rm (free)}_{{\bm p}_{\perp}, p_{\eta},s}$ satisfy the following simultaneous differential equation: 
\begin{align}
	&\frac{i}{\sqrt{m^2+{\bm p}_{\perp}^2}} \left[  \partial_{\tau}  +   \frac{i p_{\eta} + 1/2 }{ \tau }  \right]
	\begin{pmatrix}
		{}_1 \chi^{\rm (free)}_{{\bm p}_{\perp}, p_{\eta},s} \\
		{}_2 \chi^{\rm (free)}_{{\bm p}_{\perp}, p_{\eta},s}
	\end{pmatrix}
	 = 
	\begin{pmatrix}
		{}_2 \chi^{{\rm (free)}*}_{{\bm p}_{\perp}, p_{\eta},s} \\
		-{}_1 \chi^{{\rm (free)}*}_{{\bm p}_{\perp}, p_{\eta},s}
	\end{pmatrix} .  \label{eqa18}
\end{align}

We are ready to construct all the mode functions ${}_{\pm}\psi_{{\bm p}_{\perp}, p_{\eta}, s}^{\rm (free)} [\tilde{A}_{\mu}] $.  Using the definition of $\phi^{\rm (free)}$ (Eq.~(\ref{eqa4})), one has
\begin{align}
	\begin{pmatrix}
		{}_{+}\psi_{{\bm p}_{\perp}, p_{\eta}, s}^{\rm (free)} [\tilde{A}_{\mu}] \\
		{}_{-}\psi_{{\bm p}_{\perp}, p_{\eta}, s}^{\rm (free)} [\tilde{A}_{\mu}]
	\end{pmatrix}
	 \equiv
	[ i \Slash{\partial} - q \tilde{\Slash{A}} + m ]
	\begin{pmatrix}
		{}_{1}\phi^{\rm (free)}_{{\bm p}_{\perp}, p_{\eta}, s}  \\
		{}_{2}\phi^{\rm (free)}_{{\bm p}_{\perp}, p_{\eta}, s} 
	\end{pmatrix} \Gamma_s.   \label{eqa19}
\end{align}
Here, we have changed the left subscript $k=1,2$ into $\pm$ for a notational simplicity because ${}_{+}\psi_{{\bm p}_{\perp}, p_{\eta}, s}^{\rm (free)} [\tilde{A}_{\mu}] $ (${}_{-}\psi_{{\bm p}_{\perp}, p_{\eta}, s}^{\rm (free)} [\tilde{A}_{\mu}] $) corresponds to the positive (negative) frequency mode function in the $\tau$-$\eta$ coordinates as we will explain soon.  With the help of Eqs.~(\ref{eqa8}) and (\ref{eqa18}), one finds that Eq.~(\ref{eqa19}) can be more explicitly written as 
\begin{align}
	\begin{pmatrix}
		{}_{+}\psi_{{\bm p}_{\perp}, p_{\eta}, s}^{\rm (free)} [\tilde{A}_{\mu}] \\
		{}_{-}\psi_{{\bm p}_{\perp}, p_{\eta}, s}^{\rm (free)} [\tilde{A}_{\mu}]
	\end{pmatrix}
	 &=
	\Omega \left[  	\begin{pmatrix} {}_{1}\chi^{\rm (free)}_{{\bm p}_{\perp}, p_{\eta}, s} \\ {}_{2}\chi^{\rm (free)}_{{\bm p}_{\perp}, p_{\eta}, s} \end{pmatrix} V_{s,1}  + \begin{pmatrix} {}_{2}\chi^{{\rm (free)}*}_{{\bm p}_{\perp}, p_{\eta}, s} \\ -{}_{1}\chi^{{\rm (free)}*}_{{\bm p}_{\perp}, p_{\eta}, s} \end{pmatrix} V_{s,2} \right] \frac{ {\rm e}^{i {\bm p}_{\perp} \cdot {\bm x}_{\perp}} {\rm e}^{i p_{\eta} \eta}}{(2\pi)^{3/2}}. \label{eqa_22}
\end{align}
Here, we have introduced four-spinors, $V_{s,1}$ and $V_{s,2}$, by
\begin{align}
	V_{s,1} \equiv {\rm e}^{\eta/2} \frac{ - {\bm p}_{\perp} \cdot {\bm \gamma}_{\perp} + m  }{ \sqrt{ m^2 + {\bm p}_{\perp}^2  } } \Gamma_s, \ V_{s,2} \equiv {\rm e}^{-\eta/2} \gamma^t \Gamma_s, \label{eqa_23}
\end{align}
which are normalized as	
\begin{align}
	\bar{V}_{s,i} \gamma^{\tau} V_{s,j} = \delta_{ij}.  \label{eqa_21}
\end{align}
From the normalization conditions for ${}_{k} \chi^{\rm (free)}_{{\bm p}_{\perp}, p_{\eta},s}$ (Eq.~(\ref{eqa_17})) and $V_{s,i}$ (Eq.~(\ref{eqa_21})), it is evident that the mode functions ${}_{\pm}\psi_{{\bm p}_{\perp}, p_{\eta}, s}^{\rm (free)} [\tilde{A}_{\mu}] $ (Eq.~(\ref{eqa19})) satisfy the correct normalization condition for spinor fields in the $\tau$-$\eta$ coordinates (see also Eqs.~(\ref{eq_48}) and (\ref{eq_49}) in the main text) as
\begin{align}
	( {}_{\pm}\psi_{{\bm p}_{\perp}, p_{\eta}, s}^{\rm (free)} [\tilde{A}_{\mu}] | {}_{\pm} \psi_{{\bm p}'_{\perp}, p'_{\eta}, s'}^{\rm (free)} [\tilde{A}_{\mu}]  )_{\rm F}  = \delta_{ss'} \delta^2 ({\bm p}_{\perp} - {\bm p}'_{\perp}) \delta(p_{\eta}-p'_{\eta}) , \ \ ( {}_{\pm}\psi_{{\bm p}_{\perp}, p_{\eta}, s}^{\rm (free)} [\tilde{A}_{\mu}] | {}_{\mp} \psi_{{\bm p}'_{\perp}, p'_{\eta}, s'}^{\rm (free)} [\tilde{A}_{\mu}]  )_{\rm F} = 0,
\end{align}
where the fermion inner product $(\psi_1|\psi_2)_{\rm F}$ is the same as is defined in Eq.~(\ref{eq_50}).

Our mode functions ${}_{+} \psi_{{\bm p}_{\perp}, p_{\eta}, s}^{\rm (free)} [\tilde{A}_{\mu}] $ (${}_{-} \psi_{{\bm p}_{\perp}, p_{\eta}, s}^{\rm (free)} [\tilde{A}_{\mu}]$) defined in Eq.~(\ref{eqa19}) can actually be understood as the positive (negative) frequency mode function in the $\tau$-$\eta$ coordinates because it can be written as a superposition of the positive (negative) frequency mode function in the Cartesian coordinates \cite{tan11, som74}.  To see this, we use the integral representations for the Hankel functions $H^{(n)}_{\nu} (z)$ ($n=1,2$): 
\begin{align}
	H^{(1)}_{\nu} (z) = \frac{ {\rm e}^{-i \nu \pi/2}}{i\pi} \int^{\infty}_{-\infty} dt {\rm e}^{iz \cosh t - \nu t}, \ H^{(2)}_{\nu} (z) = - \frac{ {\rm e}^{i \nu \pi/2}}{i\pi} \int^{\infty}_{-\infty} dt {\rm e}^{-iz \cosh t - \nu t} \label{eqa_26}
\end{align}
to obtain 
\begin{align}
	{}_{\pm} \psi^{({\rm free})}_{{\bm p}_{\perp}, p_{\eta}, s} [ \tilde{A}_{\mu} ] = \int dp_{z} \frac{ {\rm e}^{ \pm i p_{\eta} y_{\bm p} } }{ \sqrt{2\pi \omega_{\bm p}} }  {}_{\pm} \psi^{({\rm free})}_{{\bm p}_{\perp}, p_z, s} [ \tilde{A}_{m} ].  \label{eqa_27}
\end{align}
Here, $\omega_{\bm p}$ is on-shell energy $\omega_{\bm p} = \sqrt{m^2 + {\bm p}_{\perp}^2 + p_z^2 }$ and $y_{\bm p}$ is the momentum rapidity as was introduced in Eq.~(\ref{eq_67}).  ${}_{\rm \pm} \psi^{({\rm free})}_{{\bm p}_{\perp}, p_z, s} [ \tilde{A}_{m} ]$ are the positive/negative frequency mode functions in the Cartesian coordinates, which satisfy the free field equation of motion in the Cartesian coordinates, $0 = [ i \gamma^m (\partial_m + iq \tilde{A}_m) - m ]{}_{\rm \pm} \psi^{({\rm free})}_{{\bm p}_{\perp}, p_z, s} [ \tilde{A}_{m} ]$, and that are labeled by $p_z$ being the Fourier conjugate to $z$: They are given by
\begin{align}
	\begin{pmatrix}
		{}_{+}\psi_{{\bm p}_{\perp}, p_{z}, s}^{\rm (free)} [\tilde{A}_{m}] \\
		{}_{-}\psi_{{\bm p}_{\perp}, p_{z}, s}^{\rm (free)} [\tilde{A}_{m}]
	\end{pmatrix}
	 &=
	\Omega \left[  	\begin{pmatrix} {}_{1}\chi^{\rm (free)}_{{\bm p}_{\perp}, p_{z}, s} \\ {}_{2}\chi_{{\bm p}_{\perp}, p_{z}, s} \end{pmatrix} v_{s,1} + \begin{pmatrix} {}_{2}\chi^{{\rm (free)}*}_{{\bm p}_{\perp}, p_{z}, s} \\ -{}_{1}\chi^*_{{\bm p}_{\perp}, p_{z}, s} \end{pmatrix} v_{s,2} \right]  \frac{ {\rm e}^{i {\bm p}_{\perp} \cdot {\bm x}_{\perp}} {\rm e}^{i p_{z} z}}{(2\pi)^{3/2}}, 
\end{align}
where we have defined the spin label $s$ by the direction of the background electric field $\tilde{F}_{tz} = E \gamma^t \gamma^z$ as in ${}_{\pm} \psi_{{\bm p}_{\perp}, p_{\eta}, s}^{\rm (free)} [\tilde{A}_{\mu}]$ by expanding the spinor-space in terms of the eigenvectors of the background field $\Gamma_s$.  $\Omega$ is the same as the Wilson-line gauge factor introduced in Eq.~(\ref{eqa_14}), which does not depend on a choice of coordinates 
\begin{align}
	\Omega (x) = {\rm exp}\left[ -iq \int^x dx^{\mu} \tilde{A}_{\mu} \right]  = {\rm exp}\left[ -iq \int^x d\xi^{m} \tilde{A}_{m} \right], \label{eqWil}
\end{align}
where $\xi^m$ represent the Cartesian coordinates $\xi^m = (t,x,y,z)$ with $m$ running through $t,x,y,z$ as in the main text.  The functions ${}_{k}\chi_{{\bm p}_{\perp}, p_{z}, s}$ ($k=1,2$) can be explicitly written as
\begin{align}
		{}_1 \chi^{\rm (free)}_{{\bm p}_{\perp}, p_{z},s} 
			\equiv \frac{i}{\sqrt{2}} \sqrt{1+ \frac{p_z}{\omega_{\bm p}}} {\rm e}^{-i \omega_{\bm p} t} ,\  
		{}_2 \chi^{\rm (free)}_{{\bm p}_{\perp}, p_{z},s} 
			\equiv \frac{-i}{\sqrt{2}} \sqrt{1- \frac{p_z}{\omega_{\bm p}}} {\rm e}^{ i \omega_{\bm p} t}.  
\end{align}
and the four-spinors $v_{s,1}$ and $v_{s,2}$ are given by
\begin{align}
	v_{s,1} \equiv \frac{ - {\bm p}_{\perp} \cdot {\bm \gamma}_{\perp} + m  }{ \sqrt{ m^2 + {\bm p}_{\perp}^2  } } \Gamma_s = {\rm e}^{-\eta/2} V_{s,1},\ v_{s,2} \equiv \gamma^t \Gamma_s  = {\rm e}^{\eta/2} V_{s,2}.  
\end{align} 
${}_{\rm \pm} \psi^{({\rm free})}_{{\bm p}_{\perp}, p_z, s} [ \tilde{A}_{m} ]$ are properly normalized in the Cartesian coordinates as
\begin{align}
	\int_{t={\rm const.}} d^2{\bm x}_{\perp} dz\; {}_{\pm} \bar{\psi}^{({\rm free})}_{{\bm p}_{\perp}, p_z, s} [ \tilde{A}_{m} ] \gamma^t {}_{\pm} \psi^{({\rm free})}_{{\bm p}'_{\perp}, p'_z, s'} [ \tilde{A}_{m} ]   
			&= \delta_{ss'} \delta^2 ({\bm p}_{\perp} - {\bm p}'_{\perp}) \delta (p_z - p'_z),  \\
	\int_{t={\rm const.}} d^2{\bm x}_{\perp} dz\; {}_{\pm} \bar{\psi}^{({\rm free})}_{{\bm p}_{\perp}, p_z, s} [ \tilde{A}_{m} ] \gamma^t {}_{\mp} \psi^{({\rm free})}_{{\bm p}'_{\perp}, p'_z, s'} [ \tilde{A}_{m} ] 
			&= 0.  
\end{align}

\subsubsection{Under a spatially homogeneous and constant color electric background field} \label{appB:quark}

We consider a spatially homogeneous and constant color electric background field given by Eq.~(\ref{eqconstE}), and analytically obtain all the mode functions for the equation of motion Eq.~(\ref{eqa1}), which we write $\psi^{\rm (const.)}$.  Let us begin with the differential equation Eq.~(\ref{eq11}) for $\phi^{\rm (const.)}_s$.  To solve this equation, we make an ansatz of a form: 
\begin{align}
	\phi^{\rm (const.)}_s(x) 
		\equiv \int d^2{\bm p}_{\perp} dp_{\eta}\; \phi^{\rm (const.)}_{{\bm p}_{\perp}, p_{\eta}, s}(x)
		\equiv \int d^2{\bm p}_{\perp} dp_{\eta}\; \chi^{\rm (const.)}_{{\bm p}_{\perp}, p_{\eta},s}(\tau) \frac{{\rm e}^{-\eta/2}}{\sqrt{m^2 + {\bm p}_{\perp}^2}}   \frac{{\rm e}^{i {\bm p}_{\perp}\cdot {\bm x}_{\perp} {\rm e}^{i p_{\eta} \eta} } }{ (2\pi)^{3/2} }  . \label{eqa__35}
\end{align}
As in the pure gauge case (Appendix~\ref{appA:quark}), the momentum labels ${\bm p}_{\perp}, p_{\eta}$ and the normalization factor ${\rm e}^{-\eta/2} / \sqrt{m^2+{\bm p}_{\perp}^2}$ are introduced.  We note that even if there are pure gauge potentials in addition to the electric field, the following computations do not change by simply adding the Wilson-line gauge factor $\Omega$ (Eq.~(\ref{eqWil})) into the above ansatz (Eq.~(\ref{eqa__35})).  Now, the differential equation for $\chi^{\rm (const.)}_{{\bm p}_{\perp}, p_{\eta},s}$ reads 
\begin{align}
		0 &= \left[ \partial_{\tau}^2 + \frac{\partial_{\tau}}{\tau}  +  \left(\frac{ p_{\eta} -i/2 + qE\tau^2/2}{\tau} \right)^2  + iqE + m^2 + {\bm p}_{\perp}^2 \right] \chi^{\rm (const.)}_{{\bm p}_{\perp}, p_{\eta}, s}.  \label{eqa_36} 
\end{align}
Two independent solutions of Eq.~(\ref{eqa_36}), which we write ${}_{k}\chi^{\rm (const.)}_{{\bm p}_{\perp}, p_{\eta}, s}$ ($k=1,2$), can be written in terms of the Tricomi confluent hypergeometric function $U(a;b;z)$.  Here, we consider the following particular solutions:
\begin{align}
	{}_{1}\chi^{\rm (const.)}_{{\bm p}_{\perp}, p_{\eta}, s} 
			&\equiv \frac{1}{\sqrt{\tau}} \exp \left[ - \pi \frac{m^2 + {\bm p}_{\perp}^2 }{ 4 |qE| } - i \frac{ qE \tau^2}{4}   \right]  \left( \sqrt{m^2+{\bm p}_{\perp}^2 } \tau \right)^{-i p_{\eta}}  U \left( i \frac{m^2+{\bm p}_{\perp}^2 }{ 2qE } ;  \frac{1}{2} -ip_{\eta} ; i \frac{qE \tau^2}{2}  \right) , \label{eqa_38} \\
	{}_{2}\chi^{\rm (const.)}_{{\bm p}_{\perp}, p_{\eta}, s} 
			&\equiv  \frac{-i}{2} \frac{1}{\sqrt{\tau}} \exp \left[ - \pi \frac{m^2 + {\bm p}_{\perp}^2 }{ 4 |qE| } + i \frac{ qE \tau^2}{4}   \right]  \left( \sqrt{m^2+{\bm p}_{\perp}^2 } \tau \right)^{1+i p_{\eta}}  U \left( 1-i\frac{m^2+{\bm p}_{\perp}^2 }{ 2qE } ;  \frac{3}{2} +ip_{\eta} ; -i \frac{qE \tau^2}{2}  \right), \label{eqa_39}
\end{align}
which are normalized as
\begin{align}
	|{}_{1}\chi^{\rm (const.)}_{{\bm p}_{\perp}, p_{\eta}, s}|^2+|{}_{2}\chi^{\rm (const.)}_{{\bm p}_{\perp}, p_{\eta}, s}|^2 = 1/\tau. \label{eqa_40}
\end{align}
It is useful to note that these particular solutions ${}_{k}\chi^{\rm (const.)}_{{\bm p}_{\perp}, p_{\eta}, s}$ satisfy a simultaneous differential equation, which is similar to that for the pure gauge case Eq.~(\ref{eqa18}) as
\begin{align}
	&\frac{i}{\sqrt{m^2+{\bm p}_{\perp}^2}} \left[  \partial_{\tau}  + \frac{i p_{\eta} + i q\tilde{A}_{\eta}  + 1/2 }{ \tau }  \right]
	\begin{pmatrix}
		{}_1 \chi^{\rm (const.)}_{{\bm p}_{\perp}, p_{\eta},s} \\
		{}_2 \chi^{\rm (const.)}_{{\bm p}_{\perp}, p_{\eta},s}
	\end{pmatrix}
	  = 
	\begin{pmatrix}
		{}_2 \chi^{{\rm (const.)}*}_{{\bm p}_{\perp}, p_{\eta},s} \\
		-{}_1 \chi^{{\rm (const.)}*}_{{\bm p}_{\perp}, p_{\eta},s}
	\end{pmatrix} .  \label{eqa40}
\end{align}
Thanks to this property, one finds that the mode functions ${}_{k} \psi^{\rm (const.)}_{{\bm p}_{\perp}, p_{\eta}, s}$ constructed from ${}_{k}\chi^{\rm (const.)}_{{\bm p}_{\perp}, p_{\eta}, s}$ have the same spinor structure as what we have for the plane wave solutions ${}_{\pm} \psi^{({\rm free})}_{{\bm p}_{\perp}, p_{\eta}, s}$ (Eq.~(\ref{eqa_22})) as we will see soon.

Now, one can readily construct the mode functions ${}_{k} \psi^{\rm (const.)}_{{\bm p}_{\perp}, p_{\eta}, s}$ ($k=1,2$) for the equation of motion Eq.~(\ref{eqa1}) under a spatially homogeneous and constant color electric background field $\tilde{A}_{\mu}$ (Eq.~(\ref{eqconstE})) as%
\footnote{Unlike the plane wave solutions ${}_{\pm} \psi^{({\rm free})}_{{\bm p}_{\perp}, p_{\eta}, s}$ (Eq.~(\ref{eqa19})) studied in Appendix~\ref{appA:quark}, we have not renamed the left subscript $k=1,2$ of ${}_{k}\psi_{{\bm p}_{\perp}, p_{\eta}, s}^{\rm (const.)}$ into $\pm$ because one cannot identify the positive and the negative frequency mode functions in principle when there are interactions, which mix up the positive and the negative frequency mode functions.  }
\begin{align}
	\begin{pmatrix}
		{}_{1}\psi_{{\bm p}_{\perp}, p_{\eta}, s}^{\rm (const.)} \\
		{}_{2}\psi_{{\bm p}_{\perp}, p_{\eta}, s}^{\rm (const.)} 
	\end{pmatrix}
	 \equiv
	[ i \Slash{\partial} - q \tilde{\Slash{A}} + m ]
	\begin{pmatrix}
		{}_{1}\phi_{{\bm p}_{\perp}, p_{\eta}, s}^{\rm (const.)}  \\
		{}_{2}\phi_{{\bm p}_{\perp}, p_{\eta}, s}^{\rm (const.)}
	\end{pmatrix} \Gamma_s,   \label{eqa42}
\end{align}
where we have used the definition of $\phi^{\rm (const.)}$ (Eq.~(\ref{eqa4})).  With the help of Eqs.~(\ref{eqa8}) and (\ref{eqa40}), one finds that Eq.~(\ref{eqa42}) can be more explicitly written as
\begin{align}
	\begin{pmatrix}
		{}_{1}\psi_{{\bm p}_{\perp}, p_{\eta}, s}^{\rm (const.)}  \\
		{}_{2}\psi_{{\bm p}_{\perp}, p_{\eta}, s}^{\rm (const.)} 
	\end{pmatrix}
	 &=
	\left[  	\begin{pmatrix} {}_{1}\chi^{\rm (const.)}_{{\bm p}_{\perp}, p_{\eta}, s} \\ {}_{2}\chi^{\rm (const.)}_{{\bm p}_{\perp}, p_{\eta}, s} \end{pmatrix} V_{s,1}  + \begin{pmatrix} {}_{2}\chi^{{\rm (const.)}*}_{{\bm p}_{\perp}, p_{\eta}, s} \\ -{}_{1}\chi^{{\rm (const.)}*}_{{\bm p}_{\perp}, p_{\eta}, s} \end{pmatrix} V_{s,2} \right]  \frac{ {\rm e}^{i {\bm p}_{\perp} \cdot {\bm x}_{\perp}} {\rm e}^{i p_{\eta} \eta}}{(2\pi)^{3/2}}.  \label{eqa43}
\end{align}
From the normalization conditions for ${}_{k} \chi^{\rm (const.)}_{{\bm p}_{\perp}, p_{\eta},s}$ (Eq.~(\ref{eqa_40})) and $V_{s,i}$ (Eq.~(\ref{eqa_21})), one can confirm that the mode functions ${}_{\pm}\psi_{{\bm p}_{\perp}, p_{\eta}, s}^{\rm (const.)} $ (Eq.~(\ref{eqa43})) are correctly normalized as
\begin{align}
	( {}_{\pm}\psi_{{\bm p}_{\perp}, p_{\eta}, s}^{\rm (const.)} | {}_{\pm} \psi_{{\bm p}'_{\perp}, p'_{\eta}, s'}^{\rm (const.)}   )_{\rm F}  = \delta_{ss'} \delta^2 ({\bm p}_{\perp} - {\bm p}'_{\perp}) \delta(p_{\eta}-p'_{\eta}) , \ \ ( {}_{\pm}\psi_{{\bm p}_{\perp}, p_{\eta}, s}^{\rm (const.)}  | {}_{\mp} \psi_{{\bm p}'_{\perp}, p'_{\eta}, s'}^{\rm (const.)}  )_{\rm F} = 0.  
\end{align}

\subsubsection{Under a spatially homogeneous and constant color electric background field with lifetime $T$ }\label{appC:quark}

We consider a spatially homogeneous and constant color electric background field with lifetime $T$ (Eq.~(\ref{eqEfield})), and find out all the mode functions $\psi^{\rm (finite)}$ for the equation of motion Eq.~(\ref{eqa1}).  The problem is equivalent to solving the equation of motion Eq.~(\ref{eqa1}) under a pure gauge background field for $0<\tau<\tau_0$ and $\tau_0+T<\tau$, and under a spatially homogeneous and constant color electric background field for $\tau_0 < \tau < \tau_0+T$.  All the mode functions for respective regions are already derived in Appendix~\ref{appA:quark} and Appendix~\ref{appB:quark}, respectively.  Thus, all we have to do is to connect these solutions smoothly at the boundary $\tau = \tau_0$ and $\tau = \tau_0 + T$.  Namely, we require
\begin{align}
	\left.\psi^{\rm (finite)}\right|_{\tau = \tau_0-0, \tau_0+T-0}  = \left.\psi^{\rm (finite)}\right|_{\tau = \tau_0+0, \tau_0+T+0}.  
\end{align}
In making this connection, it is useful to use a linear relation between ${}_{\pm} \psi^{({\rm free})}_{{\bm p}_{\perp}, p_{\eta}, s}$ and ${}_{k} \psi^{\rm (const.)}_{{\bm p}_{\perp}, p_{\eta}, s}$ ($k=1,2$) at fixed time $\tau = \tau_1$ described by
\begin{align}
	\begin{pmatrix}
		{}_{+}\psi_{{\bm p}_{\perp}, p_{\eta}, s}^{\rm (free)} [\tilde{A}_{\mu}(\tau_1)] \\
		{}_{-}\psi_{{\bm p}_{\perp}, p_{\eta}, s}^{\rm (free)} [\tilde{A}_{\mu}(\tau_1)]
	\end{pmatrix} 
	&=
	\sum_{s'} \int d^2{\bm p}'_{\perp} dp'_{\eta}
	\begin{pmatrix} 
		( {}_1 \psi^{\rm (const.)}_{{\bm p}'_{\perp}, p_{\eta}', s'} | {}_+ \psi^{({\rm free})}_{{\bm p}_{\perp}, p_{\eta}, s}[\tilde{A}_{\mu}(\tau_1)] )_{\rm F} & ( {}_2 \psi^{\rm (const.)}_{{\bm p}'_{\perp}, p'_{\eta}, s'} | {}_+ \psi^{({\rm free})}_{{\bm p}_{\perp}, p_{\eta}, s}[\tilde{A}_{\mu}(\tau_1)] )_{\rm F} \\
		( {}_1 \psi^{\rm (const.)}_{{\bm p}'_{\perp}, p'_{\eta}, s} | {}_- \psi^{({\rm free})}_{{\bm p}_{\perp}, p_{\eta}, s}[\tilde{A}_{\mu}(\tau_1)] )_{\rm F} & ( {}_2 \psi^{\rm (const.)}_{{\bm p}'_{\perp}, p'_{\eta}, s'} | {}_- \psi^{({\rm free})}_{{\bm p}_{\perp}, p_{\eta}, s}[\tilde{A}_{\mu}(\tau_1)] )_{\rm F}
	\end{pmatrix} 
	\begin{pmatrix}
		{}_{1}\psi_{{\bm p}'_{\perp}, p'_{\eta}, s'}^{\rm (const.)} \\
		{}_{2}\psi_{{\bm p}'_{\perp}, p'_{\eta}, s'}^{\rm (const.)}
	\end{pmatrix} \nonumber\\
	&\equiv 
	U^{\rm (q)}_{{\bm p}_{\perp}, p_{\eta}, s}(\tau_1)
	\begin{pmatrix}
		{}_{1}\psi_{{\bm p}_{\perp}, p_{\eta}-qE\tau_1^2/2, s}^{\rm (const.)} \\
		{}_{2}\psi_{{\bm p}_{\perp}, p_{\eta}-qE\tau_1^2/2, s}^{\rm (const.)}
	\end{pmatrix}. 		\label{eqa46}
\end{align}
The matrix elements are given by
\begin{align}
	A^{\rm (q)}_{{\bm p}_{\perp}, p_{\eta}, s}(\tau_1) 
		&\equiv ( U^{\rm (q)}_{{\bm p}_{\perp}, p_{\eta}, s}(\tau_1) )_{11} = \left[ ( U^{\rm (q)}_{{\bm p}_{\perp}, p_{\eta}, s}(\tau_1) )_{22} \right]^*  \nonumber\\
		&=\tau_1 \left[  {}_1 \chi^{{\rm (const.)}*}_{ {\bm p}_{\perp}, p_{\eta}-qE\tau_1^2/2, s } (\tau_1)  {}_1 \chi^{{\rm (free)}}_{ {\bm p}_{\perp}, p_{\eta}, s }(\tau_1) +   {}_2 \chi^{{\rm (const.)}}_{ {\bm p}_{\perp}, p_{\eta}-qE\tau_1^2/2, s }(\tau_1)   {}_2 \chi^{{\rm (free)}*}_{ {\bm p}_{\perp}, p_{\eta}, s }(\tau_1) \right] , \\
	B^{\rm (q)}_{{\bm p}_{\perp}, p_{\eta}, s}(\tau_1) 
		&\equiv ( U^{\rm (q)}_{{\bm p}_{\perp}, p_{\eta}, s}(\tau_1) )_{21} = \left[ - ( U^{\rm (q)}_{{\bm p}_{\perp}, p_{\eta}, s}(\tau_1) )_{12} \right]^* \nonumber\\
		&=  \tau_1 \left[  {}_1 \chi^{{\rm (const.)}*}_{ {\bm p}_{\perp}, p_{\eta}-qE\tau_1^2/2, s } (\tau_1)  {}_2 \chi^{{\rm (free)}}_{ {\bm p}_{\perp}, p_{\eta}, s }(\tau_1)  -   {}_2 \chi^{{\rm (const.)}}_{ {\bm p}_{\perp}, p_{\eta}-qE\tau_1^2/2, s }(\tau_1)   {}_1 \chi^{{\rm (free)}*}_{ {\bm p}_{\perp}, p_{\eta}, s }(\tau_1) \right] , 
\end{align} 
and are normalized as 
\begin{align}
	1 = |A^{\rm (q)}_{{\bm p}_{\perp}, p_{\eta}, s}|^2 + |B^{\rm (q)}_{{\bm p}_{\perp}, p_{\eta}, s}|^2, 
\end{align}
which means that the transformation $U^{\rm (q)}_{{\bm p}_{\perp}, p_{\eta}, s} (\tau_1) $ is unitary: $1 = U^{{\rm (q)}\dagger}_{{\bm p}_{\perp}, p_{\eta}, s}(\tau_1)  U^{\rm (q)}_{{\bm p}_{\perp}, p_{\eta}, s}(\tau_1) $.  Although the mode functions diverge at $\tau \rightarrow 0$ because of the coordinate singularity at $\tau=0$ of the $\tau$-$\eta$ coordinates, one can safely take the limit $\tau_1 \rightarrow 0$ of the transformation $U^{\rm (q)}$, i.e., the matrix elements $A^{\rm (q)}, B^{\rm (q)}$.  By using
\begin{align}
	H_{\nu}^{(1)}(z) 
		&\xrightarrow[|z| \rightarrow 0]{} 
			\left( \frac{z}{2} \right)^{-\nu} \left( \frac{\Gamma(\nu)}{i\pi} + {\mathcal O}(|z|)  \right)  + \left( \frac{z}{2} \right)^{\nu} \left( \frac{1+ i\cot (\nu\pi)}{ \Gamma(1+\nu) } + {\mathcal O}(|z|)  \right), \label{eqa49} \\
	H_{\nu}^{(2)}(z) 
		&\xrightarrow[|z| \rightarrow 0]{} 
			\left( \frac{z}{2} \right)^{-\nu} \left( -\frac{\Gamma(\nu)}{i\pi} + {\mathcal O}(|z|)  \right)  + \left( \frac{z}{2} \right)^{\nu} \left( \frac{1- i\cot (\nu\pi)}{ \Gamma(1+\nu) } + {\mathcal O}(|z|)  \right), \\
	U(a;b;z) 
		&\xrightarrow[|z| \rightarrow 0]{}
			z^{1-b} \left( \frac{\Gamma(-1+b)}{\Gamma(a)}  +  {\mathcal O}(|z|) \right) + \left( \frac{\Gamma(1-b)}{\Gamma(1+a-b)} +  {\mathcal O}(|z|) \right), \label{eqa51} 
\end{align}
one obtains
\begin{align}
	A^{\rm (q)}_{{\bm p}_{\perp}, p_{\eta}, s}(\tau) 
		&\xrightarrow[\tau \rightarrow 0]{}
			\frac{\sqrt{\pi}}{2} \frac{ {\rm e}^{ -\frac{i\pi}{4} \left( 1-\frac{qE}{|qE|} \right) } {\rm e}^{ -\frac{\pi p_{\eta}}{2} \left( 1+\frac{qE}{|qE|} \right) }  {\rm e}^{ -\pi \frac{m^2+{\bm p}_{\perp}^2}{4|qE|} }  \left( \frac{|qE|}{m^2+{\bm p}_{\perp}^2}  \right)^{-ip_{\eta}-1/2 } }{ \cosh(\pi p_{\eta}) \Gamma\left(1-i \frac{m^2+{\bm p}_{\perp}^2}{2qE} \right)  } \nonumber\\
			&\ \ \ \ \ \ \ \ \ \ \times \left[ 1  +   {\rm e}^{-\frac{i\pi}{4}\left( \frac{qE}{|qE|} -2 \right) } {\rm e}^{ \frac{\pi p_{\eta}}{2} \left( \frac{qE}{|qE|}  + 2 \right)  }  \left( \frac{2|qE|}{m^2+{\bm p}_{\perp}^2} \right)^{ip_{\eta}+1/2} \frac{ \Gamma\left( 1 - i\frac{m^2+{\bm p}_{\perp}^2}{2qE} \right) }{ \Gamma \left( \frac{1}{2} - ip_{\eta} - i\frac{m^2+{\bm p}_{\perp}^2}{2qE} \right) }   \right], \\
			B^{\rm (q)}_{{\bm p}_{\perp}, p_{\eta}, s}(\tau) 
		&\xrightarrow[\tau \rightarrow 0]{}
			\frac{\sqrt{\pi}}{2} \frac{ {\rm e}^{ \frac{i\pi}{4} \left( 1+\frac{qE}{|qE|} \right) } {\rm e}^{ \frac{\pi p_{\eta}}{2} \left( 1-\frac{qE}{|qE|} \right) }  {\rm e}^{ -\pi \frac{m^2+{\bm p}_{\perp}^2}{4|qE|} }  \left( \frac{|qE|}{m^2+{\bm p}_{\perp}^2}  \right)^{-ip_{\eta}-1/2 } }{ \cosh(\pi p_{\eta}) \Gamma\left(1-i \frac{m^2+{\bm p}_{\perp}^2}{2qE} \right)  } \nonumber\\
			&\ \ \ \ \ \ \ \ \ \ \times \left[ 1  +   {\rm e}^{-\frac{i\pi}{4}\left( \frac{qE}{|qE|} +2 \right) } {\rm e}^{ \frac{\pi p_{\eta}}{2} \left( \frac{qE}{|qE|}  - 2 \right)  }  \left( \frac{2|qE|}{m^2+{\bm p}_{\perp}^2} \right)^{ip_{\eta}+1/2} \frac{ \Gamma\left( 1 - i\frac{m^2+{\bm p}_{\perp}^2}{2qE} \right) }{ \Gamma \left( \frac{1}{2} - ip_{\eta} - i\frac{m^2+{\bm p}_{\perp}^2}{2qE} \right) }   \right].  
\end{align}

We consider two kinds of boundary conditions for the mode functions: We define mode functions ${}_{\pm} \psi^{({\rm finite; in})}_{{\bm p}_{\perp}, p_{\eta}, s}$ (${}_{\pm} \psi^{({\rm finite; out})}_{{\bm p}_{\perp}, p_{\eta}, s}$) by a boundary condition at $\tau < \tau_0$ ($\tau > \tau_0+T$) to coincide with the plane wave solutions ${}_{\pm} \psi^{({\rm free})}_{{\bm p}_{\perp}, p_{\eta}, s} [\tilde{A}_{\mu}]$.  Using the linear relation, Eq.~(\ref{eqa46}), one can easily construct such mode functions, ${}_{\pm} \psi^{({\rm finite; in})}_{{\bm p}_{\perp}, p_{\eta}, s}$ and ${}_{\pm} \psi^{({\rm finite; out})}_{{\bm p}_{\perp}, p_{\eta}, s}$, as 
\begin{align}
	\begin{pmatrix}
		{}_{+} \psi^{({\rm finite; in})}_{{\bm p}_{\perp}, p_{\eta}, s} \\
		{}_{-} \psi^{({\rm finite; in})}_{{\bm p}_{\perp}, p_{\eta}, s} 
	\end{pmatrix}
	= 
	\left\{
		\begin{array}{ll}
			\begin{pmatrix}
				{}_{+} \psi^{({\rm free})}_{{\bm p}_{\perp}, p_{\eta}, s} [\tilde{A}_{\mu}] \\
				{}_{-} \psi^{({\rm free})}_{{\bm p}_{\perp}, p_{\eta}, s} [\tilde{A}_{\mu}] 
			\end{pmatrix}	& {\rm for}\ 0<\tau<\tau_0 \\
			\\
			U^{\rm (q)}_{{\bm p}_{\perp}, p_{\eta}, s}(\tau_0)
			\begin{pmatrix}
				{}_{1} \psi^{({\rm const.})}_{{\bm p}_{\perp}, p_{\eta}-qE\tau_0^2/2, s}  \\
				{}_{2} \psi^{({\rm const.})}_{{\bm p}_{\perp}, p_{\eta}-qE\tau_0^2/2, s} 
			\end{pmatrix}	& {\rm for}\ \tau_0<\tau<\tau_0+T \\
			\\
			U^{\rm (q)}_{{\bm p}_{\perp}, p_{\eta}, s}(\tau_0)
			U^{{\rm (q)}\dagger}_{{\bm p}_{\perp}, p_{\eta}-qE\tau_0^2/2+qE(\tau_0+T)^2/2, s}(\tau_0+T) & \\
			\ \ \ \ \ \ \ \ \ \ \ \ \ \ \ \ \ \ \ \ \times
			\begin{pmatrix}
				{}_{+} \psi^{({\rm free})}_{{\bm p}_{\perp}, p_{\eta}-qE\tau_0^2/2+qE(\tau_0+T)^2/2, s} [\tilde{A}_{\mu}] \\
				{}_{-} \psi^{({\rm free})}_{{\bm p}_{\perp}, p_{\eta}-qE\tau_0^2/2+qE(\tau_0+T)^2/2, s} [\tilde{A}_{\mu}]
			\end{pmatrix}	& {\rm for}\ \tau_0+T<\tau \\
		\end{array}
	\right. ,
\end{align}
\begin{align}
	\begin{pmatrix}
		{}_{+} \psi^{({\rm finite; out})}_{{\bm p}_{\perp}, p_{\eta}, s} \\
		{}_{-} \psi^{({\rm finite; out})}_{{\bm p}_{\perp}, p_{\eta}, s} 
	\end{pmatrix}
	= 
	\left\{
		\begin{array}{ll}
			U^{{\rm (q)}}_{{\bm p}_{\perp}, p_{\eta}, s}(\tau_0+T)
			U^{{\rm (q)}\dagger}_{{\bm p}_{\perp}, p_{\eta}+qE\tau_0^2/2-qE(\tau_0+T)^2/2, s}(\tau_0) & \\
			\ \ \ \ \ \ \ \ \ \ \ \ \ \ \ \ \ \ \ \ \times
			\begin{pmatrix}
				{}_{+} \psi^{({\rm free})}_{{\bm p}_{\perp}, p_{\eta}+qE\tau_0^2/2-qE(\tau_0+T)^2/2, s} [\tilde{A}_{\mu}] \\
				{}_{-} \psi^{({\rm free})}_{{\bm p}_{\perp}, p_{\eta}+qE\tau_0^2/2-qE(\tau_0+T)^2/2, s} [\tilde{A}_{\mu}] 
			\end{pmatrix}	& {\rm for}\ 0<\tau<\tau_0 \\
			\\
			U^{{\rm (q)}}_{{\bm p}_{\perp}, p_{\eta}, s}(\tau_0+T)
			\begin{pmatrix}
				{}_{1} \psi^{({\rm const.})}_{{\bm p}_{\perp}, p_{\eta}-qE(\tau_0+T)^2/2, s}  \\
				{}_{2} \psi^{({\rm const.})}_{{\bm p}_{\perp}, p_{\eta}-qE(\tau_0+T)^2/2, s} 
			\end{pmatrix}	& {\rm for}\ \tau_0<\tau<\tau_0+T \\
			\\
			\begin{pmatrix}
				{}_{+} \psi^{({\rm free})}_{{\bm p}_{\perp}, p_{\eta}, s} [\tilde{A}_{\mu}] \\
				{}_{-} \psi^{({\rm free})}_{{\bm p}_{\perp}, p_{\eta}, s} [\tilde{A}_{\mu}]
			\end{pmatrix}	& {\rm for}\ \tau_0+T<\tau \\
		\end{array}
	\right. .
\end{align}

These two sets of mode functions are not independent but related with each other by a Bogoliubov transformation discussed in the main text (see Eq.~(\ref{eq_64})).  Now, one can analytically compute the Bogoliubov coefficients as
\begin{align}
	&( {}_+ \psi^{\rm (finite; out)}_{{\bm p}_{\perp}, p_{\eta}, s} | {}_+ \psi^{\rm (finite; in)}_{{\bm p}'_{\perp}, p'_{\eta}, s'} )_{\rm F} = \left[ ( {}_- \psi^{\rm (finite; out)}_{{\bm p}_{\perp}, p_{\eta}, s} | {}_- \psi^{\rm (finite; in)}_{{\bm p}'_{\perp}, p'_{\eta}, s'} )_{\rm F} \right]^* \nonumber\\
		&\ = \delta_{ss'} \delta^2({{\bm p}_{\perp} - {\bm p}'_{\perp}}) \delta ( p'_{\eta} - (p_{\eta} + qE\tau_0^2/2 - qE (\tau_0+T)^2/2)  )  \nonumber\\
		&\ \ \ \ \times \left[ A^{\rm (q)}_{{\bm p}_{\perp}, p_{\eta}+qE\tau_0^2/2-qE(\tau_0+T)^2/2, s}(\tau_0) A^{{\rm (q)}*}_{{\bm p}_{\perp}, p_{\eta}, s}(\tau_0+T)  + B^{{\rm (q)}*}_{{\bm p}_{\perp}, p_{\eta}+qE\tau_0^2/2-qE(\tau_0+T)^2/2, s}(\tau_0) B^{{\rm (q)}}_{{\bm p}_{\perp}, p_{\eta}, s}(\tau_0+T) \right] , \\
	&( {}_- \psi^{\rm (finite; out)}_{{\bm p}_{\perp}, p_{\eta}, s} | {}_+ \psi^{\rm (finite; in)}_{{\bm p}'_{\perp}, p'_{\eta}, s'} )_{\rm F} = \left[ -( {}_+ \psi^{\rm (finite; out)}_{{\bm p}_{\perp}, p_{\eta}, s} | {}_- \psi^{\rm (finite; in)}_{{\bm p}'_{\perp}, p'_{\eta}, s'} )_{\rm F} \right]^* \nonumber\\
		&\ = \delta_{ss'} \delta^2({{\bm p}_{\perp} - {\bm p}'_{\perp}}) \delta ( p'_{\eta} - (p_{\eta} + qE\tau_0^2/2 - qE (\tau_0+T)^2/2)  ) \nonumber\\
		&\ \ \ \ \times \left[ A^{\rm (q)}_{{\bm p}_{\perp}, p_{\eta}+qE\tau_0^2/2-qE(\tau_0+T)^2/2, s}(\tau_0) B^{{\rm (q)}*}_{{\bm p}_{\perp}, p_{\eta}, s}(\tau_0+T)  - B^{{\rm (q)}*}_{{\bm p}_{\perp}, p_{\eta}+qE\tau_0^2/2-qE(\tau_0+T)^2/2, s}(\tau_0) A^{{\rm (q)}}_{{\bm p}_{\perp}, p_{\eta}, s}(\tau_0+T) \right] .
\end{align}
\end{widetext}

\subsection{Gluon} 

We consider the Abelianized equation of motion for the gluon field $W_{\mu}$ in the $\tau$-$\eta$ coordinates (see Eq.~(\ref{eq48})): 
\begin{align}
	\left[ ( \nabla_{\rho} + iq \tilde{A}_{\rho} )^2 g^{\mu\nu} + 2iq \tilde{F}^{\mu\nu}  \right] W_{\nu} &= 0 . \label{eqb1}  
\end{align}
Here, we omit the color indices $A$ for simplicity.

In solving Eq.~(\ref{eqb1}), we first expand the gluon field $W_{\mu}$ in terms of a polarization vector in the $\tau$-$\eta$ coordinates $\varepsilon_{\mu, \sigma}$ ($\sigma=0,1,2,3$) and a scalar amplitude $\phi_{\sigma}$ as
\begin{align}
	W_{\mu} \equiv \sum_{\sigma=1}^4  \varepsilon_{\mu, \sigma} \phi_{\sigma}.  \label{eqb2}
\end{align}
By noting that the choice of the polarization vector is arbitrary in principle, we assume here that the polarization vector $\varepsilon_{\mu, \sigma}$ are constructed from a constant vector $\tilde{\varepsilon}_{m, \sigma} $ by contracting the viervein matrix $e^{\mu}_{\;\;m}$ as
\begin{align}
	\varepsilon_{\mu, \sigma} = e^{m}_{\ \ \mu} \tilde{\varepsilon}_{m, \sigma}.  
\end{align}
Under this assumption, the covariant derivative of the polarization vector $\varepsilon_{\nu, \sigma}$ vanish as $\nabla_{\mu} \varepsilon_{\nu, \sigma} = 0$.

In this Appendix, we only consider the cases where a constant color electric field pointing to the $z$-direction is present at most.  For such cases, it is convenient to choose $\varepsilon_{\mu, \sigma}$ ($\tilde{\varepsilon}_{m, \sigma} $) to be an eigenvector of the background field strength tensor $\tilde{F}_{\mu\nu}$ ($\tilde{F}_{mn}$) as 
\begin{align}
	\tilde{F}_{\mu}^{\;\;\nu} \varepsilon_{\nu, \sigma}  =  \Lambda_{\sigma} \varepsilon_{\mu, \sigma},\ \Leftrightarrow \ \tilde{F}_{m}^{\;\;n} \tilde{\varepsilon}_{n, \sigma}  =  \Lambda_{\sigma} \tilde{\varepsilon}_{m, \sigma},  
\end{align}
where four eigenvalues $\Lambda_{\sigma}$ are given by
\begin{align}
	\Lambda_{0} = -E,\ 
	\Lambda_{1} = \Lambda_{2} = 0,\ 
	\Lambda_{3} = E.  \label{eqb61}
\end{align}
In other words, we have defined the polarization of gluons by the direction of the background field.  We also normalize the polarization vector $\varepsilon_{\mu, \sigma}$ ( $\tilde{\varepsilon}_{\mu, \sigma}$) as 
\begin{align}
	g^{\mu\nu} \varepsilon_{\mu, \sigma}^{*} \varepsilon_{\nu, \sigma'} = -\xi_{\sigma \sigma'}  ,\ 
	\sum_{\sigma,\sigma'} \xi_{\sigma \sigma'} \varepsilon_{\mu, \sigma}^{*} \varepsilon_{\nu, \sigma'} = -g_{\mu\nu}, \label{eqb36} 
\end{align}	
and
\begin{align}
	\eta^{mn} \tilde{\varepsilon}_{m, \sigma}^{*} \tilde{\varepsilon}_{n, \sigma'} = -\xi_{\sigma \sigma'}  ,\ 
	\sum_{\sigma,\sigma'} \xi_{\sigma \sigma'} \tilde{\varepsilon}_{m, \sigma}^{*} \tilde{\varepsilon}_{n, \sigma'} = -\eta_{mn},  \label{eqbb36}
\end{align}
where $\xi_{\sigma \sigma'}$ is the indefinite metric introduced in Eq.~(\ref{eq76}).

Now, one obtains a differential equation for $\phi_{\sigma}$ as
\begin{align}
	0 &= \left[ ( \partial_{\mu} + iq \tilde{A}_{\mu} )^2 + \frac{ \partial_{\tau} + iq\tilde{A}_{\tau} }{ \tau } + 2iq \Lambda_{\sigma}   \right] \phi_{\sigma} . \label{eqb38}
\end{align}

\subsubsection{Under a pure gauge background field (plane wave solutions)} \label{appA:gluon}

In order to construct all the mode functions for the equation of motion Eq.~(\ref{eqb1}) under a pure gauge background field $\tilde{A}_{\mu}$ (Eq.~(\ref{eqpure})), which we write $W_{\mu}^{\rm (free)}$, we first consider to solve the differential equation for $\phi_{\sigma}^{\rm (free)}$ (Eq.~(\ref{eqb38})).  For this, we make an ansatz of a form: 
\begin{align}
	\phi_{\sigma}^{\rm (free)}(x) 
		&\equiv \int d^2{\bm p}_{\perp} dp_{\eta} \phi_{{\bm p}_{\perp}, p_{\eta}, \sigma}^{\rm (free)} (x), \\
	\phi_{{\bm p}_{\perp}, p_{\eta}, \sigma}^{\rm (free)}
		&\equiv \Omega(x) \chi^{\rm (free)}_{{\bm p}_{\perp}, p_{\eta}, \sigma}(\tau) \frac{ {\rm e}^{i{\bm p}_{\perp}\cdot {\bm x}_{\perp} } {\rm e}^{i p_{\eta} \eta } }{ (2\pi)^{3/2} }.  
\end{align}
Here, the momentum labels ${\bm p}_{\perp}, p_{\eta}$ are introduced.  $\Omega$ is the Wilson-line gauge factor which is the same as what we have defined in Eq.~(\ref{eqa_14}).  One readily finds that $\chi^{\rm (free)}_{{\bm p}_{\perp}, p_{\eta}, \sigma}$ satisfies the Bessel differential equation:
\begin{align}
	0 	&= \Biggl[ \tau^2 \partial_{\tau}^2 + \tau \partial_{\tau} + \left\{ \left( \sqrt{m^2+{\bm p}_{\perp}^2} \tau \right)^2 - (ip_{\eta})^2  \right\}     \Biggr] \chi_{{\bm p}_{\perp}, p_{\eta}, \sigma}^{\rm (free)}.  \label{eqb41}
\end{align}
Since Eq.~(\ref{eqb41}) is a second order differential equation, there are two independent solutions, which we write ${}_{k}\chi^{\rm (free)}_{{\bm p}_{\perp}, p_{\eta}, \sigma}$ ($k=1,2$).  In this Appendix, we consider
\begin{align}
	{}_1 \chi_{{\bm p}_{\perp}, p_{\eta}, \sigma}^{\rm (free)}
		&\equiv \frac{ \sqrt{\pi} }{ 2i } {\rm e}^{\pi p_{\eta}/2} H^{(2)}_{ip_{\eta}} ( |{\bm p}_{\perp}| \tau ), \nonumber\\
	{}_2 \chi_{{\bm p}_{\perp}, p_{\eta}, \sigma}^{\rm (free)}
		&\equiv ({}_1 \chi_{{\bm p}_{\perp}, p_{\eta}, \sigma})^* , 
\end{align}
where we have normalized the solutions ${}_{k}\chi^{\rm (free)}_{{\bm p}_{\perp}, p_{\eta}, \sigma}$ by
\begin{align}
	\frac{1}{ \tau} 
		&= \sum_{\sigma'} \xi_{\sigma \sigma'} \left[ i  \left( {}_1 \chi^{{\rm (free)}*}_{{\bm p}_{\perp}, p_{\eta}, \sigma} \overset{\leftrightarrow}{\partial_{\tau}} {}_1 \chi^{\rm (free)}_{{\bm p}_{\perp}, p_{\eta}, \sigma'}  \right) \right] \nonumber\\
		&= -\sum_{\sigma'} \xi_{\sigma \sigma'} \left[ i \left( {}_2 \chi_{{\bm p}_{\perp}, p_{\eta}, \sigma}^{{\rm (free)}*} \overset{\leftrightarrow}{\partial_{\tau}} {}_2 \chi_{{\bm p}_{\perp}, p_{\eta}, \sigma'}^{\rm (free)}  \right) \right],   \label{eqb44} \\
	0 
		&= \sum_{\sigma'} \xi_{\sigma \sigma'} \left[ i  \left( {}_1 \chi_{{\bm p}_{\perp}, p_{\eta}, \sigma}^{{\rm (free)}*} \overset{\leftrightarrow}{\partial_{\tau}} {}_2 \chi_{{\bm p}_{\perp}, p_{\eta}, \sigma'}^{\rm (free)}  \right) \right]  \nonumber\\
		&= \sum_{\sigma'} \xi_{\sigma \sigma'} \left[  i  \left( {}_2 \chi_{{\bm p}_{\perp}, p_{\eta}, \sigma}^{{\rm (free)}*} \overset{\leftrightarrow}{\partial_{\tau}} {}_1 \chi_{{\bm p}_{\perp}, p_{\eta}, \sigma'}^{\rm (free)}  \right) \right].   \label{eqb_46}  
\end{align}

Now, we are ready to construct all the mode functions ${}_{\pm} W^{({\rm free})}_{\mu, {\bm p}_{\perp}, p_{\eta}, \sigma} [ \tilde{A}_{\nu} ]$.  By using the definition of $\phi^{\rm (free)}_{\sigma}$ (Eq.~(\ref{eqb2})), one can construct the ${}_{\pm} W^{({\rm free})}_{\mu, {\bm p}_{\perp}, p_{\eta}, \sigma} [ \tilde{A}_{\nu} ]$ as 
\begin{align}
	\begin{pmatrix}
		{}_{+} W^{({\rm free})}_{\mu, {\bm p}_{\perp}, p_{\eta}, \sigma} [ \tilde{A}_{\nu} ] \\
		{}_{-} W^{({\rm free})}_{\mu, {\bm p}_{\perp}, p_{\eta}, \sigma} [ \tilde{A}_{\nu} ]
	\end{pmatrix}
		&\equiv 
	\varepsilon_{\mu, \sigma}
	\begin{pmatrix}
		{}_{1} \phi_{{\bm p}_{\perp}, p_{\eta}, \sigma}^{\rm (free)} \\
		{}_{2} \phi_{{\bm p}_{\perp}, p_{\eta}, \sigma}^{\rm (free)} 
	\end{pmatrix}\nonumber\\
		&=
	\varepsilon_{\mu, \sigma} \Omega 
	\begin{pmatrix}
		{}_{1} \chi_{{\bm p}_{\perp}, p_{\eta}, \sigma}^{\rm (free)} \\
		{}_{2} \chi_{{\bm p}_{\perp}, p_{\eta}, \sigma}^{\rm (free)} 
	\end{pmatrix}
	\frac{{\rm e}^{i {\bm p}_{\perp}\cdot {\bm x}_{\perp}} {\rm e}^{ip_{\eta} \eta} }{ (2\pi)^{3/2} }.  \label{eqb47}
\end{align}
Here, we have changed the left subscript $k=1,2$ into $\pm$ for a notational simplicity because ${}_{+} W^{({\rm free})}_{\mu, {\bm p}_{\perp}, p_{\eta}, \sigma} [ \tilde{A}_{\nu} ]$ (${}_{-} W^{({\rm free})}_{\mu, {\bm p}_{\perp}, p_{\eta}, \sigma} [ \tilde{A}_{\nu} ]$) corresponds to the positive (negative) frequency mode function in the $\tau$-$\eta$ coordinates as we will explain soon.  With the help of the normalization conditions Eq.~(\ref{eqb36}) for $\varepsilon_{\mu, \sigma}$ and Eqs.~(\ref{eqb44}) and (\ref{eqb_46}) for ${}_{k} \chi^{\rm (free)}_{{\bm p}_{\perp}, p_{\eta}, \sigma}$, one can easily check that, in the temporal gauge $\tilde{A}_{\tau} = 0$, the mode functions satisfy the correct normalization condition for vector fields in the $\tau$-$\eta$ coordinates (see also Eqs.~(\ref{eq_75}) and (\ref{eq_76}) in the main text): 
\begin{align}
	&-g^{\mu\nu} ( {}_{\pm} W^{({\rm free})}_{\mu, {\bm p}_{\perp}, p_{\eta}, \sigma} [ \tilde{A}_{\rho} ]  |  {}_{\pm} W^{({\rm free})}_{\nu, {\bm p}'_{\perp}, p'_{\eta}, \sigma' } [ \tilde{A}_{\rho} ]  )_{\rm B} \nonumber\\ 
	&\ \ \ \ \ \ \ \ \ \ \ \ \ \ \ \ \ \ \ \ \ = \pm \xi_{\sigma \sigma'} \delta^2({\bm p}_{\perp} - {\bm p}'_{\perp}) \delta (p_{\eta} - p'_{\eta}), \\
	&-g^{\mu\nu} ( {}_{\pm} W^{({\rm free})}_{\mu, {\bm p}_{\perp}, p_{\eta}, \sigma} [ \tilde{A}_{\rho} ] |  {}_{\mp} W^{({\rm free})}_{\nu, {\bm p}'_{\perp}, p'_{\eta}, \sigma'} [ \tilde{A}_{\rho} ] )_{\rm B} = 0.  
\end{align}

The mode function ${}_{+} W^{({\rm free})}_{\mu, {\bm p}_{\perp}, p_{\eta}, \sigma} [ \tilde{A}_{\nu} ]$ (${}_{-} W^{({\rm free})}_{\mu, {\bm p}_{\perp}, p_{\eta}, \sigma} [ \tilde{A}_{\nu} ]$) can be written as a superposition of the positive (negative) frequency mode function in the Cartesian coordinates, and hence one can understand that ${}_{+} W^{({\rm free})}_{\mu, {\bm p}_{\perp}, p_{\eta}, \sigma} [ \tilde{A}_{\nu} ]$ (${}_{-} W^{({\rm free})}_{\mu, {\bm p}_{\perp}, p_{\eta}, \sigma} [ \tilde{A}_{\nu} ]$) can be understood as the positive (negative) frequency mode function in the $\tau$-$\eta$ coordinates.  In order to see this, we again use the integral representations for the Hankel functions $H^{(n)}_{\nu}(z)$ ($n=1,2$) (Eq.~(\ref{eqa_26})) to find
\begin{align}
	{}_{\pm} W^{({\rm free})}_{\mu, {\bm p}_{\perp}, p_{\eta}, \sigma} [\tilde{A}_{\nu}] = e^{m}_{\;\; \mu} \int dp_z \frac{ {\rm e}^{\pm i p_{\eta} y_{\bm p}} }{ \sqrt{2\pi \omega_{\bm p}}} {}_{\rm \pm} W^{({\rm free})}_{m, {\bm p}_{\perp}, p_{z}, \sigma }[\tilde{A}_{n}].   \label{eqb80}
\end{align}
Here, ${}_{+} W^{({\rm free})}_{m, {\bm p}_{\perp}, p_{z}, \sigma} [ \tilde{A}_{n} ]$ (${}_{-} W^{({\rm free})}_{m, {\bm p}_{\perp}, p_{z}, \sigma} [ \tilde{A}_{n} ]$) is the positive (negative) frequency mode function in the Cartesian coordinates satisfying the free field equation of motion in the Cartesian coordinates as $0 =  (\partial_l + iq\tilde{A}_l )^2 {}_{\rm \pm} W^{({\rm free})}_{m, {\bm p}_{\perp}, p_{\eta}, \sigma} [ \tilde{A}_{n} ]$ labeled by $p_z$ conjugate to $z$: 
\begin{align}
		{}_{\pm} W^{({\rm free})}_{m, {\bm p}_{\perp}, p_{z}, \sigma} [ \tilde{A}_{n} ] 
		=
	\tilde{\varepsilon}_{m, \sigma} \Omega 
	\frac{ {\rm e}^{\mp i\omega_{\bm p} t} }{\sqrt{2\omega_{\bm p}} }
	\frac{{\rm e}^{i {\bm p}_{\perp}\cdot {\bm x}_{\perp}} {\rm e}^{ip_{z} z} }{ (2\pi)^{3/2} }.  
\end{align}
These mode functions are properly normalized in the Cartesian coordinates.  In the temporal gauge $\tilde{A}_t = 0$, it reads
\begin{align}
	&-\eta^{mn} \int_{t={\rm const.}} d^2{\bm x}_{\perp} dz\nonumber\\
		&\ \ \ \ \ \ \ \ \ \ \times{}_{\pm} W^{({\rm free})*}_{m, {\bm p}_{\perp}, p_{z}, \sigma} [ \tilde{A}_{l} ] \overset{\leftrightarrow}{\partial}_t {}_{\pm} W^{({\rm free})}_{n, {\bm p}'_{\perp}, p'_{z}, \sigma'} [ \tilde{A}_{l} ] \nonumber\\
		&\ \ \ \ \ \ \ \ \ \ \ \ \ \ \ \ \ \ \ \ \ \ \ \ \ \ \ = \pm \xi_{\sigma \sigma'} \delta^2({\bm p}_{\perp} - {\bm p}'_{\perp}) \delta (p_{z} - p'_z), \\
	&-\eta^{mn} \int_{t={\rm const.}} d^2{\bm x}_{\perp} dz\nonumber\\
		&\ \ \ \ \ \ \ \ \ \ \times {}_{\pm} W^{({\rm free})*}_{m, {\bm p}_{\perp}, p_{z}, \sigma} [ \tilde{A}_{l} ] \overset{\leftrightarrow}{\partial}_t {}_{\mp} W^{({\rm free})}_{n, {\bm p}'_{\perp}, p'_{z}, \sigma'} [ \tilde{A}_{l} ] 
		= 0. 
\end{align}

\subsubsection{Under a spatially homogeneous and constant color electric background field} \label{appB:gluon}

We consider a spatially homogeneous and constant color electric background field (Eq.~(\ref{eqconstE})), and construct all the mode functions for the equation of motion Eq.~(\ref{eqb1}), which we write $W^{\rm (const.)}_{\mu}$.  First, we solve Eq.~(\ref{eqb38}) for $\phi_{\sigma}^{\rm (const.)}$ by making an ansatz: 
\begin{align}
	\phi^{\rm (const.)}_{\sigma}(x) 
		&\equiv \int d^2{\bm p}_{\perp} dp_{\eta}\; \phi^{\rm (const.)}_{{\bm p}_{\perp}, p_{\eta}, \sigma}(x), \\
	\phi^{\rm (const.)}_{{\bm p}_{\perp}, p_{\eta}, \sigma}(x) \label{eqb__35}
		&\equiv \chi^{\rm (const.)}_{{\bm p}_{\perp}, p_{\eta},s}(\tau)   \frac{{\rm e}^{i {\bm p}_{\perp}\cdot {\bm x}_{\perp} {\rm e}^{i p_{\eta} \eta} } }{ (2\pi)^{3/2} }  .
\end{align}
As in the pure gauge case (Appendix~\ref{appA:gluon}), the momentum labels ${\bm p}_{\perp}, p_{\eta}$ are introduced.  We note that even if there are pure gauge potentials in addition to the electric field, the following computations do not change by simply adding the Wilson-line gauge factor $\Omega$ (Eq.~(\ref{eqWil})) into the above ansatz (Eq.~(\ref{eqb__35})).  Now, the differential equation for $\chi^{\rm (const.)}_{{\bm p}_{\perp}, p_{\eta}, \sigma}$ becomes  
\begin{align}
		0 = \left[ \partial_{\tau}^2 + \frac{\partial_{\tau}}{\tau}  +  \left(\frac{ p_{\eta} + qE\tau^2/2}{\tau} \right)^2  + {\bm p}_{\perp}^2 + 2iq\Lambda_{\sigma} \right] \chi^{\rm (const.)}_{{\bm p}_{\perp}, p_{\eta}, \sigma},  \label{eqb_36} 
\end{align}
where the eigenvalue $\Lambda_{\sigma}$ are given by Eq.~(\ref{eqb61}).  Two independent solutions of Eq.~(\ref{eqb_36}), which we write ${}_{k}\chi^{\rm (const.)}_{{\bm p}_{\perp}, p_{\eta}, \sigma}$ ($k=1,2$), can be written in terms of the Tricomi confluent hypergeometric function $U(a;b;z)$.  Here, we consider the following particular solutions:
\begin{align}
	{}_{1}\chi^{\rm (const.)}_{{\bm p}_{\perp}, p_{\eta}, \sigma} 
			&\equiv \frac{1}{\sqrt{2}}   \exp \left[ - \frac{\pi}{2} \left( \frac{{\bm p}_{\perp}^2}{2|qE|} + p_{\eta} + i \frac{q\Lambda_{\sigma}}{qE}   \right)  - i \frac{|qE|\tau^2}{4} \right] \nonumber\\
			& \ \ \ \times\left( \frac{|qE|\tau^2}{2} \right)^{i p_{\eta}/2 } \nonumber\\
			& \ \ \ \times U \left( \frac{1}{2} + i\frac{{\bm p}_{\perp}^2 }{ 2|qE| } + ip_{\eta} - \frac{q \Lambda_{\sigma}}{qE} ;  1 + ip_{\eta} ; i \frac{|qE| \tau^2}{2}  \right) , \label{eqb_38} \\
	{}_{2}\chi^{\rm (const.)}_{{\bm p}_{\perp}, p_{\eta}, \sigma} &\equiv \sum_{ \sigma'} \xi_{\sigma \sigma'} {}_{1}\chi^{{\rm (const.)}*}_{{\bm p}_{\perp}, p_{\eta}, \sigma'} ,  \label{eqb_39}
\end{align}
which are normalized as 
\begin{align}
	\frac{1}{ \tau} 
		&= \sum_{\sigma'} \xi_{\sigma \sigma'} \left[ i \left( {}_1 \chi^{{\rm (const.)}*}_{{\bm p}_{\perp}, p_{\eta}, \sigma} \overset{\leftrightarrow}{\partial_{\tau}} {}_1 \chi^{\rm (const.)}_{{\bm p}_{\perp}, p_{\eta}, \sigma'}  \right) \right] \nonumber\\
		&= -\sum_{\sigma'} \xi_{\sigma \sigma'} \left[ i \left( {}_2 \chi_{{\bm p}_{\perp}, p_{\eta}, \sigma}^{{\rm (const.)}*} \overset{\leftrightarrow}{\partial_{\tau}} {}_2 \chi_{{\bm p}_{\perp}, p_{\eta}, \sigma'}^{\rm (const.)}  \right) \right], \label{eqb_44} \\
	0 
		&= \sum_{\sigma'} \xi_{\sigma \sigma'} \left[ i \left( {}_1 \chi_{{\bm p}_{\perp}, p_{\eta}, \sigma}^{{\rm (const.)}*} \overset{\leftrightarrow}{\partial_{\tau}} {}_2 \chi_{{\bm p}_{\perp}, p_{\eta}, \sigma'}^{\rm (const.)}  \right) \right] \nonumber\\
		&= \sum_{\sigma'} \xi_{\sigma \sigma'} \left[  i \left( {}_2 \chi_{{\bm p}_{\perp}, p_{\eta}, \sigma}^{{\rm (const.)}*} \overset{\leftrightarrow}{\partial_{\tau}} {}_1 \chi_{{\bm p}_{\perp}, p_{\eta}, \sigma'}^{\rm (const.)}  \right) \right].     \label{eqb___46}  
\end{align}

Now, one readily obtains the mode functions ${}_{k} W^{({\rm const.})}_{\mu, {\bm p}_{\perp}, p_{\eta}, \sigma}$ ($k=1,2$) as
\begin{align}
	\begin{pmatrix}
		{}_{1} W^{({\rm const.})}_{\mu, {\bm p}_{\perp}, p_{\eta}, \sigma} [ \tilde{A}_{\nu} ] \\
		{}_{2} W^{({\rm const.})}_{\mu, {\bm p}_{\perp}, p_{\eta}, \sigma} [ \tilde{A}_{\nu} ]
	\end{pmatrix}
		&\equiv 
	\varepsilon_{\mu, \sigma}
	\begin{pmatrix}
		{}_{1} \phi_{{\bm p}_{\perp}, p_{\eta}, \sigma}^{\rm (const.)} \\
		{}_{2} \phi_{{\bm p}_{\perp}, p_{\eta}, \sigma}^{\rm (const.)} 
	\end{pmatrix}
		\nonumber\\
		&=
	\varepsilon_{\mu, \sigma}  
	\begin{pmatrix}
		{}_{1} \chi_{{\bm p}_{\perp}, p_{\eta}, \sigma}^{\rm (const.)} \\
		{}_{2} \chi_{{\bm p}_{\perp}, p_{\eta}, \sigma}^{\rm (const.)} 
	\end{pmatrix}
	\frac{{\rm e}^{i {\bm p}_{\perp}\cdot {\bm x}_{\perp}} {\rm e}^{ip_{\eta} \eta} }{ (2\pi)^{3/2} },  \label{eqb_47}
\end{align}
where the definition of $\phi^{\rm (const.)}$ (Eq.~(\ref{eqb2})) is used.  With the normalization conditions Eq.~(\ref{eqb36}) for $\varepsilon_{\mu, \sigma}$, and Eqs.~(\ref{eqb_44}) and (\ref{eqb___46}) for ${}_{k} \chi^{\rm (const.)}_{{\bm p}_{\perp}, p_{\eta}, \sigma}$, one finds that the mode functions ${}_{k} W^{({\rm const.})}_{\mu, {\bm p}_{\perp}, p_{\eta}, \sigma} $ are correctly normalized as
\begin{align}
	&-g^{\mu\nu} ( {}_{1} W^{({\rm const.})}_{\mu, {\bm p}_{\perp}, p_{\eta}, \sigma}   |  {}_{1} W^{({\rm const.})}_{\nu, {\bm p}'_{\perp}, p'_{\eta}, \sigma' }   )_{\rm B} \nonumber\\
	&\ \ \ \ \ \ \ \ = +g^{\mu\nu} ( {}_{2} W^{({\rm const.})}_{\mu, {\bm p}_{\perp}, p_{\eta}, \sigma}   |  {}_{2} W^{({\rm const.})}_{\nu, {\bm p}'_{\perp}, p'_{\eta}, \sigma' }   )_{\rm B} \nonumber\\
	&\ \ \ \ \ \ \ \ \ \ \ \ \ = \xi_{\sigma \sigma'} \delta^2({\bm p}_{\perp} - {\bm p}'_{\perp}) \delta (p_{\eta} - p'_{\eta}), \\
	&-g^{\mu\nu} ( {}_{1} W^{({\rm const.})}_{\mu, {\bm p}_{\perp}, p_{\eta}, \sigma}   |  {}_{2} W^{({\rm const.})}_{\nu, {\bm p}'_{\perp}, p'_{\eta}, \sigma' }  )_{\rm B} \nonumber\\
	&\ \ \ \ \ \ \ \ =  -g^{\mu\nu} ( {}_{2} W^{({\rm const.})}_{\mu, {\bm p}_{\perp}, p_{\eta}, \sigma}   |  {}_{1} W^{({\rm const.})}_{\nu, {\bm p}'_{\perp}, p'_{\eta}, \sigma' }   )_{\rm B} =0.  
\end{align}

\subsubsection{Under a spatially homogeneous and constant color electric background field with lifetime $T$ }\label{appC:gluon}

We consider a constant color electric background field with lifetime $T$ (Eq.~(\ref{eqEfield})), and find out all the mode functions $W^{\rm (finite)}_{\mu}$ for the equation of motion Eq.~(\ref{eqb1}).  The problem is equivalent to solving the equation of motion Eq.~(\ref{eqb1}) under a pure gauge background field for $0<\tau<\tau_0$ and $\tau_0+T<\tau$, and under a spatially homogeneous and constant color electric background field for $\tau_0 < \tau < \tau_0+T$.  All the mode functions for respective regions are already derived in Appendix~\ref{appA:gluon} and Appendix~\ref{appB:gluon}, respectively.  Thus, all we have to do is to connect these solutions smoothly at the boundary $\tau = \tau_0$ and $\tau = \tau_0 + T$.  Namely, we require
\begin{align}
	\left. W^{\rm (finite)}_{\mu} \right|_{\tau = \tau_0-0, \tau_0+T-0}  
		&= \left. W^{\rm (finite)}_{\mu}  \right|_{\tau = \tau_0+0, \tau_0+T+0}, \\
	\left. \nabla_{\tau} W^{\rm (finite)}_{\mu} \right|_{\tau = \tau_0-0, \tau_0+T-0}  
		&= \left. \nabla_{\tau} W^{\rm (finite)}_{\mu}  \right|_{\tau = \tau_0+0, \tau_0+T+0}.  
\end{align}
In making this connection, it is useful to use a linear relation between ${}_{\pm} W^{({\rm free})}_{\mu, {\bm p}_{\perp}, p_{\eta}, \sigma}$ and ${}_{k} W^{\rm (const.)}_{\mu, {\bm p}_{\perp}, p_{\eta}, \sigma}$ ($k=1,2$) at fixed time $\tau = \tau_1$ described by
\begin{widetext}
\begin{align}
	&\begin{pmatrix}
		{}_{+}W_{\mu, {\bm p}_{\perp}, p_{\eta}, \sigma}^{\rm (free)} [\tilde{A}_{\rho}(\tau_1)] \\
		{}_{-}W_{\mu, {\bm p}_{\perp}, p_{\eta}, \sigma}^{\rm (free)} [\tilde{A}_{\rho}(\tau_1)]
	\end{pmatrix} \nonumber\\
	&=
	-g^{\nu \lambda} \sum_{\sigma' \sigma''} \xi_{\sigma' \sigma''} \int d^2{\bm p}'_{\perp} dp'_{\eta} \nonumber\\
	&\ \ \ \ \ \times
	\begin{pmatrix} 
		( {}_1 W^{\rm (const.)}_{\nu, {\bm p}'_{\perp}, p_{\eta}', \sigma''} | {}_+ W^{({\rm free})}_{\lambda, {\bm p}_{\perp}, p_{\eta}, \sigma}[\tilde{A}_{\rho}(\tau_1)] )_{\rm B} & -( {}_2 W^{\rm (const.)}_{\nu, {\bm p}'_{\perp}, p'_{\eta}, \sigma''} | {}_+ W^{({\rm free})}_{\lambda, {\bm p}_{\perp}, p_{\eta}, \sigma}[\tilde{A}_{\rho}(\tau_1)] )_{\rm B} \\
		( {}_1 W^{\rm (const.)}_{\nu, {\bm p}'_{\perp}, p'_{\eta}, \sigma''} | {}_- W^{({\rm free})}_{\lambda, {\bm p}_{\perp}, p_{\eta}, \sigma}[\tilde{A}_{\rho}(\tau_1)] )_{\rm B} & -( {}_2 W^{\rm (const.)}_{\nu, {\bm p}'_{\perp}, p'_{\eta}, \sigma''} | {}_- W^{({\rm free})}_{\lambda, {\bm p}_{\perp}, p_{\eta}, \sigma}[\tilde{A}_{\rho}(\tau_1)] )_{\rm B}
	\end{pmatrix}
	\begin{pmatrix}
		{}_{1} W_{\mu, {\bm p}'_{\perp}, p'_{\eta}, \sigma'}^{\rm (const.)} \\
		{}_{2} W_{\mu, {\bm p}'_{\perp}, p'_{\eta}, \sigma'}^{\rm (const.)}
	\end{pmatrix} \nonumber\\
	&\equiv 
	U^{\rm (g)}_{{\bm p}_{\perp}, p_{\eta}, \sigma}(\tau_1)
	\begin{pmatrix}
		{}_{1} W_{{\bm p}_{\perp}, p_{\eta}-qE\tau_1^2/2, \sigma}^{\rm (const.)} \\
		{}_{2} W_{{\bm p}_{\perp}, p_{\eta}-qE\tau_1^2/2, \sigma}^{\rm (const.)}
	\end{pmatrix} .		\label{eqb__46}
\end{align}
The matrix elements are given by
\begin{align}
	A^{\rm (g)}_{{\bm p}_{\perp}, p_{\eta}, \sigma}(\tau_1) 
		&\equiv ( U^{\rm (g)}_{{\bm p}_{\perp}, p_{\eta}, \sigma}(\tau_1) )_{11} = \sum_{\sigma'} \xi_{\sigma \sigma'} \left[ ( U^{\rm (g)}_{{\bm p}_{\perp}, p_{\eta}, \sigma'}(\tau_1) )_{22} \right]^* = \sum_{\sigma'} \xi_{\sigma \sigma'} \left[  i \tau_1  \left. \left(  {}_1 \chi^{{\rm (const.)}*}_{ {\bm p}_{\perp}, p_{\eta}-qE\tau_1^2/2, \sigma' } \overset{\leftrightarrow}{\partial_{\tau}} {}_1 \chi^{{\rm (free)}}_{ {\bm p}_{\perp}, p_{\eta}, \sigma } \right) \right|_{\tau = \tau_1} \right], \\
		B^{\rm (g)}_{{\bm p}_{\perp}, p_{\eta}, \sigma}(\tau_1) 
		&\equiv ( U^{\rm (g)}_{{\bm p}_{\perp}, p_{\eta}, \sigma}(\tau_1) )_{21} = \sum_{\sigma'} \xi_{\sigma \sigma'}\left[ ( U^{\rm (g)}_{{\bm p}_{\perp}, p_{\eta}, \sigma'}(\tau_1) )_{12} \right]^* = \sum_{\sigma'} \xi_{\sigma \sigma'} \left[  i \tau_1  \left. \left(  {}_1 \chi^{{\rm (const.)}*}_{ {\bm p}_{\perp}, p_{\eta}-qE\tau_1^2/2, \sigma' } \overset{\leftrightarrow}{\partial_{\tau}} {}_2 \chi^{{\rm (free)}}_{ {\bm p}_{\perp}, p_{\eta}, \sigma } \right) \right|_{\tau = \tau_1} \right] , 
\end{align} 
and are normalized as
\begin{align}
	1 = \sum_{\sigma'} \xi_{\sigma \sigma'} \left[  A^{\rm (g)}_{{\bm p}_{\perp}, p_{\eta}, \sigma} [A^{\rm (g)}_{{\bm p}_{\perp}, p_{\eta}, \sigma'}]^* - B^{\rm (g)}_{{\bm p}_{\perp}, p_{\eta}, \sigma} [ B^{\rm (g)}_{{\bm p}_{\perp}, p_{\eta}, \sigma'}]^* \right]
\end{align}
so that $\det U^{\rm (g)}_{{\bm p}_{\perp}, p_{\eta}, \sigma} (\tau_1) =1 $ holds.  Although the mode functions diverge at $\tau \rightarrow 0$ because of the coordinate singularity at $\tau=0$ of the $\tau$-$\eta$ coordinates, one can safely take the limit $\tau_1 \rightarrow 0$ of the transformation $U^{\rm (g)}$, i.e., the coefficients $A^{\rm (g)}, B^{\rm (g)}$.  By using the asymptotic formulas for the special functions Eqs.~(\ref{eqa49})-(\ref{eqa51}), one finds 
\begin{align}
	A^{\rm (g)}_{{\bm p}_{\perp}, p_{\eta}, \sigma}(\tau) 
		&\xrightarrow[\tau \rightarrow 0]{} \sum_{\sigma'} \xi_{\sigma \sigma'}
			\left\{ -\sqrt{\frac{\pi}{2}} \frac{ \left( \frac{2|qE|}{{\bm p}_{\perp}^2} \right)^{-ip_{\eta}/2} \exp \left[ -\frac{\pi}{2} \left( \frac{{\bm p}^2_{\perp}}{2|qE|} - i \frac{q\Lambda_{\sigma'}}{qE} \right)  \right] }{ \sinh(\pi p_{\eta}) \Gamma\left( \frac{1}{2} - i \frac{{\bm p}_{\perp}^2}{2|qE|} - \frac{q\Lambda_{\sigma'}}{qE} \right) } {\rm e}^{-\pi p_{\eta}} \right.\nonumber\\
			&\ \ \ \ \ \ \ \ \ \ \ \ \ \ \ \ \ \ \ \ \ \left. \times \left[ 1 - \left( \frac{2|qE|}{{\bm p}_{\perp}^2} \right)^{i p_{\eta}} {\rm e}^{3\pi p_{\eta}/2} \frac{ \Gamma \left( \frac{1}{2} - i \frac{{\bm p}_{\perp}^2}{2|qE|} - \frac{q\Lambda_{\sigma'}}{qE} \right) }{ \Gamma \left( \frac{1}{2} - i \frac{{\bm p}_{\perp}^2}{2|qE|} -ip_{\eta} - \frac{q\Lambda_{\sigma'}}{qE} \right)}   \right] \right\}	, \\
			B^{\rm (g)}_{{\bm p}_{\perp}, p_{\eta}, \sigma}(\tau) 
		&\xrightarrow[\tau \rightarrow 0]{} \sum_{\sigma'} \xi_{\sigma \sigma'}
			\left\{ -\sqrt{\frac{\pi}{2}} \frac{ \left( \frac{2|qE|}{{\bm p}_{\perp}^2} \right)^{-ip_{\eta}/2} \exp \left[ -\frac{\pi}{2} \left( \frac{{\bm p}^2_{\perp}}{2|qE|} - i \frac{q\Lambda_{\sigma'}}{qE} \right)  \right] }{ \sinh(\pi p_{\eta}) \Gamma\left( \frac{1}{2} - i \frac{{\bm p}_{\perp}^2}{2|qE|} - \frac{q\Lambda_{\sigma'}}{qE} \right) } \right.\nonumber\\
			&\ \ \ \ \ \ \ \ \ \ \ \ \ \ \ \ \ \ \ \ \ \left. \times \left[ 1 - \left( \frac{2|qE|}{{\bm p}_{\perp}^2} \right)^{i p_{\eta}} {\rm e}^{-\pi p_{\eta}/2} \frac{ \Gamma \left( \frac{1}{2} - i \frac{{\bm p}_{\perp}^2}{2|qE|} - \frac{q\Lambda_{\sigma'}}{qE} \right) }{ \Gamma \left( \frac{1}{2} - i \frac{{\bm p}_{\perp}^2}{2|qE|} -ip_{\eta} - \frac{q\Lambda_{\sigma'}}{qE} \right)}   \right] \right\}.  
\end{align}

We consider two kinds of boundary conditions for the mode functions: We define mode functions ${}_{\pm} W^{({\rm finite; in})}_{\mu, {\bm p}_{\perp}, p_{\eta}, \sigma}$ (${}_{\pm} W^{({\rm finite; out})}_{\mu, {\bm p}_{\perp}, p_{\eta}, \sigma}$) by a boundary condition at $\tau < \tau_0$ ($\tau > \tau_0+T$) to coincide with the plane wave solutions ${}_{\pm} W^{({\rm free})}_{\mu, {\bm p}_{\perp}, p_{\eta}, \sigma} [\tilde{A}_{\nu}]$.  With the linear relation, Eq.~(\ref{eqb__46}), one can construct such mode functions ${}_{\pm} W^{({\rm finite; in})}_{\mu, {\bm p}_{\perp}, p_{\eta}, \sigma}$ and ${}_{\pm} W^{({\rm finite; out})}_{\mu, {\bm p}_{\perp}, p_{\eta}, \sigma}$ as 
\begin{align}
	\begin{pmatrix}
		{}_{+} W^{({\rm finite; in})}_{\mu, {\bm p}_{\perp}, p_{\eta}, \sigma} \\
		{}_{-} W^{({\rm finite; in})}_{\mu, {\bm p}_{\perp}, p_{\eta}, \sigma} 
	\end{pmatrix}
	= 
	\left\{
		\begin{array}{ll}
			\begin{pmatrix}
				{}_{+} W^{({\rm free})}_{\mu, {\bm p}_{\perp}, p_{\eta}, \sigma} [\tilde{A}_{\nu}] \\
				{}_{-} W^{({\rm free})}_{\mu, {\bm p}_{\perp}, p_{\eta}, \sigma} [\tilde{A}_{\nu}] 
			\end{pmatrix}	& {\rm for}\ 0<\tau<\tau_0 \\
			\\
			U^{\rm (g)}_{{\bm p}_{\perp}, p_{\eta}, \sigma}(\tau_0)
			\begin{pmatrix}
				{}_{1} W^{({\rm const.})}_{\mu, {\bm p}_{\perp}, p_{\eta}-qE\tau_0^2/2, \sigma}  \\
				{}_{2} W^{({\rm const.})}_{\mu, {\bm p}_{\perp}, p_{\eta}-qE\tau_0^2/2, \sigma} 
			\end{pmatrix}	& {\rm for}\ \tau_0<\tau<\tau_0+T \\
			\\
			U^{\rm (g)}_{{\bm p}_{\perp}, p_{\eta}, \sigma}(\tau_0)
			U^{{\rm (g)}-1}_{{\bm p}_{\perp}, p_{\eta}-qE\tau_0^2/2+qE(\tau_0+T)^2/2, \sigma}(\tau_0+T) & \\
			\ \ \ \ \ \ \ \ \ \ \ \ \ \ \ \ \ \ \ \ \times
			\begin{pmatrix}
				{}_{+} W^{({\rm free})}_{\mu, {\bm p}_{\perp}, p_{\eta}-qE\tau_0^2/2+qE(\tau_0+T)^2/2, \sigma} [\tilde{A}_{\nu}] \\
				{}_{-} W^{({\rm free})}_{\mu, {\bm p}_{\perp}, p_{\eta}-qE\tau_0^2/2+qE(\tau_0+T)^2/2, \sigma} [\tilde{A}_{\nu}]
			\end{pmatrix}	& {\rm for}\ \tau_0+T<\tau \\
		\end{array}
	\right. , 
\end{align}
\begin{align}
	\begin{pmatrix}
		{}_{+} W^{({\rm finite; out})}_{\mu, {\bm p}_{\perp}, p_{\eta}, \sigma} \\
		{}_{-} W^{({\rm finite; out})}_{\mu, {\bm p}_{\perp}, p_{\eta}, \sigma} 
	\end{pmatrix}
	= 
	\left\{
		\begin{array}{ll}
			U^{{\rm (g)}}_{{\bm p}_{\perp}, p_{\eta}, \sigma}(\tau_0+T)
			U^{{\rm (g)}-1}_{{\bm p}_{\perp}, p_{\eta}+qE\tau_0^2/2-qE(\tau_0+T)^2/2, \sigma}(\tau_0) & \\
			\ \ \ \ \ \ \ \ \ \ \ \ \ \ \ \ \ \ \ \ \times
			\begin{pmatrix}
				{}_{+} W^{({\rm free})}_{\mu, {\bm p}_{\perp}, p_{\eta}+qE\tau_0^2/2-qE(\tau_0+T)^2/2, \sigma} [\tilde{A}_{\nu}] \\
				{}_{-} W^{({\rm free})}_{\mu, {\bm p}_{\perp}, p_{\eta}+qE\tau_0^2/2-qE(\tau_0+T)^2/2, \sigma} [\tilde{A}_{\nu}] 
			\end{pmatrix}	& {\rm for}\ 0<\tau<\tau_0 \\
			\\
			U^{{\rm (g)}}_{{\bm p}_{\perp}, p_{\eta}, \sigma}(\tau_0+T)
			\begin{pmatrix}
				{}_{1} W^{({\rm const.})}_{\mu, {\bm p}_{\perp}, p_{\eta}-qE(\tau_0+T)^2/2, \sigma}  \\
				{}_{2} W^{({\rm const.})}_{\mu, {\bm p}_{\perp}, p_{\eta}-qE(\tau_0+T)^2/2, \sigma} 
			\end{pmatrix}	& {\rm for}\ \tau_0<\tau<\tau_0+T \\
			\\
			\begin{pmatrix}
				{}_{+} W^{({\rm free})}_{\mu, {\bm p}_{\perp}, p_{\eta}, \sigma} [\tilde{A}_{\nu}] \\
				{}_{-} W^{({\rm free})}_{\mu, {\bm p}_{\perp}, p_{\eta}, \sigma} [\tilde{A}_{\nu}]
			\end{pmatrix}	& {\rm for}\ \tau_0+T<\tau \\
		\end{array}
	\right. .
\end{align}

These two sets of mode functions are not independent but related with each other by a Bogoliubov transformation discussed in the main text (see Eq.~(\ref{eq_85})).  Now, one can analytically compute the Bogoliubov coefficients as
\begin{align}
	& \sum_{\sigma''} \xi_{\sigma \sigma''} (-g^{\mu\nu}) ( {}_+ W^{\rm (finite; out)}_{\mu, {\bm p}_{\perp}, p_{\eta}, \sigma''} | {}_+ W^{\rm (finite; in)}_{\nu, {\bm p}'_{\perp}, p'_{\eta}, \sigma'} )_{\rm B} 
	= \sum_{\sigma'''} \xi_{\sigma \sigma'''} \left[ -\sum_{\sigma''} \xi_{\sigma''' \sigma''}^{-1} (-g^{\mu\nu})   ( {}_- W^{\rm (finite; out)}_{\mu, {\bm p}_{\perp}, p_{\eta}, \sigma''} | {}_- W^{\rm (finite; in)}_{\nu, {\bm p}'_{\perp}, p'_{\eta}, \sigma'} )_{\rm B} \right]^* \nonumber\\
		&= \delta_{\sigma \sigma'} \delta^2({{\bm p}_{\perp} - {\bm p}'_{\perp}}) \delta ( p'_{\eta} - (p_{\eta} + qE\tau_0^2/2 - qE (\tau_0+T)^2/2)  )  \nonumber\\
		&\ \ \ \times \sum_{\sigma''} \xi_{\sigma \sigma''} \left[ A^{{\rm (g)}}_{{\bm p}_{\perp}, p_{\eta}+qE\tau_0^2/2-qE(\tau_0+T)^2/2, \sigma}(\tau_0) A^{{\rm (g)}*}_{{\bm p}_{\perp}, p_{\eta}, \sigma''}(\tau_0+T) - B^{{\rm (g)}*}_{{\bm p}_{\perp}, p_{\eta}+qE\tau_0^2/2-qE(\tau_0+T)^2/2, \sigma''}(\tau_0) B^{{\rm (g)}}_{{\bm p}_{\perp}, p_{\eta}, \sigma}(\tau_0+T) \right] , \\
	& - \sum_{\sigma''} \xi_{\sigma \sigma''} (-g^{\mu\nu}) ({}_- W^{\rm (finite; out)}_{\mu, {\bm p}_{\perp}, p_{\eta}, \sigma''} | {}_+ W^{\rm (finite; in)}_{\nu, {\bm p}'_{\perp}, p'_{\eta}, \sigma'} )_{\rm B} = \sum_{\sigma'''} \xi_{\sigma \sigma'''}  \left[ \sum_{\sigma''} \xi_{\sigma''' \sigma''} (-g^{\mu\nu})   ( {}_+ W^{\rm (finite; out)}_{\mu, {\bm p}_{\perp}, p_{\eta}, \sigma''} | {}_- W^{\rm (finite; in)}_{\nu, {\bm p}'_{\perp}, p'_{\eta}, \sigma'} )_{\rm B} \right]^* \nonumber\\
		&= \delta_{\sigma \sigma'} \delta^2({{\bm p}_{\perp} - {\bm p}'_{\perp}}) \delta ( p'_{\eta} - (p_{\eta} + qE\tau_0^2/2 - qE (\tau_0+T)^2/2)  ) \nonumber\\
		&\ \ \ \times \sum_{\sigma''} \xi_{\sigma \sigma''} \left[ - A^{\rm (g)}_{{\bm p}_{\perp}, p_{\eta}+qE\tau_0^2/2-qE(\tau_0+T)^2/2, \sigma}(\tau_0) B^{{\rm (g)}*}_{{\bm p}_{\perp}, p_{\eta}, \sigma''}(\tau_0+T)  + B^{{\rm (g)}*}_{{\bm p}_{\perp}, p_{\eta}+qE\tau_0^2/2-qE(\tau_0+T)^2/2, \sigma''}(\tau_0) A^{{\rm (g)}}_{{\bm p}_{\perp}, p_{\eta}, \sigma}(\tau_0+T) \right] .
\end{align}
\end{widetext}

\subsection{Ghost} 

We consider the Abelianized equation of motion for ghost and anti-ghost fields, $C$ and $\bar{C}$, in the $\tau$-$\eta$ coordinates (see Eq.~(\ref{eq50})).  It is sufficient for this purpose to consider a differential equation of a type: 
\begin{align}
	0 	&= ( \nabla_{\mu} + iq \tilde{A}_{\mu} )^2 \Theta \nonumber\\
		&= \left[ ( \partial_{\mu} + iq \tilde{A}_{\mu} )^2 + \frac{\partial_{\tau} + iq\tilde{A}_{\tau} }{ \tau }   \right] \Theta.  \label{eqc__1}
\end{align}
This equation is exactly the same as Eq.~(\ref{eqb38}) for the gluon fields $\phi_{\sigma}$ for $\Lambda_{\sigma} = 0$ and so one can solve Eq.~(\ref{eqc__1}) in the same way as what we did in Appendix~\ref{appA:gluon}.   Therefore, we just write down the results without repeating the derivation and/or discussions in the following.

\subsubsection{Under a pure gauge background field (plane wave solutions)} \label{appA:ghost}

Under a pure gauge background field $\tilde{A}_{\mu}$ given by Eq.~(\ref{eqpure}), the positive and the negative frequency mode functions ${}_{\pm} \Theta^{({\rm free})}_{{\bm p}_{\perp}, p_{\eta}} [\tilde{A}_{\mu}]$ are given by
\begin{align}
	{}_{+} \Theta^{({\rm free})}_{{\bm p}_{\perp}, p_{\eta}}[\tilde{A}_{\mu}] 
		&= \Omega {}_{1} \chi^{({\rm free})}_{{\bm p}_{\perp}, p_{\eta}} \frac{ {\rm e}^{i{\bm p}_{\perp} \cdot {\bm x}_{\perp} } {\rm e}^{i p_{\eta} \eta} }{ (2\pi)^{3/2} }   ,\nonumber\\ 
	{}_{-} \Theta^{({\rm free})}_{{\bm p}_{\perp}, p_{\eta}}[\tilde{A}_{\mu}] 
		&= \Omega {}_{2} \chi^{({\rm free})}_{{\bm p}_{\perp}, p_{\eta}} \frac{ {\rm e}^{i{\bm p}_{\perp} \cdot {\bm x}_{\perp} } {\rm e}^{i p_{\eta} \eta} }{ (2\pi)^{3/2} },   \label{eqc57}
\end{align}
where 
\begin{align}
	{}_{1} \chi^{({\rm free})}_{{\bm p}_{\perp}, p_{\eta}} = \frac{\sqrt{\pi}}{2i} {\rm e}^{\pi p_{\eta}/2} H^{(2)}_{ip_{\eta}} (|{\bm p}_{\perp}| \tau),\ 
	{}_{2} \chi^{({\rm free})}_{{\bm p}_{\perp}, p_{\eta}} = [{}_{-} \chi^{({\rm free})}_{{\bm p}_{\perp}, p_{\eta}}]^* .
\end{align}
The mode functions satisfy the correct normalization conditions for scalar fields in the $\tau$-$\eta$ coordinates (see also Eqs.~(\ref{eq_96}) and (\ref{eq_97}) in the main text).  For temporal gauge $\tilde{A}_{\tau} = 0$, it reads 
\begin{align}
	( {}_{\pm} \Theta_{{\bm p}_{\perp}, p_{\eta} }^{({\rm free})}[\tilde{A}_{\mu}] |  {}_{\pm} \Theta_{{\bm p}'_{\perp}, p'_{\eta} }^{({\rm free})}[\tilde{A}_{\mu}] )_{\rm B} &= \pm \delta^2({\bm p}_{\perp} - {\bm p}'_{\perp}) \delta (p_{\eta} - p'_{\eta}),  \\
	( {}_{\pm} \Theta_{ {\bm p}_{\perp}, p_{\eta} }^{({\rm free})}[\tilde{A}_{\mu}]  |  {}_{\mp} \Theta_{ {\bm p}'_{\perp}, p'_{\eta} }^{({\rm free})} [\tilde{A}_{\mu}] )_{\rm B} &= 0.  
\end{align}

To see the mode functions ${}_{\pm} \Theta^{({\rm free})}_{{\bm p}_{\perp}, p_{\eta}} [\tilde{A}_{\mu}]$ defined in Eq.~(\ref{eqc57}) are actually the positive/negative frequency mode functions in the $\tau$-$\eta$ coordinates, we again use the integral representation for the Hankel functions $H^{(n)}_{\nu} (z)$ ($n=1,2$) (Eq.~(\ref{eqa_26})) to get the same integral relation as that for quarks Eq.~(\ref{eqa_27}) and for gluons Eq.~(\ref{eqb80}) as
\begin{align}
	{}_{\pm} \Theta^{({\rm free})}_{ {\bm p}_{\perp}, p_{\eta}} [\tilde{A}_{\mu}] = \int dp_z \frac{ {\rm e}^{\pm i p_{\eta} y_{\bm p}} }{ \sqrt{2\pi \omega_{\bm p}}} {}_{\pm} \Theta^{({\rm free})}_{ {\bm p}_{\perp}, p_{z}} [\tilde{A}_{m}] .  
\end{align}
Every notations are the same as in the previous two cases except ${}_{\pm} \Theta^{({\rm free})}_{ {\bm p}_{\perp}, p_{z}} [\tilde{A}_{m}]$ being the positive/negative frequency mode functions in the Cartesian coordinates:
\begin{align}
		{}_{\pm} \Theta^{({\rm free})}_{ {\bm p}_{\perp}, p_{z}} [ \tilde{A}_{m} ] 
		= 
	\frac{ {\rm e}^{\mp i\omega_{\bm p} t} }{\sqrt{2\omega_{\bm p}} }
	\frac{{\rm e}^{i {\bm p}_{\perp}\cdot {\bm x}_{\perp}} {\rm e}^{ip_{z} z} }{ (2\pi)^{3/2} }, 
\end{align}
which are properly normalized in the Cartesian coordinates with the temporal gauge condition $\tilde{A}_{t} = 0$ as 
\begin{align}
	&\int_{t={\rm const.}} d^2{\bm x}_{\perp} dz\; {}_{\pm} \Theta^{({\rm free})*}_{ {\bm p}_{\perp}, p_{z} } [ \tilde{A}_{m} ] \overset{\leftrightarrow}{\partial}_t {}_{\pm} \Theta^{({\rm free})}_{ {\bm p}'_{\perp}, p'_{z} } [ \tilde{A}_{m} ]  \nonumber\\
		&\ \ \ \ \ \ \ \ \ \ \ \ \ \ \ \ \ \ \ \ \ \ \ \ \ \ \ \ \ \ \ \ \ \ \  = \pm  \delta^2({\bm p}_{\perp} - {\bm p}'_{\perp}) \delta (p_{z} - p'_z), \\
	&\int_{t={\rm const.}} d^2{\bm x}_{\perp} dz\;  {}_{\pm} \Theta^{({\rm free})*}_{ {\bm p}_{\perp}, p_{z} } [ \tilde{A}_{m} ] \overset{\leftrightarrow}{\partial}_t {}_{\mp} \Theta^{({\rm free})}_{ {\bm p}'_{\perp}, p'_{z} } [ \tilde{A}_{m} ] 
		= 0. 
\end{align}

\subsubsection{Under a spatially homogeneous and constant color electric background field} \label{appB:ghost}

Under a spatially homogeneous and constant color electric field (Eq.~(\ref{eqconstE})), the positive/negative frequency mode functions ${}_{\pm} \Theta^{({\rm const.})}_{{\bm p}_{\perp}, p_{\eta}} $ are given by
\begin{align}
	{}_{1} \Theta^{({\rm const.})}_{{\bm p}_{\perp}, p_{\eta}} 
		&= {}_{1} \chi^{({\rm const.})}_{{\bm p}_{\perp}, p_{\eta}} \frac{ {\rm e}^{i{\bm p}_{\perp} \cdot {\bm x}_{\perp} } {\rm e}^{i p_{\eta} \eta} }{ (2\pi)^{3/2} }   ,\nonumber\\
	{}_{2} \Theta^{({\rm const.})}_{{\bm p}_{\perp}, p_{\eta}} 
		&= {}_{2} \chi^{({\rm const.})}_{{\bm p}_{\perp}, p_{\eta}} \frac{ {\rm e}^{i{\bm p}_{\perp} \cdot {\bm x}_{\perp} } {\rm e}^{i p_{\eta} \eta} }{ (2\pi)^{3/2} },   \label{eqc_57}
\end{align}
where 
\begin{align}
	{}_{1} \chi^{({\rm const.})}_{{\bm p}_{\perp}, p_{\eta}} 
			&= \frac{1}{\sqrt{2}}   \exp \left[ - \frac{\pi}{2} \left( \frac{{\bm p}_{\perp}^2}{2|qE|} + p_{\eta}   \right)  - i \frac{|qE|\tau^2}{4} \right] \nonumber\\
			&\ \ \ \times \left( \frac{|qE|\tau^2}{2} \right)^{i p_{\eta}/2 } \nonumber\\
			&\ \ \ \times U \left( \frac{1}{2} + i\frac{{\bm p}_{\perp}^2 }{ 2|qE| } + ip_{\eta}  ;  1 + ip_{\eta} ; i \frac{|qE| \tau^2}{2}  \right), \\ 
	{}_{2} \chi^{({\rm const.})}_{{\bm p}_{\perp}, p_{\eta}} 
			&= [{}_{-} \chi^{({\rm const.})}_{{\bm p}_{\perp}, p_{\eta}}]^* .
\end{align}
The mode functions are correctly normalized as
\begin{align}
	&\delta^2({\bm p}_{\perp} - {\bm p}'_{\perp}) \delta (p_{\eta} - p'_{\eta})	\nonumber\\
		& \ \ = ( {}_{1} \Theta_{{\bm p}_{\perp}, p_{\eta} }^{({\rm const.})} |  {}_{1} \Theta_{{\bm p}'_{\perp}, p'_{\eta} }^{({\rm const.})} )_{\rm B} = -( {}_{2} \Theta_{{\bm p}_{\perp}, p_{\eta} }^{({\rm const.})} |  {}_{2} \Theta_{{\bm p}'_{\perp}, p'_{\eta} }^{({\rm const.})} )_{\rm B}, \\
	&0 = ( {}_{1} \Theta_{{\bm p}_{\perp}, p_{\eta} }^{({\rm const.})} |  {}_{2} \Theta_{{\bm p}'_{\perp}, p'_{\eta} }^{({\rm const.})} )_{\rm B} = ( {}_{2} \Theta_{{\bm p}_{\perp}, p_{\eta} }^{({\rm const.})} |  {}_{1} \Theta_{{\bm p}'_{\perp}, p'_{\eta} }^{({\rm const.})} )_{\rm B}.   
\end{align}

\subsubsection{Under a spatially homogeneous and constant color electric background field with lifetime $T$ }\label{appC:ghost}

For a spatially homogeneous and constant color electric background field with lifetime $T$ (Eq.~(\ref{eqEfield})), the mode functions ${}_{\pm} \Theta^{\rm (finite; as)}_{{\bm p}_{\perp}, p_{\eta}}$ (${\rm as}={\rm in, out}$) are given by 
\begin{widetext}
\begin{align}
	\begin{pmatrix}
		{}_{+} \Theta^{({\rm finite; in})}_{{\bm p}_{\perp}, p_{\eta}} \\
		{}_{-} \Theta^{({\rm finite; in})}_{{\bm p}_{\perp}, p_{\eta}} 
	\end{pmatrix}
	= 
	\left\{
		\begin{array}{ll}
			\begin{pmatrix}
				{}_{+} \Theta^{({\rm free})}_{{\bm p}_{\perp}, p_{\eta}} [\tilde{A}_{\mu}] \\
				{}_{-} \Theta^{({\rm free})}_{{\bm p}_{\perp}, p_{\eta} } [\tilde{A}_{\mu}] 
			\end{pmatrix}	& {\rm for}\ 0<\tau<\tau_0 \\
			\\
			U^{\rm (gh)}_{{\bm p}_{\perp}, p_{\eta} }(\tau_0)
			\begin{pmatrix}
				{}_{1} \Theta^{({\rm const.})}_{ {\bm p}_{\perp}, p_{\eta}-qE\tau_0^2/2}  \\
				{}_{2} \Theta^{({\rm const.})}_{{\bm p}_{\perp}, p_{\eta}-qE\tau_0^2/2} 
			\end{pmatrix}	& {\rm for}\ \tau_0<\tau<\tau_0+T \\
			\\
			U^{\rm (gh)}_{{\bm p}_{\perp}, p_{\eta}}(\tau_0)
			U^{{\rm (gh)}-1}_{{\bm p}_{\perp}, p_{\eta}-qE\tau_0^2/2+qE(\tau_0+T)^2/2}(\tau_0+T) & \\
			\ \ \ \ \ \ \ \ \ \ \ \ \ \ \ \ \ \ \ \ \times
			\begin{pmatrix}
				{}_{+} \Theta^{({\rm free})}_{{\bm p}_{\perp}, p_{\eta}-qE\tau_0^2/2+qE(\tau_0+T)^2/2} [\tilde{A}_{\mu}] \\
				{}_{-} \Theta^{({\rm free})}_{ {\bm p}_{\perp}, p_{\eta}-qE\tau_0^2/2+qE(\tau_0+T)^2/2} [\tilde{A}_{\mu}]
			\end{pmatrix}	& {\rm for}\ \tau_0+T<\tau \\
		\end{array}
	\right. ,
\end{align}
\begin{align}
	\begin{pmatrix}
		{}_{+} \Theta^{({\rm finite; out})}_{ {\bm p}_{\perp}, p_{\eta} } \\
		{}_{-} \Theta^{({\rm finite; out})}_{ {\bm p}_{\perp}, p_{\eta} } 
	\end{pmatrix}
	= 
	\left\{
		\begin{array}{ll}
			U^{{\rm (gh)}}_{{\bm p}_{\perp}, p_{\eta} }(\tau_0+T)
			U^{{\rm (gh)}-1}_{{\bm p}_{\perp}, p_{\eta}+qE\tau_0^2/2-qE(\tau_0+T)^2/2 }(\tau_0) & \\
			\ \ \ \ \ \ \ \ \ \ \ \ \ \ \ \ \ \ \ \ \times
			\begin{pmatrix}
				{}_{+} \Theta^{({\rm free})}_{ {\bm p}_{\perp}, p_{\eta}+qE\tau_0^2/2-qE(\tau_0+T)^2/2 } [\tilde{A}_{\mu}] \\
				{}_{-} \Theta^{({\rm free})}_{ {\bm p}_{\perp}, p_{\eta}+qE\tau_0^2/2-qE(\tau_0+T)^2/2 } [\tilde{A}_{\mu}] 
			\end{pmatrix}	& {\rm for}\ 0<\tau<\tau_0 \\
			\\
			U^{{\rm (gh)}}_{{\bm p}_{\perp}, p_{\eta} }(\tau_0+T)
			\begin{pmatrix}
				{}_{1} \Theta^{({\rm const.})}_{{\bm p}_{\perp}, p_{\eta}-qE(\tau_0+T)^2/2}  \\
				{}_{2} \Theta^{({\rm const.})}_{ {\bm p}_{\perp}, p_{\eta}-qE(\tau_0+T)^2/2 } 
			\end{pmatrix}	& {\rm for}\ \tau_0<\tau<\tau_0+T \\
			\\
			\begin{pmatrix}
				{}_{+} \Theta^{({\rm free})}_{{\bm p}_{\perp}, p_{\eta}} [\tilde{A}_{\mu}] \\
				{}_{-} \Theta^{({\rm free})}_{ {\bm p}_{\perp}, p_{\eta}} [\tilde{A}_{\mu}]
			\end{pmatrix}	& {\rm for}\ \tau_0+T<\tau \\
		\end{array}
	\right. .  
\end{align}
Here, the matrix $U^{(\rm gh)}$ is given by
\begin{align}
	U^{\rm (gh)}_{{\bm p}_{\perp}, p_{\eta} } 
		= 
		\begin{pmatrix}
			A^{\rm (gh)}_{{\bm p}_{\perp}, p_{\eta} } & B^{{\rm (gh)}*}_{{\bm p}_{\perp}, p_{\eta} } \\
			B^{\rm (gh)}_{{\bm p}_{\perp}, p_{\eta} } & A^{{\rm (gh)}*}_{{\bm p}_{\perp}, p_{\eta} }
		\end{pmatrix}, 
\end{align}
where the matrix elements $A^{\rm (gh)}, B^{\rm (gh)}$ are 
\begin{align}
		A^{\rm (gh)}_{{\bm p}_{\perp}, p_{\eta} } (\tau_1)
			&=i \tau_1  \left. \left(  {}_1 \chi^{{\rm (const.)}*}_{ {\bm p}_{\perp}, p_{\eta}-qE\tau_1^2/2 } \overset{\leftrightarrow}{\partial_{\tau}} {}_1 \chi^{{\rm (free)}}_{ {\bm p}_{\perp}, p_{\eta} } \right) \right|_{\tau = \tau_1},\  
		B^{\rm (gh)}_{{\bm p}_{\perp}, p_{\eta} } (\tau_1)
			&=i \tau_1  \left. \left(  {}_1 \chi^{{\rm (const.)}*}_{ {\bm p}_{\perp}, p_{\eta}-qE\tau_1^2/2 } \overset{\leftrightarrow}{\partial_{\tau}} {}_2 \chi^{{\rm (free)}}_{ {\bm p}_{\perp}, p_{\eta} } \right) \right|_{\tau = \tau_1} .  
\end{align}
The normalization condition for $A^{\rm (gh)}, B^{\rm (gh)}$ is 
\begin{align}
	1 = | A^{\rm (gh)}_{{\bm p}_{\perp}, p_{\eta} } |^2 + | B^{\rm (gh)}_{{\bm p}_{\perp}, p_{\eta} } |^2
\end{align}
so that $\det U^{\rm (gh)}_{{\bm p}_{\perp}, p_{\eta} } = 1 $ holds.  In the limit of $\tau \rightarrow 0$, $A^{\rm (gh)}, B^{\rm (gh)}$ behaves as 
\begin{align}
	A^{\rm (gh)}_{{\bm p}_{\perp}, p_{\eta}}(\tau) 
		&\xrightarrow[\tau \rightarrow 0]{} 
			 -\sqrt{\frac{\pi}{2}} \frac{ \left( \frac{2|qE|}{{\bm p}_{\perp}^2} \right)^{-ip_{\eta}/2} \exp \left[ -\frac{\pi}{2} \left( \frac{{\bm p}^2_{\perp}}{2|qE|}  \right)  \right] }{ \sinh(\pi p_{\eta}) \Gamma\left( \frac{1}{2} - i \frac{{\bm p}_{\perp}^2}{2|qE|}  \right) } {\rm e}^{-\pi p_{\eta}}  \left[ 1 - \left( \frac{2|qE|}{{\bm p}_{\perp}^2} \right)^{i p_{\eta}} {\rm e}^{3\pi p_{\eta}/2} \frac{ \Gamma \left( \frac{1}{2} - i \frac{{\bm p}_{\perp}^2}{2|qE|}  \right) }{ \Gamma \left( \frac{1}{2} - i \frac{{\bm p}_{\perp}^2}{2|qE|} -ip_{\eta}  \right)}   \right] 	, \\
			B^{\rm (gh)}_{{\bm p}_{\perp}, p_{\eta} }(\tau) 
		&\xrightarrow[\tau \rightarrow 0]{} 
			 -\sqrt{\frac{\pi}{2}} \frac{ \left( \frac{2|qE|}{{\bm p}_{\perp}^2} \right)^{-ip_{\eta}/2} \exp \left[ -\frac{\pi}{2} \left( \frac{{\bm p}^2_{\perp}}{2|qE|} \right)  \right] }{ \sinh(\pi p_{\eta}) \Gamma\left( \frac{1}{2} - i \frac{{\bm p}_{\perp}^2}{2|qE|}  \right) }  \left[ 1 - \left( \frac{2|qE|}{{\bm p}_{\perp}^2} \right)^{i p_{\eta}} {\rm e}^{-\pi p_{\eta}/2} \frac{ \Gamma \left( \frac{1}{2} - i \frac{{\bm p}_{\perp}^2}{2|qE|}  \right) }{ \Gamma \left( \frac{1}{2} - i \frac{{\bm p}_{\perp}^2}{2|qE|} -ip_{\eta} - \frac{q\Lambda_{\sigma'}}{qE} \right)}   \right] .  
\end{align}

The Bogoliubov coefficients between the two sets of mode functions (see Eq.~(\ref{eq_106})) are given by
\begin{align}
	&( {}_+ \Theta^{\rm (finite; out)}_{ {\bm p}_{\perp}, p_{\eta} } | {}_+ \Theta^{\rm (finite; in)}_{ {\bm p}'_{\perp}, p'_{\eta} } )_{\rm B} =  \left[ -  ( {}_- \Theta^{\rm (finite; out)}_{ {\bm p}_{\perp}, p_{\eta} } | {}_- \Theta^{\rm (finite; in)}_{ {\bm p}'_{\perp}, p'_{\eta} } )_{\rm B} \right]^* \nonumber\\
		&=  \delta^2({{\bm p}_{\perp} - {\bm p}'_{\perp}}) \delta ( p'_{\eta} - (p_{\eta} + qE\tau_0^2/2 - qE (\tau_0+T)^2/2)  )  \nonumber\\
		&\ \ \ \ \times \left[ A^{{\rm (gh)}}_{{\bm p}_{\perp}, p_{\eta}+qE\tau_0^2/2-qE(\tau_0+T)^2/2 }(\tau_0) A^{{\rm (gh)}*}_{{\bm p}_{\perp}, p_{\eta} }(\tau_0+T)  - B^{{\rm (gh)}*}_{{\bm p}_{\perp}, p_{\eta}+qE\tau_0^2/2-qE(\tau_0+T)^2/2 }(\tau_0) B^{{\rm (gh)}}_{{\bm p}_{\perp}, p_{\eta} }(\tau_0+T) \right] , \\
	&- ({}_- \Theta^{\rm (finite; out)}_{ {\bm p}_{\perp}, p_{\eta} } | {}_+ \Theta^{\rm (finite; in)}_{ {\bm p}'_{\perp}, p'_{\eta} } )_{\rm B} =   \left[ ( {}_+ \Theta^{\rm (finite; out)}_{ {\bm p}_{\perp}, p_{\eta} } | {}_- \Theta^{\rm (finite; in)}_{ {\bm p}'_{\perp}, p'_{\eta}} )_{\rm B} \right]^* \nonumber\\
		&=  \delta^2({{\bm p}_{\perp} - {\bm p}'_{\perp}}) \delta ( p'_{\eta} - (p_{\eta} + qE\tau_0^2/2 - qE (\tau_0+T)^2/2)  ) \nonumber\\
		&\ \ \ \ \times  \left[ - A^{\rm (gh)}_{{\bm p}_{\perp}, p_{\eta}+qE\tau_0^2/2-qE(\tau_0+T)^2/2 }(\tau_0) B^{{\rm (gh)}*}_{{\bm p}_{\perp}, p_{\eta} }(\tau_0+T)  + B^{{\rm (gh)}*}_{{\bm p}_{\perp}, p_{\eta}+qE\tau_0^2/2-qE(\tau_0+T)^2/2 }(\tau_0) A^{{\rm (gh)}}_{{\bm p}_{\perp}, p_{\eta} }(\tau_0+T) \right] .
\end{align}
\end{widetext}

\end{document}